%% file: toda-c.tex
\documentclass[reqno,a4paper]{amsart}
\usepackage{amssymb,graphicx,ifthen,calc,wrapfig,psfrag,comment,eepic,array,color,rotating}
\usepackage[all]{xy}
\xyoption{import}
\usepackage{hyperref}
\usepackage{caption}
\usepackage{longtable}

\def\shortfile{6}
\def\showkeys{0}
\ifthenelse{\equal{\showkeys}{0}}{}{\usepackage{showkeys}}
\input{definitions.tex}

\begin{document}

\title[Entropy and Toda Lattices]{Positive-Entropy Integrable Systems and the Toda Lattice, II}
\author{Leo T. Butler}
\date{\today}

\thanks{}

\begin{abstract}
This note constructs completely integrable convex Hamiltonians on the
cotangent bundle of certain $\T^k$ bundles over $\T^l$. A central role
is played by the Lax representation of a \BT\ lattice. The
classification of these systems, up to iso-energetic topological
conjugacy, is related to the classification of abelian groups of
Anosov toral automorphisms by their topological entropy function.

  \keywords{Geodesic flows, entropy, solvmanifolds, nonintegrability,
    integrable systems, convex hamiltonians,
    {\it AMS MSC(2000) classification}: 58F17, 53D25, 37D40}

\end{abstract}

\maketitle

\section{Introduction}
\label{int}

Say that a smooth flow $\varphi : M \times \R \to M$ is {\em
  integrable} if there is an open dense subset $L \subset M$ such that
  $L$ is fibred by $b$-dimensional tori and the smooth bundle
  coordinate charts $(I,\phi) : U \to {\bf D}^a \times \T^b$ conjugate
  $\varphi$ to a smooth translation-type flow $t\cdot(I,\phi) =
  (I,\phi + t \xi(I))$ on the fibres of $L$. This local form,
  classically known as action-angle coordinates, suggests that
  integrable flows are dynamically uninteresting. The example of the
  geodesic flow of a compact $3$-dimensional $Sol$ manifold which is
  completely integrable and has positive topological entropy, due to
  Bolsinov and Taimanov~\cite{BT:2000a}, is proof that this is not the
  case. The present paper generalises the examples
  of~\cite{BT:2000a,B7}. First, it shows how to construct integrable
  convex hamiltonian systems on cotangent bundles of certain
  solmanifolds in higher dimensions that are analogues to the $Sol$
  geometric $3$-manifolds when the monodromy group is not $\R$-split;
  second, it shows that Lax representations of \BT\ lattices are
  essential to construct these integrable systems, and moreover, the
  double-bracket Lax representations are essential to understand the
  dynamics on the singular set; third, the Lax map of a \BT\ lattice
  and the `momentum map' of a natural ${\mathcal F}$-structure on the
  solmanifold form a dual pair; and, finally, the topological
  classification of these integrable systems can be resolved by
  classifying abelian groups of Anosov toral automorphisms by the
  topological entropy function.

This appears to be a novel and interesting phenomenon: the
  construction of these integrable systems uses the machinery of Lax
  representations and R-matrices, while their dynamical classification
  uses machinery developed to understand hyperbolic dynamical systems.

Let us now sketch the constructions and results of the present paper.

\subsection{The $Sol$-manifolds} \label{ssec:cf} 
Let $A$ be a torsion-free, abelian group of diffeomorphisms of
$\T^b$. The group $A$ acts on $\T^b \times A_\R$, where
$A_\R = A \otimes_{\Z} \R$, via the diagonal action
\begin{align} \label{al:Aact}
  \forall \alpha\in A, y\in\T^b, x\in A_\R: && \alpha\star(y,x) :=
  (\alpha(y), x+\alpha \otimes 1 ).
\end{align}
This action is free and proper. The compact, smooth quotient is
denoted by $\Sigma$ or $\Sigma_A$. The fibring of $\Sigma$ by the tori
$\T^b$ equips $\Sigma$ with a natural ${\mathcal F}$-structure.

Henceforth it is assumed that $A < \GL{b;\Z}$ is an abelian group of
semi-simple elements, hence contained in a Cartan subgroup of
$\GL{b;\C}$, and therefore an exponential subgroup of
$\GL{b;\C}$. When $A$ is an exponential subgroup of $\GL{b;\C}$, the
universal cover $\ucSigma$ of $\Sigma$ admits the structure of a
solvable Lie group as follows: When $A$ is an exponential subgroup of
$\GL{b;\C}$, $A_\R$ is naturally identified with an abelian subgroup
of $\GL{b;\R}$. From this, there is a natural Lie group structure on
$\R^b \star A_\R =: \Smat$ and $\Z^b \star A =: \Smat_\Z$ is a lattice
subgroup of $\Smat$, \cf \ref{ssec:cons-sol-nt}. In the general case,
$A$ contains a finite-index exponential subgroup $A_2$~\cite[Theorem
4.28]{Raghunathan}. The finite covering $\Sigma_2$ of $\Sigma$ induced
by $A_2$ has a universal cover with a solvable Lie group structure; in
this case, the fundamental group of $\Sigma$ need not embed as a
subgroup in this universal cover, although it does act as a free and
proper group of deck transformations~\cite[pp.s
70-71]{Raghunathan}. An elementary argument shows that if $\Gamma$ is
a finite group of deck transformations and $\varphi$ is an integrable
flow on $M$ that is $\Gamma$-invariant, then the induced flow on
$M/\Gamma$ is integrable, also. So, to simplify the discussion in this
introduction, without losing generality, it will be assumed that $A$
is an exponential subgroup of $\GL{b;\C}$.

\subsection{Integrable geodesic flows} \label{ssec:in} 
Let $y_i$ be coordinates on $\C^n$ which diagonalise $A$. Define
complex-valued differential $1$-forms on $\Sigma$ by
\begin{align} \label{al:1f}
  \nu_i &= \exp(-\left\langle \ell_i , x \right\rangle)\, \d y_i,
  &\text{and \ }
  \eta_i &= \d x_i
\end{align}
where $\ell_i \in \Hom{A_\R;\R}$ is the linear form which maps $x \in
A_\R$ to the logarithm of the modulus of its $i$-th eigenvalue and
$x_i = \left\langle \ell_i , x \right\rangle$. A riemannian metric on
$\Sigma$ can be defined by
\begin{equation} \label{eq:rm}
  \metric = \ds\sum_{i,j} Q_{ij}\ \nu_i \cdot \nu_j + \ds\sum_{i,j} R_{ij}\
  \eta_i \cdot \eta_j + \ds\sum_{i,j} S_{ij}\ \eta_i \cdot \nu_j,
\end{equation}
where $Q, R$ and $S$ are constant, complex symmetric matrices chosen
so that $\metric$ is a real, symmetric, positive-definite
$(0,2)$-tensor. The metric $\metric$ is the general form of a
left-invariant metric on $\Smat=\R^b \star A_\R$. When the
off-diagonal term $S$ vanishes, the subgroups $\R^b$ and $A_\R$ are
orthogonal, totally geodesic and flat. By left-invariance of
$\metric$, each left translate of these two subgroups share these
properties.

\begin{question} \label{qu:1}
  Which metrics $\metric$ have a completely integrable geodesic flow?
\end{question}

Some answers are known. The examples of Bolsinov and Taimanov shows
that when $A$ is a cyclic group, then the geodesic flow is completely
integrable for $\metric$ with $S=0$ and $Q,R$ arbitrary
\cite{BT:2000a,BT:2000b}. The present author showed that when $A$ has
rank $b-1$ (so $A$ is $\R$-split), $Q_{ij} = \delta_{ij}
\epsilon_i^2$, $R$ is of a special form and $S=0$, then the geodesic
flow is completely integrable. To explain the special form of $R$,
write the Hamiltonian of $\metric$ in canonical coordinates:
\begin{equation} \label{eq:hrm}
  2H_\metric = \ds \sum_{ij} Q_*^{ij}\ \exp(\left\langle \ell_i+\ell_j
  ,x\right\rangle) + \ds \sum_{ij} R^{ij}\ p_{x_i}p_{x_j},
\end{equation}
where $Q_*^{ij}= Q^{ij}\,\ p_{y_i}p_{y_j}$ (no sum). Because $y_i$ is
a cyclic variable, $p_{y_i}$ is a first integral. $H_\metric$ reduces
to a family of \BT-like Hamiltonians in the canonical variables
$(x,p_x)$. If one diagonalises $Q$, then the complete integrability of
the \BT\ Hamiltonian dictates the form of $R$. The introduction of
\cite{B7} has an explicit example.%
\footnote{This is also referred to as the Toda lattice or the
  Kostant-Toda lattice, but Kostant in \cite{Kostant} attributes to
  Bogoyavlenskij~\cite{Bogo} the recognition of the role
  played by root systems of semisimple Lie algebras. }
The work of Adler \& Van Moerebeke and Kozlov \& Treschev suggests
that when $S=0$ the only completely integrable Hamiltonians
$H_\metric$ arise from \BT\ lattices or their deformations
\cite{AvM3,KT,Kozbook}.

The preceding argument glosses over a subtlety: the cyclic variables
$p_{y_i}$ are defined only on the universal cover. In the above cases,
one can construct smooth integrals that descend to the quotient; this
is true in general, but the difficulty lies in choosing $R$. This is
related to Lax representations.

\subsection{Lax Representations and momentum maps} \label{ssec:lax} 
\begin{center}
\begin{figure}[htb]
{
\psfrag{regular}{regular orbits}
\psfrag{singular}{\textcolor{red}{singular orbits}}
\psfrag{mod}{$\bmod \sim$}
\psfrag{reduced}{reduced space: $\Lie{\T^b}^*/ \sim$}
\psfrag{unreduced}{unreduced space: $\Lie{\T^b}^*$}
\includegraphics[width=10cm, height=7cm]{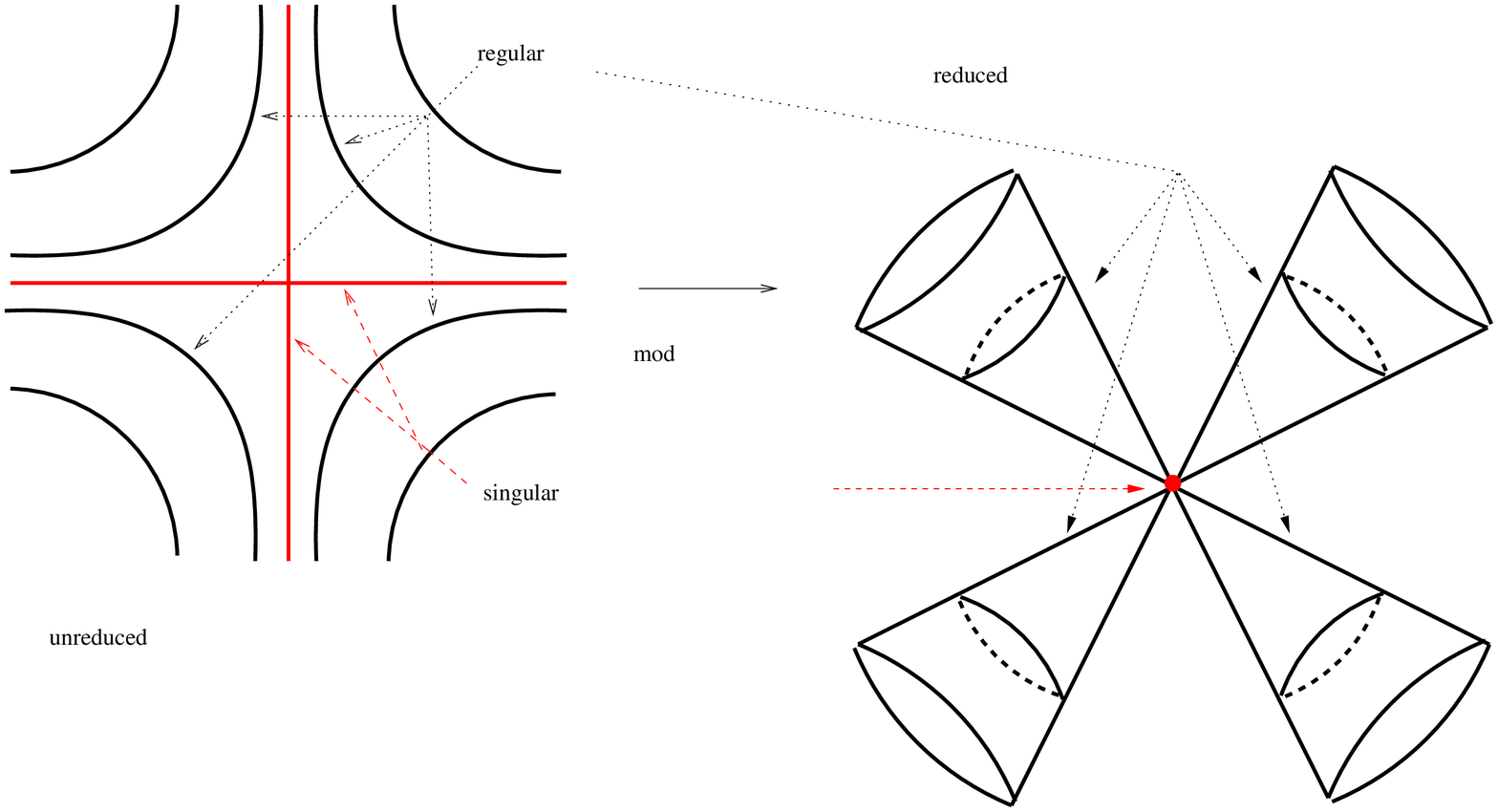}
\caption{The quotient map from $\Lie{\T^b}^* \to \Lie{\T^b}^*/ \sim$. The
regular points are points with a non-zero component in each eigenspace of $A$;
the singular set is the complement. } \label{fig:orbits}
}
\end{figure}
\end{center}

On the covering $\hat{\Sigma} = \T^b \times A_\R$, there is the
obvious free action of $\T^b$. This induces an $\F$-structure on
$\Sigma$, which one may think of as a locally-defined free action of
$\T^b$. The momentum map $\hat f$ of the $\T^b$-action induces a map
$f$ by equivariance, as illustrated in the right-hand side of
\eqref{eq:mmap}:
\begin{equation} \label{eq:mmap}
  \xymatrix{
    \L_R^* \ar[d] & T^* \hat{\Sigma} \ar[d]^{\hat{\Pi}} \ar[l]_{\hat\lax} \ar[r]^{\hat{f}} &
    \Lie{\T^b}^* \ar[d]_{\pmod A} \ar[rd]^{\pmod\sim}\\
    \L_R^* & T^* \Sigma \ar[l]_{\lax} \ar@/_5mm/[rr]^>>>>{f} \ar[r] & \Lie{\T^b}^*/A
    \ar[r]^{{\tiny\rm coll.}} & \Lie{\T^b}^*/\sim.
  }
\end{equation}
$\Lie{\T^b}^*/A$ is neither a smooth manifold nor a Hausdorff space
but it does contain an open and dense subspace that is a smooth
manifold. One can collapse the singular set of $\Lie{\T^b}^*/A$ to a
single point to define a Hausdorff topological space
$\Lie{\T^b}^*/\sim$, which is a smooth manifold outside of a single
point, as illustrated in figure \ref{fig:orbits}. Since the collapse
is $A$-invariant, the map $f$ is defined naturally. The map $f$ is a
first integral of $H_\metric$ and one can loosely think of $f$ as the
momentum map of the locally-defined $\T^b$ action on $T^*\Sigma$.

On the left of (\ref{eq:mmap}) is a map $\lax$, called a Lax matrix,
that is implicit in the identification of $H_\metric$ with a \BT\
Hamiltonian. The construction of a Poisson Lax map that Poisson
commutes with $f$ is the key difficulty in proving the complete
integrability of $H_\metric$.

\begin{question} \label{qu:2}
  What conditions on $A$ imply the existence of a Poisson Lax map $\lax$
  such that the $\T^b$-momentum map $f$ and $\lax$ form a dual pair?
\end{question}

Implicit in the two papers of Bolsinov and Taimanov is the fact that
if $A$ is cyclic, then this question is trivially
soluble. In~\cite[p. 529]{B7}, the present author shows that there is
a Poisson map that Poisson commutes with the $\T^b$-momentum map $f$
when $A < \GL{b;\Z}$ is $\R$-split and of finite index in its
centraliser (the relation to Lax maps is hinted at in the remark on
\cite[p. 529]{B7}). To generalise that construction, it appears
necessary to use the machinery of Lax representations.

\subsubsection{Positive topological entropy} \label{sssec:pe}
\begin{wrapfigure}[12]{r}[0pt]{0pt}
{
\psfrag{Tb0}{{\small $\T^b \times 0$}}
\psfrag{Tb1}{{\small $\T^b \times v$}}
\psfrag{geodesic}{\hspace{-1mm}{\small geodesic}}
\psfrag{vy}{\hspace{-3mm}\vspace{-2mm}{\small $v\cdot y$}}
\psfrag{y}{{\small $y$}}
  \includegraphics[width=3cm, height=3cm]{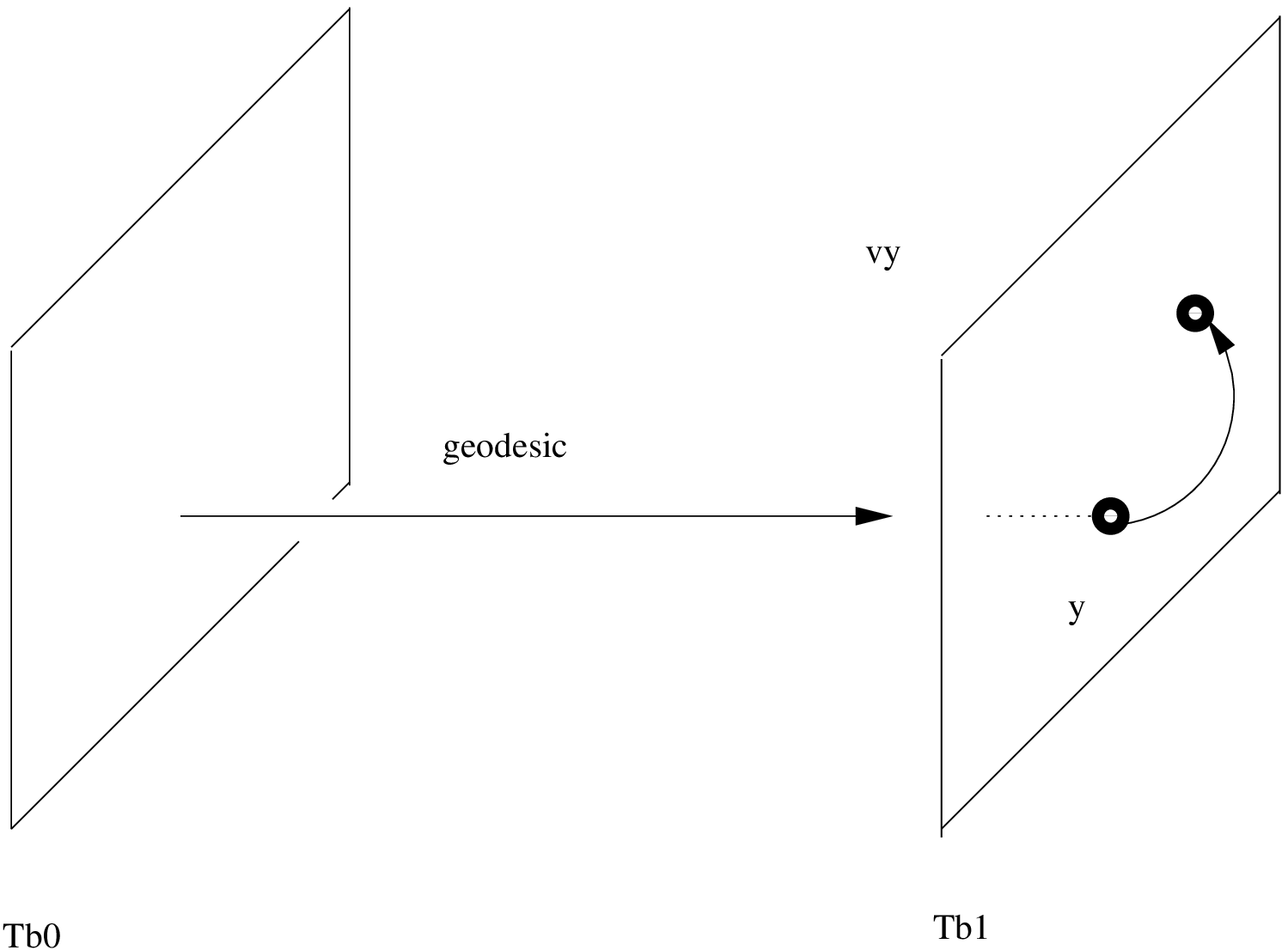}
}
\caption{A return map.} \label{fig:an}
\end{wrapfigure}The geodesic flow of $\metric$ must have positive topological entropy,
since $\pi_1(\Sigma)$ has exponential word
growth~\cite{Dinaburg}. When $S=0$, there is a direct proof of this:
since $A_\R$ is flat and totally geodesic in $\ucSigma$, as are all
its left-translates, each curve $t \mapsto tv+y$ for $v\in A_\R$,
$y\in\R^b$ is a geodesic. On $\Sigma$, for $v\in A$, the geodesic is
periodic and one sees that the geodesic flow induces the return map on
$\T^b$ defined by $y \mapsto v \cdot y$ -- which is a partially
hyperbolic, and generally Anosov, automorphism of $\T^b$ (see figure
\ref{fig:an}).

The appearance of such `subsystems' heavily constrains the topological
conjugacy class of a completely integrable geodesic flow of the form
of $\metric$.

\medskip

\subsection{Results} \label{sec:res} 
Let us sketch the main theorems of this paper. 

\subsubsection{Complete integrability} \label{sssec:res-1} 
\begin{definition} \label{de:max}
  A torsion-free abelian subgroup $A < \GL{b;\Z}$ is {\em maximal} if
  its elements are semisimple and it is of finite index in its
  centraliser.
\end{definition}

{{
\def\r{{\sf r}}
\def\c{{\sf c}}
\begin{theorem} \label{thm:1}
  If $A<\GL{b;\Z}$ is a maximal subgroup, then there is a Poisson Lax
  map $\lax$ such that (\ref{eq:mmap}) describes a dual pair, see
  (\ref{eq:L0-2}). In particular, there is a reversible {\em Finsler}
  metric on $\Sigma_A$ whose geodesic flow is completely integrable.

If, in addition, an irreducible element in $A$ has exactly $\r$ real
eigenvalues and $2\c$ non-real eigenvalues, then the geodesic flow of
the riemannian metric $\metric$ (\ref{eq:rm}) is completely integrable
when its Hamiltonian is defined as in (\ref{eq:HL0}) with root system
$\g^{(m)}$ in the cases described by Table \ref{tab:eig}.

In all cases, the singular set is a real-analytic
variety. Consequently, the integrable systems are {\em semi-simple} in
the sense of \cite{B5}.
\begin{center}
\begin{table}[!h]
{{
\setlength{\extrarowheight}{5pt}
\begin{tabular}{|p{1cm}p{1cm}p{1.7cm}l|p{1cm}p{1cm}p{1.7cm}|}
\hline
\r  & \c  & $\g^{(m)}$               && \r  & \c  & $\g^{(m)}$\\\hline\hline
$*$ & $0$ & $ \Ars{n}, \Dtwors{n+1}$ && $2$ & $1$ & $ \Dthreers$ \\
$0$ & $*$ & $ \Ars{n}, \Dtwors{n+1}$ && $2$ & $*$ & $ \Brs{n}, \Crs{n}$ \\   
$1$ & $*$ & $ \Atwors{2n} $          && $3$ & $*$ & $ \Atwors{2n-1} $ \\
    &     &                          && $4$ & $*$ & $ \Drs{n}$ \\\hline
\end{tabular}
}}
\caption[]{Conditions on eigenvalues and root systems which produce a
  riemannian metric with integrable geodesic flow ($*$ is an
  arbitrary positive integer).} \label{tab:eig}
\end{table}
\end{center}
\end{theorem}
}}

The first row in the upper left corner of the table summarises the
result of \cite{B7}. In the cases not covered in the table, it is
uncertain if $\Sigma_A$ admits a riemannian metric with completely
integrable geodesic flow -- the construction here yields only
completely integrable geodesic flows of reversible Finslers. If not,
one would have the first example of a compact smooth manifold that
admits a completely integrable reversible Finsler, but not a
riemannian, geodesic flow.

\subsubsection{Topological Entropy and Iso-energetic Topological Conjugacy} \label{sssec:res-2}
The tangent spaces to the $\T^b$-fibres of $\Sigma$ form a sub-bundle
$\Vert \subset T \Sigma$. Let $\Hstar \subset T^* \Sigma $ be the
annihilator of $\Vert$. The subspace $\Hstar$ is invariant under the
geodesic flow of Theorem \ref{thm:1} and that geodesic flow has
positive topological entropy on $\Hstar$. We prove

\begin{theorem} \label{thm:2}
Let $\Sigma=\Sigma_A$ be as in Theorem \ref{thm:1} and let $\varphi$
be the Hamiltonian flow on $T^*\Sigma$ induced by the Hamiltonian
$\Htoda$ defined in (\ref{eq:HL0}). Then
\begin{enumerate}
\item $\Hstar$ is a weakly normally hyperbolic invariant manifold. Its
  stable and unstable manifolds coincide and equal the pre-image of the
  equivalence class of $0$ in $\Lie{\T^b}^*/\sim$ under the
  $\T^b$-momentum map $f$;
\item the topological entropy of $\varphi | \Htoda^{-1}(\frac{1}{2})$
  equals that of $\varphi | \Hstar \cap \Htoda^{-1}(\frac{1}{2})$, when $\Htoda$
  is induced by the $\Ars{n}$ \BT\ lattice;
\item in all cases, the topological entropy of $\varphi | \Hstar \cap
  \Htoda^{-1}(\frac{1}{2})$ is calculable (see Table \ref{tab:entropy}).
\end{enumerate}
\end{theorem}

The construction of the Lax map in theorem \ref{thm:1} is unique up to
the action of a permutation group. If $\phi_1, \phi_2$ are two such
permutations and $\varphi_1, \varphi_2$ are the resulting geodesic
flows, an interesting question is whether these flows are
topologically distinct. A topological invariant, namely the marked
homology spectrum, does distinguish these flows in many cases even
when topological entropy cannot. To explain, let
\begin{align} \label{al:h}
  \hr(v) &= h_{top}(v) && \hr : A \to \R
\end{align}
be the entropy function, where $v \in A$ is viewed as a
$\T^b$-automorphism.

\begin{theorem} \label{thm:3}
  If there is a topological conjugacy of $\varphi_1, \varphi_2$ on their
  respective unit sphere bundles, then there is an automorphism $\f : A
  \to A$ such that
  \begin{equation} \label{eq:f}
    \hr \circ \f = \hr.
  \end{equation}
If the number-theoretic closure of $A$ is a group of Anosov
automorphisms, then $\f$ is induced by a Galois automorphism.
\end{theorem}

In many cases, the group of Galois automorphisms is trivial, which
implies that each of the constructed Hamiltonian flows must be
topologically non-conjugate. In general, one should not expect the
number-theoretic closure of $A$ to be a group of Anosov automorphisms,
though.

\begin{question} \label{qu:3}
  Which automorphisms of $A$ fix the entropy function $\hr$?
\end{question}

This leads to a further question, whose formulation is somewhat
technical and is deferred to section \ref{sec:an}, question
\ref{qu:6}. Finally, theorem \ref{thm:summary} provides information on
the number of distinct topological conjugacy classes of integrable
Hamiltonian flows provided by theorem \ref{thm:1}.

Question \ref{qu:3} is a rigidity question: to what extent does the
entropy of an action determine that action. An approach to this
question is to ask which embeddings of $\Z^a \cong A$ into $\GL{b;\Z}$
have equal entropies. Katok, Katok and Schmidt give examples of
iso-entropic actions of $\Z^2$ on $\T^3$ by maximal subgroups of
$\GL{3;\Z}$ that are conjugate in $\GL{3;\Q}$ but not conjugate in
$\GL{3;\Z}$~\cite{KKS}. However, the suspension manifolds of these
actions are not homotopy equivalent. Indeed, if $A' < \GL{b;\Z}$ is
not conjugate to $A$ in $\GL{b;\Z}$, then $\pi_1(\Sigma_{A'})$ is not
isomorphic to $\pi_1(\Sigma_A)$. %
%
% To see this claim, it suffices to prove that if $A$ and $A'$ act
% strongly irreducibly on $\Z^b$, then $\Z^b$ is the maximal normal
% abelian subgroup of $\Z^b \star A$ (resp. '). Therefore, a group
% isomorphism between the two groups must map $\Z^b$ to itself, which
% implies that the subgroups $A$ and $A'$ are conjugate over $\GL{b;\Z}$.
%
Thus, question \ref{qu:3} is somewhat finer than the iso-entropic
rigidity problem examined in \cite{KKS}.

\subsection{Two additional questions} \label{ssec:coq}

Let $\Sigma$ be a torus bundle over a torus of the type described in
section \ref{ssec:cf}. $\Sigma$ is aspherical and the fundamental
group of $\Sigma$ is a poly-$\Z$ group%
\footnote{That is, there is a sequence of subgroups $0=\D_m \lhd
  \D_{m-1} \lhd \cdots \lhd \D_0 = \D$ such that $\D_i/\D_{i+1} \cong
  \Z$ for all $i$.}, so Theorem 15B.1 of~\cite{Wall} implies that
$\pi_1(\Sigma)$ determines $\Sigma$ up to homeomorphism. The {\em
  standard} smooth structure on $\Sigma$ is defined by the above
construction. In general, the topological manifold $\Sigma$ may
admit several inequivalent smooth structures. 

\begin{question} \label{qu:4}
  Which smooth structures on the topological manifold $\Sigma$ admit a
  riemannian or Finsler metric whose geodesic flow is completely
  integrable?
\end{question}

This question is already quite interesting when $A=1$ and $\Sigma$ is
a torus since \cite{B8} shows that the integrals cannot all be
real-analytic if the smooth structure is non-standard. It is unknown
if there are analogous obstructions when $A \neq 1$.

And, finally,

\begin{question} \label{qu:5}
  What conditions on $A<\GL{b;\Z}$ imply that $\Sigma_A$ admits a
  riemannian or Finsler metric whose geodesic flow is completely
  integrable?
\end{question}

Theorem \ref{thm:cas} shows that there are natural examples of groups
$A$ that are not maximal, yet $\Sigma_A$ admits a completely
integrable Finsler. These examples are constructed using symmetries
provided by number-theoretic considerations. The difficulty in the
general case, where there are no obvious symmetries, is the
construction of the Lax map $\lax[]$ appears to break down.

\ifthenelse{\shortfile=1}{\end{document}}{}

\section{Notation and Preliminary Definitions}
\subsection{Integrability} \label{sec:pm}
{ 
\def\F{{\mathcal A}} 

The present paper's definition of complete integrability follows that
of \cite{BolsinovJovanovic:2001a,B7}. 

Let $\Sigma$ be a real-analytic manifold. The set of smooth functions
on the cotangent bundle of $\Sigma$, $C^{\infty}(T^* \Sigma)$, has two
canonical algebraic structures: it is an abelian algebra when equipped
with the natural operations of point-wise addition and multiplication;
and, coupled with the canonical {\em Poisson bracket}, $\{,\}$, $(
C^{\infty}(T^*\Sigma), \{,\} )$ is a Lie algebra of derivations of the
algebra $C^{\infty}(T^*\Sigma)$. A hamiltonian $H \in C^{\infty}(T^*
\Sigma)$ induces a vector field $Y_H := \{\,, H\}$. For $\F \subset
C^{\infty}(T^* \Sigma)$ and $P \in T^* \Sigma$, let $d\F_P = {\rm
span\, } \{df_P\ :\ f \in \F\}$ and let $Z(\F) = \{ f \in \F\ :\
\{\F,f\} \equiv 0\}$. Let $k = \sup_P \dim d\F_P$, $l = \sup_P \dim
dZ(\F)_P$. Let us say $P \in T^* \Sigma$ is $\F$-{\em regular} if
there exist $f_1,\ldots,f_k \in \F$ such that $P$ is a regular value
for the map $F = (f_1,\ldots,f_k)$ and $f_1,\ldots,f_l \in Z(\F)$; if
$P$ is not $\F$-regular then it is $\F$-{\em critical}. Let $L(\F)$ be
the set of $\F$-regular points. $H$ is assumed to be proper.

\begin{definition}[\cf \cite{BolsinovJovanovic:2001a}] \label{def:int}
$H \in C^{\infty}(T^* \Sigma)$ is {\em integrable} if there
is a Lie subalgebra $\F \subset C^{\infty}(T^*\Sigma)$ such that:
\begin{enumerate}
\item $H \in Z(\F)$;
\item $k + l = {\rm dim}\, T^* \Sigma$ and $L(\F)$ is an open and dense
subset of $T^* \Sigma$.
\end{enumerate}
If $k=l=\dim \Sigma$, we will say that $H$ is {\em completely
integrable}.
\end{definition}

Bolsinov and Jovanovic~\cite{BolsinovJovanovic:2001a} introduced this
definition of complete integrability. The standard definition of
complete integrability (resp. non-commutative integrability) are
special cases of Definition~\ref{def:int} with $\F = {\rm span}\ \{
f_1,\ldots,f_k\}$ and $l=k$ (resp. $l \leq k$) and the regular-point
set of $F = (f_1,\ldots,f_k)$ is dense. Definition~\ref{def:int} is
both more intrinsic, and more suited to the examples of the present
paper. Note that the present definition of integrability is equivalent
to that of Dazord \& Delzant~\cite{DD}.  }

\subsection{Construction of the solmanifolds and number theory} \label{ssec:cons-sol-nt}

There is a well-known correspondence between abelian subgroups of
$\GL{b;\Z}$ and groups of units in algebraic number fields of degree
$d$ dividing $b$~\cite{KKS}. The present paper exploits this
correspondence extensively. The following section establishes notation
that is used throughout. In terms of the terminology in the
introduction, we use the following translation table:
\begin{center}
  \begin{tabular}{p{4cm}lp{7cm}}
    abelian $A < \GL{b;\Z}$ & $\to$ & a group of units in the field
    generated by the eigenvalues of all $a \in A$;\\
    $\Z^b$ & $\to$ & a direct sum of copies of a subgroup of the integers of a number field.
  \end{tabular}
\end{center}

\subsubsection{Preliminaries}

Let 
$$\Q \subset F \stackrel{\iota}{\subset} E$$ be an inclusion of
algebraic number fields. For a field extension $E/F$ let the set of
embeddings of $E$ into $\C$ which fix $F$ be denoted by $\G{E/F}$; we
adopt the convention $/\Q$ is omitted. Define vector spaces
\begin{equation} \label{eq:W}
  \W{E} = \sum_{\sigma \in \G{E} } \C\sigma,
\end{equation}
and
\begin{equation} \label{eq:V}
  \V{E} = \{ x \in \W{E}\ :\ x_{\bar{\sigma}} = \bar{x}_{\sigma}\
  \forall \sigma\in\G{E}\ \},
\end{equation}
where $\bar{\ }$ denotes complex conjugation and $\bar{\sigma}$ is the
embedding $\sigma$ {\em followed} by complex conjugation. We also define
\begin{equation} \label{eq:V0}
  \Vo{E} = \{ x \in \V{E}\ :\ \sum_{\sigma \in \G{E}} x_{\sigma} =
  0,\ \&\ x_{\sigma} = x_{\bar{\sigma}}\ \forall \sigma\in\G{E}\ \}.
\end{equation}
$\G{E}$ is a basis of $\V{E}$ which induces the dual basis $\G{E}^*$
of $\V{E}^*$. An element in the dual basis shall be denoted by $\hat
\sigma$ for $ \sigma \in \G{E}$. The basis and dual basis establish a
linear isomorphism between $\V{E}$ and $\V{E}^*$ which shall be
denoted by the circumflex operator, $\V{E} \to \V{E}^*:\ x \mapsto
\hat x$, whose inverse is $\V{E}^* \to \V{E}:\ x \mapsto \check x$.

One obtains a basis of $\Vo{E}^*$ as follows: note that $\t =
\frac{1}{|\G{E}|} \sum_{ \sigma \in \G{E} } \hat\sigma$ and $\hat
\sigma - \hat{\bar{\sigma}}$ vanish on $\Vo{E}$ for all $\sigma \in
\G{E}$. If one defines $\G{E}^r$ to be the set of real embeddings of $E$ and
$\G{E}^c$ to be one-half of the non-real embeddings such that
$\G{E}^c$ is disjoint from its complex conjugate, then one observes
that
\begin{align} \label{al:voe}
  \begin{split}
    \Vo{E}^{\perp} &= \R\cdot \t \oplus \sum_{\sigma \in \G{E}^c}
    \R\cdot(\hat \sigma - \hat{\bar{\sigma}}),\\
    \Vo{E}^* &= \sum_{\sigma\in\B{F}} \R \cdot \hat \sigma|_{\Vo{E}} 
  \end{split}
\end{align} 
where $\B{E} = \G{E}^r \cup \G{E}^c$. 

The inclusion $F \stackrel{\iota}{\subset} E$ induces
\begin{align}  \label{al:1}
  \begin{array}{llccrcl}
    & \xymatrix{ \V{E} \ar@{->>}[r]^{\iota^*} & \V{F} } & \hspace{10mm} &
    \textrm{where\ } &  \iota^*( \sigma ) &=&
    \sigma|_F,\\
    \textrm{and\ } & \xymatrix{\V{E}^* \ar@{<-^{)}}[r]^{\iota} & \V{F}^*} & &
    \textrm{where\ } & \iota( \hat \tau  ) &=&
    \displaystyle\sum_{ \sigma \in \G{E}, \sigma|_F = \tau } \hat\sigma.
  \end{array}
\end{align}
Finally, define a map $\xymatrix{ \V{E}^* \ar@{->>}[r]^(.4){ \alpha } &
  \Vo{F}^*}$ by
\begin{alignat}{2} \label{al:alpha}
  \alpha &= \jj^* \, \hat{\iota}^*, &\hspace{5mm} &\hat \sigma \mapsto
  \left. \hat \tau\right|_{\Vo{F}}
  \textrm{where\ } \tau=\sigma|_F,
\end{alignat}
$\jj^*$ is the adjoint of the inclusion map $\Vo{F}
\stackrel{\jj}{\subset} \V{F}$ and $\hat\iota^* = \hat{\ }\iota^*\check{\
}$. This allows one to define a pairing between $\V{E}^*$ and
$\Vo{F}$, denoted as follows
\begin{align} \label{al:pairing}
  \begin{array}{rclcl}
    \left\langle \hat \sigma  , x  \right\rangle\hspace{-2mm} &:=& \hspace{-2mm}\left\langle
    \alpha(\hat \sigma ) , x \right\rangle &\hspace{20mm}& \forall \sigma
    \in \G{E}, x \in \Vo{F},\\
    &\ =& \hspace{-2mm}\left\langle \hat \tau  , x  \right\rangle &\hspace{20mm}&
    \textrm{where\ } \tau=\sigma|_F.
  \end{array}
\end{align}
Since $x=\sum_{ \tau \in \G{F} } x_{ \tau } \cdot \tau$, it is
apparent that
\begin{equation} \label{eq:pairing2}
  \left\langle \hat \sigma , x \right\rangle =
  x_{\left(\sigma|_F\right) } \hspace{20mm} \forall \sigma \in \G{E}, x
  \in \Vo{F},
\end{equation}
so the notation is natural.

\begin{comment}

  \medskip
  \noindent{\bf Example.} Let $\zeta$ be a $7$-th root of unity and let
  $E=\Q(\zeta)$ and $F=\Q(\zeta+\bar{\zeta})$ be the maximal totally
  real subfield of $E$. Then $\G{E} = \{ \sigma^n\ :\
  n=\pm 1,\pm 2, \pm 3\}$, where $\sigma(\zeta) := \zeta^5$ and
  $\bar{\sigma}=\sigma^{-1}$, while $\G{F} = \{ \tau,\tau^2,\tau^3 \}$,
  where $\tau=\left.\sigma\right|_F$. We then have that
  \begin{equation} \label{eq:ExV}
    \V{E} = \left\{ x = \sum_{n=1}^3 x_{ \sigma^n } \sigma^n + \bar{x}_{
      \sigma^n } \sigma^{-n}\ : \ x_{ \sigma^n } \in \C\   \right\}, 
  \end{equation}
  and
  \begin{equation} \label{eq:ExV0}
    \Vo{F} = \left\{ x = 2 x_{ \tau }  s_{ \tau } + 2x_{
      \tau^2 } s_{ \tau^2 } + 2x_{ \tau^3 } s_{ \tau^3 } :\ x_{ \tau^n }
    \in \R,\ \sum_{n=1}^3 x_{ \tau^n }  = 0\
    \right\},
  \end{equation}
  which is linearly isomorphic to $\Vo{E}$.
  \medskip

\end{comment}

\subsection{ An embedding of $\O_E$ in $\V{E}$ }
Let $\O_E$ be the ring of integers of $E$, and let $\U_E$ be the group
of multiplicative units of $\O_E$. Define a map $\eta : \O_E \to
\V{E}$ by
\begin{equation} \label{eq:eta}
  \eta( \alpha ) := \sum_{ \sigma \in \G{E} } \sigma( \alpha ) \cdot
  \sigma,
\end{equation}
for each $ \alpha \in \O_E$. 

\begin{lemma} \label{lem:eta}
  The map $\eta$ is an embedding whose image---call it $\Eta{E}$---is a discrete, cocompact
  subgroup of $\V{E}$.
\end{lemma}

\begin{proof}
  This is standard.
\end{proof}

Let $\Tf{E} = \V{E}/\Eta{E}$ be the resulting torus. $\Tf{E}$ is
equipped with a canonical affine structure from $\V{E}$ and the group
$\U_E$ acts by automorphisms of $\Tf{E}$ defined by
\begin{equation} \label{eq:uact}
  u\cdot\y = \sum_{ \sigma \in \G{E} } \sigma(u) \cdot \y_{\sigma}
  \cdot \sigma \ +\ \Eta{E},
\end{equation}
where $\y = \sum_{ \sigma \in \G{E} } \y_{\sigma} \cdot \sigma \ +\
\Eta{E}$ is an element in $\Tf{E}$ and $u \in \U_E$. The action in
(\ref{eq:uact}) is well-defined since $\Eta{E}$ is mapped to itself by $\U_E$.
{\it A fortiori}, equation (\ref{eq:uact}) also defines an action of
$\U_F \subset \U_E$ as an abelian group of automorphisms of $\Tf{E}$.

\subsection{ An embedding of $\U^+_F$ in $\Vo{F}$ }
Define a map $\ell : \U_F \to \V{F}$ by
\begin{equation} \label{eq:ell}
  \ell(u) = \sum_{ \sigma \in \G{F} } \ln| \sigma(u) | \cdot \sigma.
\end{equation}
Since $\bar{ \sigma }$ is $\sigma$ followed by complex conjugation, it
is clear that $\logu{F}=:\im \ell \subset \Vo{F}$. Dirichlet's theorem on the
group of units of an algebraic number field characterises the image of
$\ell$ as a discrete, cocompact subgroup of $\Vo{F}$, while $\ker
\ell =: \kerl_F$ is the set of units all of whose conjugates lie on the unit
circle. Stated otherwise, there is an unnatural splitting of $\U_F$
via a commutative diagram
$$
\xymatrix{
  \kerl_F \ar@{^{(}->}[r]\ar[d]^{=} & \U_F \ar@{->>}[r]\ar[d]^{=} & \U_F/\kerl_F
  \ar[r]^{\ell}_{\cong}\ar[d]^{\cong} & \logu{F}\ar[d]^{\cong},\\
  \kerl_F \ar@{^{(}->}[r] & \kerl_F \oplus \U^+_F \ar@{->>}[r] & \U^+_F 
  \ar[r]^{\cong} & \Z^{r+c-1},\\
}
$$ where $r$ (resp. $2c$) is the number of real (resp. non-real)
embeddings of $F$. When $F$ has a real embedding, which one may take
to be the identity embedding $F\subset\C$, then $\kerl_F = \{\pm 1\}$
and $\U^+_F$ may be taken to be the multiplicative group of positive
units in $\U_F$ --- hence the notation. To summarise

\begin{lemma} \label{lem:ell}
  The image of the map $\ell : \U^+_F \to \Vo{F}$ --- call it $\logu{F}$
  --- is a discrete, cocompact subgroup of $\Vo{F}$ isomorphic to $\U^+_F$.
\end{lemma}

\subsection{ An action of $\U^+_F$ on $\Tf{E} \times \Vo{F}$} 
For $\y \in \Tf{E}$, $\x \in \Vo{F}$ and $u \in \U^+_F$ define 
\begin{equation} \label{eq:udact}
  u \cdot (\y,\x) := (u \cdot \y, \x + \ell(u)).
\end{equation}
This action is clearly free and proper. Let $ \Sigma $
denote the compact manifold obtained by quotienting $\Tf{E} \times
\Vo{F}$ by this action of $\U^+_F$. 

\begin{lemma} \label{lem:sigma}
  There is a commutative diagram of natural maps
  \begin{equation} \label{eq:a}
    \begin{split}
      \xymatrix@!R@R=6pt{
        \V{E} \times \Vo{F} \ar@{->>}[r] \ar@{^{(}->>}[dd]^{=} & \Tf{E} \times
        \Vo{F} \ar@{->>}[r] \ar@{^{(}->>}[dd]^{=} & ( \Tf{E} \times
        \Vo{F})/\U^+_F \ar@{^{(}->>}[dd]^{=} \\
        \\
        \tilde{ \Sigma } \ar@{->>}[r]^{\tilde{\pi}} & \hat{ \Sigma } \ar@{->>}[r]^{\hat\pi} & \Sigma. 
      }
    \end{split}
  \end{equation}
  Therefore, $\pi_1( \Sigma )$ is naturally isomorphic to the
  semi-direct product $ \Delta = \U^+_F \star \O_E$, while there is a
  natural fibring of $\Sigma$ by tori over a torus
  \begin{equation} \label{eq:tof}
    \xymatrix@!R@R=6pt{
      \Tf{E}\ \ar@{^{(}->}[r] & \Sigma \ar@{->>}[r]^(.4){\p} & \Tf{o,F},
    }
  \end{equation}
  where $\Tf{o,F} = \Vo{F}/\logu{F}$.
\end{lemma}

\begin{proof}
  Naturality of the construction implies the lemma.
\end{proof}

\subsection{ The cotangent bundle $T^* \Sigma$}
The vector space structures on $\V{E}$ and $\Vo{F}$ give a
tautological trivialisation of their cotangent bundles. Lemma
\ref{lem:sigma} therefore implies that there is a commutative diagram
\begin{equation} \label{eq:cot_sigma}
  \begin{split}
    \xymatrix@!R@R=6pt{
      \V{E}^* \times \V{E} \times \Vo{F}^* \times \Vo{F} 
      \ar@{->>}[r] \ar@{^{(}->>}[dd]^{=} & 
      \V{E}^* \times \Tf{E} \times \Vo{F}^* \times \Vo{F} 
      \ar@{->>}[r] \ar@{^{(}->>}[dd]^{=} & 
      (\V{E}^* \times \Tf{E} \times
      \Vo{F}^* \times \Vo{F})/\U^+_F \ar@{^{(}->>}[dd]^{=} \\
      \\
      T^*\tilde{ \Sigma } \ar@{->>}[r]^{\tilde{\Pi}} & T^*\hat{ \Sigma } \ar@{->>}[r]^{\tilde\Pi} & T^*\Sigma,
    }
  \end{split}
\end{equation}
where $\hat\Pi$ is the covering map induced by $\hat\pi$, etc. Let us
introduce coordinates on $T^* \hat\Sigma $ by
$$P \in T^* \hat\Sigma \iff  P = (\py, \y + \Eta{E}, \px, \x) \in
\V{E}^* \times \Tf{E} \times \Vo{F}^* \times \Vo{F}.$$ The action of
$\U^+_F$ on $T^*\hat \Sigma$ is the natural lift of the action on $\Sigma$ 
\begin{equation} \label{eq:up}
  u \cdot P = ( u \cdot \py, u \cdot \y + \Eta{E}, \px, \x + \ell(u))
\end{equation}
where $u\cdot \y$ is defined in equation (\ref{eq:uact}) and $u
\cdot \py = \sum_{ \sigma \in \G{E} } \py_{ \sigma }\cdot
\sigma(u)^{-1} \cdot \hat \sigma$ is the induced contragredient
action.

\subsection{Functions on $T^*\Sigma$}
The function $P \mapsto \px$ is $\U^+_F$-invariant, so one may view
$\px$ as a submersion $\onto[]{T^* \Sigma}{\Vo{F}^*}$. 

Fix a positive integer $b_{\sigma}$ for each $\sigma \in \G{E}$ and
define the function
\begin{equation} \label{eq:gamma}
  \gamma_{\sigma}(P) := \exp(b_{\sigma}\cdot\left\langle \hat \sigma , \x \right\rangle ) \times \left|\py_{\sigma}\right|^{b_{\sigma}}
\end{equation}
where the pairing $\left\langle \hat \sigma , \x \right\rangle$ is
defined in equation (\ref{al:pairing}). 

\begin{lemma} \label{lem:gamma}
  The function $\gamma_\sigma$ is $\U_F$-invariant and it is
  real-analytic if $b_\sigma$ is even.
\end{lemma}
\begin{proof}
  From equation (\ref{eq:up}), we know that for each $u \in \U_F$
  \begin{equation} \label{eq:gamma1}
    \gamma_{\sigma}(u\cdot P) = \gamma_{\sigma}(P) \times \exp(b_{\sigma}
    \ln|\sigma(u)|) \times |\sigma(u)|^{-b_{\sigma}} = \gamma_{\sigma}(P).
  \end{equation}
  It is clear that $\exp(b_{\sigma}\cdot\left\langle \hat \sigma , \x
  \right\rangle )$ is real-analytic, and $|\py_\sigma|^{b_\sigma}$ is
  real-analytic if $b_\sigma$ is a positive even integer.
\end{proof}

\begin{remark} \label{re:mom}
  {\rm

    Fix even integers $b_\sigma$ as in lemma \ref{lem:gamma}. One may
    define a {\em momentum-like} map $\lambda : T^*\Sigma \to \Vo{F}^*
    \oplus \V{E}^*$ by
    \begin{align} \label{al:momap}
      \lambda(P) &= \px \oplus \sum_{\sigma \in \G{E}}
      \gamma_{\sigma}(P)\cdot \hat{\sigma}, &\forall P\in T^*\Sigma.
    \end{align}
    When the universal cover $\tilde{ \Sigma }$ of $\Sigma$ admits the
    structure of a solvable Lie group with $\Delta$ as a lattice subgroup,
    $\Vo{F}^* \oplus \V{E}^*$ -- as the dual of a Lie algebra -- admits a
    canonical Poisson structure. In this case, the map $\lambda$ is
    left-invariant and Poisson and therefore mimics the properties of the
    classical momentum map.

  }
\end{remark}

\ifthenelse{\shortfile=2}{\end{document}}{}

\section{Lax Representations} \label{sec:lax}

\subsection{Real split affine Lie algebras} \label{ssec:split}
Let us briefly recall the construction underlying the Lax
representation of periodic \BT\  lattices. This
discussion follows that in~\cite{RSTS,AvM1,AvM2}. Let $\g$ be a simple
real Lie algebra with the real split Cartan sub-algebra $\h$; $\g$ is
also known as the real normal form of the simple complex Lie algebra
$\g\otimes\C$. The Cartan-Killing form of $\g$ is denoted by
$\left\langle\langle , \right\rangle\rangle$ when viewed as a bilinear
form on $\g$, and it is denoted by $\kappa$ when viewed as a linear
isomorphism of $\g$ with $\g^*$. Recall that $\left\langle\langle ,
\right\rangle\rangle$ is non-degenerate on $\h$. As $\h$ is a real
split Cartan sub-algebra, $\g$ decomposes as
\begin{equation} \label{eq:rsd}
  \g = \h + \sum_{r \in \Psi_*} \g_r
\end{equation}
where $\Rs_* \subset \h^*$ is the set of roots and $\g_r$ is the root
space associated with $r$, $\g_r = \left\{ x \in \g\ :\ \ad{h}x =
\left\langle r , h \right\rangle\, x \ \ \ \forall h\in\h
\right\}$. There is a set of simple roots $\Rs_0 \subset \Rs_*$ such
that every root is an integer linear combination of the roots in $\Rs_0$
with entirely non-negative or non-positive coefficients. The height of
a root is the sum of these coefficients; there is a unique root,
$\minrt$, of minimal height. Let $\Rs$ be $\Rs_0 \cup \left\{ \minrt
\right\}.$

Define $\L$ to be the set of Laurent polynomials in the variable
$\lambda$ with coefficients in $\g$; $\L$ inherits an obvious Lie
algebra structure. Let $\d = \lambda\, \frac{\partial\ }{\partial
  \lambda}$ be a derivation; define $[\d, x\cdot \lambda^n ] = n x \cdot
\lambda^n$ for all integers $n$ and $x \in\g$. Then $\hat\g = \L + \R
\cdot \d$ is a real split Lie algebra with Cartan sub-algebra $\hat\h =
\h + \R \cdot \d$. The Cartan sub-algebra induces a weight-space
decomposition of $\L$ as
\begin{equation} \label{eq:wsg}
  \L = \h + \sum_{\r \in \RsL_*} \L_{\r}
\end{equation}
where $\RsL_* = \left\{ \r \in \hat\h^* :\ {\left. \r \right|_{\h}}
\in \Rs_* \cup \{0\}, \left\langle \r , \d \right\rangle \in \Z, \r
\neq 0 \right\}$.  The weight set $\RsL_*$ has a basis of simple
weights $\RsL = \Rs \cup \left\{ \bminrt \right\}$, where
$\left. \bminrt \right|_{\h}=\minrt$ and $\left\langle \bminrt ,\d
\right\rangle=1$.  Each $\r \in \RsL_*$ is an integer linear
combination of roots in $\RsL$. By defining the height of $\r$ as the
sum of these coefficients one obtains the principal grading
\begin{equation} \label{eq:pg}
  \L = \sum_{n \in \Z} \L_n, 
\end{equation}
where $\L_0=\h$, $\L_n = \sum_{\htt{\r} = n} \L_{\r}$
otherwise, and $[\L_n,\L_m] \subseteq \L_{m+n}$ for all $m,n$. 
It is observed that
\begin{equation} \label{eq:lone}
  \L_{\pm 1} = \g_{\minrt} \lambda^{\pm 1} + \sum_{r \in \Rs} \g_{\pm r}
  = \sum_{\r \in \RsL} \R \cdot \e_{\pm \r}
\end{equation}
the same sign appearing throughout, and $\e_{\r}$ is a vector
normalised so that \newline $\kappa\cdot[\e_{\r}, \e_{-\r}] \in \Rs_0$. The
sub-algebras $\L_+ = \sum_{n\geq 0} \L_n, \L_- = \sum_{n<0} \L_n$
permit the definition of a second Lie algebra structure on $\L$,
defined by
\begin{equation} \label{eq:rbracket}
  [x,y]_R := [x_+,y_+] - [x_-,y_-]
\end{equation}
for $x=x_+ + x_+,y=y_-+y_+ \in \L_- \oplus \L_+$. The Cartan-Killing
form $\kappa$ allows one to identify $\L_n^* = \L_{-n}$ for all $n$,
in such a way that $\kappa(\e_{\r}) = \e_{-\r}$ or
$\left\langle\langle \e_{\r} , \e_{-\r[s]} \right\rangle\rangle =
\delta_{{\r},{\r[s]}}$. Indeed, note that $\e_{\r} = \e_r \cdot
\lambda^n$ where $r=\left. \r \right|_{\h}$ and $n=\left\langle \r ,
\d \right\rangle$ for all roots $\r \in \RsL$, so it suffices to find
a suitable basis of $\g$ in order to define the vectors $\e_{\r}$. One
also knows that
\begin{lemma} \label{lem:pss}
  For each $\mu \in \L_{-1}^*$, the affine subspace $\mu + \L_0^* +
  \L_1^*$ is a Poisson subspace of $\L^*_R$. The Casimirs of $\L^*$ are
  in involution on $\L^*_R$.
\end{lemma}
\begin{proof}
  See references~\cite{RSTS,AvM1}.
\end{proof}

\subsection{A second splitting} \label{ssec:car}
Let $\Lnot = \h + \sum_{\r \in \RsL_*} \R\cdot \e_{\r}$, a sub-algebra
of the loop algebra $\L$ on which the Cartan-Killing form is
non-degenerate. One can distinguish two sub-algebras $\Lnot_{\pm}$ such
that $\Lnot = \Lnot_{-}\oplus \Lnot_{+}$ as a vector space:
\begin{equation} \label{eq:split}
  \left.
  \begin{array}{ll}
    \Lnot_{-}=\sum_{\r \in \RsL_+} \R\cdot(\e_{\r}-\e_{-\r}),&
    \Lnot_{+} = \h + \sum_{\r \in \RsL_+} \R\cdot\e_{\r},\hspace{3mm}\textrm{so}\\
    \Lnot_{-}^* \equiv \Lnot_{+}^{\perp} = \sum_{\r \in \RsL_+} \R\cdot\e_{\r}, &
    \Lnot_{+}^* \equiv \Lnot_{-}^{\perp} = \h + \sum_{\r \in \RsL_+} \R\cdot(\e_{\r}+\e_{-\r}),
  \end{array}
  \right.
\end{equation}
where $\RsL_+\subset \RsL_*$ is the set of positive roots. One can
define a grading on both $\Lnot_{\pm}$ by defining the height of a
root $\r\in\RsL_+$ to be $\htt{\r}=\htt{r}+(1+k)\left\langle \r , \d
\right\rangle$ where $k$ is the height of the maximal root of
$\g$. With this grading, a basis of $\Lnot_{+1}$ (resp. $\Lnot_{-1}$)
is $\left\{ \e_{\r}\,:\,\r\in\RsL \right\}$ (resp. $\left\{
\e_{\r}-\e_{-\r}\,:\,\r\in\RsL \right\}$) while a basis of
$\Lnot_{+1}^*$ (resp. $\Lnot_{-1}^*$) is $\left\{
\e_{\r}+\e_{-\r}\,:\,\r\in\RsL \right\}$ (resp. $\left\{
\e_{-\r}\,:\,\r\in\RsL \right\}$). One therefore knows that $\Lnot$
admits an $R$-bracket analogous to that defined in (\ref{eq:rbracket})
and that Lemma \ref{lem:pss} also holds for $\Lnot_{R}$.

\begin{remark}
  {\rm
    If $\alpha$ is an automorphism of the graded
    Lie algebra $\L$ that fixes $\h$, then the fixed point set of $\alpha$
    is a sub-algebra that inherits a grading, splitting and a root space
    decomposition from $\L$. The constructions of both subsections
    \ref{ssec:split} and \ref{ssec:car} are applicable in this case,
    too. The automorphism $\alpha$ satisfies $\alpha(x \cdot \lambda^n) =
    \alpha(x) \cdot (\epsilon \lambda)^n$ for all $x\in\g$ and $n$, where
    $\epsilon$ is a primitive order($\alpha$) root of unity. This
    construction yields the so-called twisted loop algebras.  The twisted
    loop algebra is traditionally denoted by $\g^{(m)}$ where $m$ is the
    order of the automorphism $\alpha$; when $m=1$, one has the usual loop
    algebra $\L$.
  }
\end{remark}

\subsection{Examples} \label{ssec:exsplit} 
Let $\g = A_2 = \spl{3;\R}$. For $\h$ one can take the sub-algebra of
trace zero diagonal matrices and for the basis of positive roots of
$\g$ one can take the roots $r_1$ and $r_2$ with the minimal root $\minrt$:
\begin{xalignat}{3} \label{xa:rts}
  r_1 &= \begin{bmatrix}
    1 & 0 & 0\\ 0 & -1 & 0\\ 0 & 0 & 0
  \end{bmatrix},
  &
  r_2 &= \begin{bmatrix}
    0 & 0 & 0\\ 0 & 1 & 0\\ 0 & 0 & -1
  \end{bmatrix},
  &
  \minrt = - r_3 &= \begin{bmatrix}
    -1 & 0 & 0\\ 0 & 0 & 0\\ 0 & 0 & 1
  \end{bmatrix},
\end{xalignat}
which satisfy the linear relation $r_1+r_2+\minrt=0$. A root
$\r\in\RsL$ may be written formally as $\r = \pm r_i + n$ where
$n=\left\langle \r , \d \right\rangle$. The height of $\r$ is then
computed to be $3n\pm 1$ for $i=1,2$ and $3n\pm2$ for $i=3$. From
this, one can see that the graded pieces of $\L$, as in (\ref{eq:pg}),
are $\L_0=\h$ and
\begin{xalignat}{2} \label{xa:splita3}
  \L_{+1} &= \left\{ \begin{bmatrix}
    \lambda a_1 & \alpha_1 & 0\\ 0 & \lambda a_2 & \alpha_2\\ \lambda
    \alpha_3 & 0 & \lambda a_3
  \end{bmatrix}  \right\},
  &
  \L_{+2} &= \left\{ \begin{bmatrix}
    \lambda^2 a_1 & 0 & \alpha_3\\ \lambda \alpha_1 & \lambda^2 a_2 & 0\\ 0 & \lambda
    \alpha_2 & \lambda^2 a_3
  \end{bmatrix}  \right\},
  \notag\\
  \L_{-1} &= \left\{ \begin{bmatrix}
    \lambda^{-1} a_1 & 0 & \lambda^{-1}\alpha_3\\ \alpha_1 & \lambda^{-1} a_2 & 0\\ 
    0 & \alpha_2 & \lambda^{-1} a_3
  \end{bmatrix}  \right\},
  &
  \L_{-2} &= \left\{ \begin{bmatrix}
    \lambda^{-2} a_1 & \lambda^{-1} \alpha_1 & 0\\ 0 & \lambda^{-2} a_2 & \lambda^{-1}\alpha_2\\ \alpha_3 & 0 & \lambda^{-2} a_3
  \end{bmatrix}  \right\},
\end{xalignat}
where $a_i,\alpha_i$ are real numbers and $\sum a_i=0$. 

The splitting in (\ref{eq:split}) of $\Lnot$ implies that the root
spaces of height $\pm1$ and their duals are
\begin{xalignat}{2} \label{xa:splita3-2}
  \Lnot_{-1} &=
  \left\{ \begin{bmatrix}
    0 & \alpha_1 & -\alpha_3 \lambda^{-1}\\ 
    -\alpha_1 & 0 & \alpha_2\\
    \alpha_3 \lambda & -\alpha_2 & 0
  \end{bmatrix}  \right\},
  &
  \Lnot_{-1}^* &=
  \left\{ \begin{bmatrix}
    0 & \alpha_1 & 0\\
    0 & 0 & \alpha_2\\
     \alpha_3 \lambda & 0 & 0
  \end{bmatrix}  \right\},
  \notag\\
  \Lnot_{+1} &=
  \left\{ \begin{bmatrix}
    0 & \alpha_1 & 0\\
    0 & 0 & \alpha_2\\
     \alpha_3 \lambda & 0 & 0
  \end{bmatrix}  \right\},
  &
  \Lnot_{+1}^* &=
  \left\{ \begin{bmatrix}
    0 & \alpha_1 & \alpha_3 \lambda^{-1}\\ 
    \alpha_1 & 0 & \alpha_2\\
    \alpha_3 \lambda & \alpha_2 & 0
  \end{bmatrix}  \right\},
\end{xalignat}
where the $\alpha_i$ are real.

\subsection{Bijections}
Assume that $F/\Q$ is an algebraic field of degree $m$ with $r$ real
embeddings and $2c$ non-real embeddings such that $r+c=n$, and that
$\g$ is a real split affine Lie algebra of rank $n-1$. Since $\B{F}$,
the set of real embeddings of $F$ plus one-half the set of complex
embeddings of $F$, has $n$ elements and the simple roots of $\L$,
$\RsL$, have $n$ elements, the sets are isomorphic.
\begin{definition} \label{def:bi}
  Let $\Bi$ be the set of bijections $\B{F} \to \RsL$.
\end{definition}

Each $\rho \in \Bi$ can be extended to a map $\G{F} \to \RsL$ by
$\rho(\bar \sigma) := \rho(\sigma)$ for all $\sigma\in\B{F}$. This
extension shall be understood throughout.

Additionally, each $\rho \in \Bi$ naturally induces a linear
isomorphism $\phi=\phi_{\rho} : \Vo{F}^* \to \h^*$. To define $\phi$
let us recall two things. First, note that the projection
$\onto[]{\hat\h^*}{\h^*}$ that is dual to the inclusion
$\into[]{\h}{\hat\h}$ induces the bijection $\bij[]{\RsL}{\Rs}$ : $\r
\mapsto r=\left. \r \right|_{\h}$. Second, there are unique positive
integers $\omega_r$ such that
\begin{equation} \label{eq:omegar}
  \sum_{r \in \Rs} \omega_r r =  0,\hspace{20mm} \gcd(\omega_r\ :\ r \in
  \Rs) = 1.
\end{equation}
%Let $\omega = \lcm(\omega_r\ :\ r \in \Rs),$ and define the positive
%integers $w_r = \omega/\omega_r$. 
For each $\tau \in \B{F}$, define $n_{\tau}$ to be $1$ if $\tau$ is a
real embedding; and $2$ if not. Then, define
\begin{equation} \label{eq:phi}
  \phi(\left. \hat\tau \right|_{\Vo{F}} ) = n_{\tau}^{-1}\, \omega_r r \hspace{10mm} {\rm
    where\ } \r=\rho(\tau).
\end{equation}
Since $\hat \tau$ equals $\hat{\bar{\tau}}$ when restricted to
$\Vo{F}$, the sole linear dependence relation amongst the set $\left\{
\left. \hat\tau \right|_{\Vo{F}} \ :\ \tau \in \B{F} \right\}$ is the
relation $$\sum_{\tau \in \B{F}} \left. n_{\tau}\, \hat\tau
\right|_{\Vo{F}} = \sum_{\tau \in \G{F}} \left. \hat\tau
\right|_{\Vo{F}} = 0.$$ Thus, equation (\ref{eq:omegar}) implies that
$\phi$ extends to a linear isomorphism.

\subsection{Lax representations} \label{ssec:laxr}
Fix $\rho \in \Bi$ and let $\phi=\phi_{\rho}$ be the induced linear
isomorphism. Let $\Phi : \Vo{F}^* \to \h^*$ be a linear map and let
$g_{\pm} : \V{E}^* \times \G{E} \to \L^*_{\pm}$ be smooth maps. Define a map
$\lax[]=\lax[\rho,\Phi] : T^* \tilde \Sigma \to \L^*$ by
\begin{equation} \label{eq:L0}
\kappa^{-1} \cdot \lax[](P) =\sum_{\sigma \in \G{E}}
  g_{-,\sigma}(\py)\cdot \e_{\r} \ + \ \Phi(\px) \ + \
  \sum_{\sigma \in \G{E}} g_{+,\sigma}(\py)\cdot\exp( b_{\sigma} \cdot
  \left\langle \hat \sigma , \x \right\rangle )\cdot \e_{-\r}
\end{equation}
where it is understood that $\r = \rho(\sigma |_F)$ in the sums and
$\L^*$ is identified with $\L$ via the Cartan-Killing form $\kappa$.

There are several choices of Lax representation that are
    useful. The first is (in all cases, $\r=\rho(\sigma|_F)$ is
    understood)
    \begin{equation} \label{eq:L0-1}
      \kappa^{-1} \cdot \lax[](P) = \sum_{\sigma \in \G{E}} |\py_{\sigma}|^{b_{\sigma}}\cdot
      \e_{\r} \ + \ \Phi(\px) \ + \ \frac{1}{2} \times \sum_{\sigma \in \G{E}}
      \exp( b_{\sigma} \cdot \left\langle \hat \sigma , \x \right\rangle
      )\cdot \e_{-\r},
    \end{equation}
    while the second is
    \begin{equation} \label{eq:L0-2}
      \kappa^{-1} \cdot \lax[](P) = \sum_{\sigma \in \G{E}} \e_{\r} \ + \ \Phi(\px) \ + \ 
      \frac{1}{2} \times \sum_{\sigma \in \G{E}} |\py_{\sigma}|^{b_{\sigma}}\cdot\exp(
      b_{\sigma} \cdot \left\langle \hat \sigma , \x \right\rangle )\cdot
      \e_{-\r},
    \end{equation}
    and a third is
    \begin{align} \label{eq:L0-3}
      \kappa^{-1} \cdot \lax[](P) &= \Phi(\px) \ + \ 
      \frac{1}{\sqrt{2}} \times \sum_{\sigma \in \G{E}} |\py_{\sigma}|^{\frac{1}{2}b_{\sigma}}\cdot\exp(
      \frac{1}{2}b_{\sigma} \cdot \left\langle \hat \sigma , \x
      \right\rangle )(\e_{\r} +
      \e_{-\r}).
    \end{align}
    Note that the Lax representations in
    (\ref{eq:L0-1}--\ref{eq:L0-2}) are related to the splitting of the
    loop algebra in section \ref{ssec:split}; the final Lax
    representation in (\ref{eq:L0-3}) is related to the splitting in
    section \ref{ssec:car}.  In all cases, the pullback of the Casimir
    $x \mapsto \frac{1}{2} \times \kappa(x,x)$ on $\L^*$ by any of the
    three Lax matrices in (\ref{eq:L0-1}--\ref{eq:L0-3}) $\lax[]$ is
    equal to
    \begin{equation} \label{eq:HL0}
      \Htoda := \frac{1}{2} \times \left\langle \Qm\cdot\px , \px
      \right\rangle + \frac{1}{2} \,\sum_{\sigma\in \G{E}}
      |\py_{\sigma}|^{b_{\sigma}}\cdot\exp( b_{\sigma} \cdot \left\langle
      \hat \sigma , \x \right\rangle ),
    \end{equation}
    where $\Qm : \Vo{F}^* \to \Vo{F}$ is defined by $\Qm = \Phi^*
    \kappa \Phi$.  This Hamiltonian is fibre-wise quadratic --- hence,
    induced by a riemannian metric --- iff $b_{\sigma}=2$ for all
    $\sigma$; in all cases, it is fibre-wise convex. The next theorem
    implies that there are constraints on $F$ if $\Htoda$ is
    fibre-wise quadratic.

As a second step, recall that $\slr{2}$ has a basis $h,\e_+,\e_-$ such
that $[h,\e_{\pm}]=\pm \e_{\pm}$, and $[\e_+,\e_-]=2h$.  The
Cartan-Killing form identifies the dual basis as $h,\e_-,\e_+$. For
each $\sigma \in \G{E}$, let $\slr{2}_{\sigma}$ be a copy of $\slr{2}$
and let $h_{\sigma}, \e_{\pm,\sigma}$ be copies of
$h,\e_{\pm}$. Define $\lax[1] : T^*\tilde \Sigma \to \g_1^*$, $\g_1 =
\displaystyle\sum_{\sigma \in \G{E}} \slr{2}_{\sigma}$ by
\begin{equation} \label{eq:L1}
  \lax[1](P) = \sum_{\sigma \in \G{E}} \py_{\sigma} \cdot h_{\sigma} +
  \exp(\left\langle \hat \sigma , \y \right\rangle)
  \cdot \e_{-,\sigma}.
\end{equation}

\begin{thm} \label{thm:lax}
  $\lax[1]$ is a Poisson map. $\lax[]=\lax[\rho,\Phi] : T^*\tilde \Sigma \to
  \L^*_R$ is a Poisson map iff there is a $c\in\frac{1}{2}\Z^+$ such that:
  \begin{enumerate}

  \item for all $\sigma\in\G{E}$ and $\r\in\RsL$ with
    $\rho(\left. \sigma \right|_F) = \r$, one has
    $n_{\left(\sigma|_F\right)}^{-1} b_{\sigma} \omega_{r} = c$; and
  \item $\Phi = c^{-1} \times \phi_{\rho}$.

  \end{enumerate}
  The map $\lax[2] = \lax[]+\lax[1] : T^*\tilde \Sigma \to \L^* +
  \g_1^*$ is a Poisson embedding if either $g_{+}$ or $g_-$ is an
  embedding and $E=F$.
\end{thm}

\begin{proof}%[Theorem \ref{thm:lax}]
  The proof shall assume that $\lax[]$ is defined by equation
  (\ref{eq:L0-1}); the remaining cases are not significantly
  different.  To prove that $\lax[1]$ is a Poisson map, one needs to
  prove that
  \begin{equation} \label{eq:pL1}
    \left\{ f \comp \lax[1] , g \comp \lax[1] \right\}_{T^*\tilde \Sigma } =
    \left\{ f,g  \right\}_{\h_1^*} \comp \lax[1],
  \end{equation}
  for all smooth functions $f,g$ on $\h_1^*$. It suffices to verify
  equation (\ref{eq:pL1}) holds for linear functions $f,g$, for a single
  copy of $\slr{2}$, and a single pair of conjugate variables $\py$ and
  $\y.$ For $f=h$ and $g=\e_+$ one sees that
  \begin{equation} \label{eq:pb1}
    \left\{ h,\e_+  \right\}_{\slr{2}^*} \comp \lax[1] = -\left\langle\langle
    \lax[1] , [h,\e_+] \right\rangle \rangle = -e^\y,
  \end{equation}
  while
  \begin{equation} \label{eq:pb2}
    \left\{ h\comp \lax[1], \e_+ \comp \lax[1]  \right\}_{T^*\tilde \Sigma} =
    \left\{ \py, e^\y  \right\} = -e^\y.
  \end{equation}
  Since $h$ and $\e_+$ are functionally independent at almost all points
  on almost all co-adjoint orbits, this proves that $\lax[1]$ is a Poisson map.

  \vspace{3mm}

  To prove the claim concerning $\lax[]$, one needs to prove that
  \begin{equation} \label{eq:pL0}
    \left\{ f \comp \lax[] , g \comp \lax[] \right\}_{T^*\tilde \Sigma } =
    \left\{ f,g  \right\}_{\L_R^*} \comp \lax[],
  \end{equation}
  for all $f,g \in C^{\infty}(\L^*)$. As above, it suffices to verify
  equation (\ref{eq:pL0}) holds for all $f,g \in \L_R$. Given the
  bracket relations on $\L_R$, it suffices to prove the equation for all
  $f,g\in \L_{-1} + \L_0 + \L_{+1}$. Let us break this into cases:
  \begin{enumerate}

  \item If $f,g \in \L_0$ or $f \in \L_{-1}$ and $g \in \L_{+1}$ or $f
    \in \L_0$ and $g \in \L_{-1}$, then $[f,g]_R = 0$ so $\left\{ f,g
    \right\}_{\L_R^*} \comp \lax[] = 0$.  On the other hand, $\left\{ f
    \comp \lax[] , g \comp \lax[] \right\}_{T^*\tilde \Sigma } = 0$ since
    functions of $\px$ alone, or a function of $\py$ and a function of
    $\x$, or a function of $\py$ and a function of $\px$ alone Poisson
    commute.

  \item If $f,g \in \L_{-1}$ or $f,g \in \L_{+1}$, then
    $[f,g]_R \in \L_{\pm 2}$. Therefore,
    \begin{equation} \label{eq:pL0-1}
      \left\{ f,g  \right\}_{\L_R^*} \comp \lax[] = -\left\langle\langle \lax[] ,[f,g]_R
      \right\rangle\rangle = 0
    \end{equation}
    since $\lax[]$ lies in $\L_{-1} + \L_0 + \L_{+1}$. On the other hand,
    $f\comp \lax[]$ and $g \comp \lax[]$ are either functions of $\py$ or $\x$
    alone. In either case, they Poisson commute on $T^*\tilde \Sigma $.

  \item If $f \in \L_0$ and $g \in \L_{+1}$, then it suffices to assume
    that $g = \e_{\r}$ for some $\r \in \RsL$. In this case,
    \begin{align} \label{eq:pL0-2}
      \left\{ f,g \right\}_{\L_R^*} \comp \lax[] &= -\left\langle \lax[]
      ,[f,\e_{\r}]_R \right\rangle = -\left\langle \r , f \right\rangle
      \times \left\langle\langle \lax[] , \e_{\r} \right\rangle\rangle \\
      \nonumber &= -\left\langle r , f \right\rangle \times \sum_{\sigma \in
        \G{E}\ {\rm s.t.\ } \rho(\left. \sigma \right|_F) = \r}
      \exp(b_{\sigma}\, \left\langle \hat \sigma , \x
      \right\rangle).
    \end{align}

    On the other hand, 
    \begin{align}
      f \comp \lax[] &= \left\langle \px , \Phi^* f \right\rangle,\\
      g \comp \lax[] &= \sum_{\sigma \in \G{E}\ {\rm s.t.\ }  \rho(\left. \sigma
        \right|_F) = \r} \exp(b_{\sigma}\, \left\langle \hat \sigma , \x
      \right\rangle),
    \end{align}
    so the Poisson bracket of these functions is
    \begin{equation} \label{eq:pL0-4}
      \left\{ f \comp \lax[] , g \comp \lax[] \right\}_{T^*\tilde \Sigma } = -
      \sum_{\sigma \in \G{E}\ {\rm s.t.\ }  \rho(\left. \sigma
        \right|_F) = \r} \exp(b_{\sigma}\, \left\langle \hat \sigma , \x
      \right\rangle) \times b_{\sigma}\, \left\langle \hat \sigma , \Phi^*f \right\rangle
    \end{equation}
    Because $\rho$ is a bijection of $\B{F}$ with $\RsL$, there is a
    unique $\tau \in \B{F}$ such that $\rho(\tau)=\r$. Therefore, due to
    the way that $\rho$ is extended to $\G{F}$, the $\sigma$'s involved in
    the above summations all satisfy $\sigma|_F = \tau$ or
    $\bar\tau$. Therefore, $\left\langle \hat \sigma , \x \right\rangle =
    \left\langle \hat \tau , \x \right\rangle$ for all $\sigma \in \G{E}$
    such that $\rho(\sigma|_F) = \r$.

    This fact about the $\sigma$'s also implies that $\left\langle \hat
    \sigma , \Phi^*f \right\rangle = \left\langle \Phi( \hat\tau|_{\Vo{F}} ) ,
    f \right\rangle$. Since $\hat \tau$ will only appear when it acts on
    $\Vo{F}$, the notation $|_{\Vo{F}}$ will be suppressed.

    Therefore, if the right-hand sides of equations (\ref{eq:pL0-2}) and
    (\ref{eq:pL0-4}) are equated for all $f \in \L_0$, then one concludes
    that
    \begin{equation} \label{eq:pL0-5}
      \sum_{\sigma \in \G{E}\ {\rm s.t.\ } \rho(\left. \sigma
        \right|_F) = \r}  \exp(b_{\sigma}\, \left\langle \hat \tau , \x
      \right\rangle) \times \left[ b_{\sigma}\, \Phi(\hat \tau) - r \right] =
      0.
    \end{equation}
    The functions $u \mapsto e^{au}, e^{bu}$ are linearly independent if
    $a\neq b$. If the $b_{\sigma}$'s in the above sum are not constant, then
    the sum in equation (\ref{eq:pL0-5}) contains two linearly independent
    functions. Therefore, the coefficients on these two functions must
    vanish. But this forces $\Phi(\hat \tau)$ to equal two different
    multiples of $r$. Absurd. Therefore, the $b_{\sigma}$'s in the sum
    must be constant. This implies that $b_{\sigma}$ is determined by $\r$
    alone, or equivalently, by $\tau$ alone.
  \end{enumerate}

  As cases (1---3) are the only independent cases to be considered, one
  concludes that if $\lax[]$ is a Poisson map, then there are integers
  $b_{\tau}$, $\tau \in \B{F}$, such that the integers $b_{\sigma}$,
  $\sigma\in\G{E}$, satisfy
  \begin{equation} \label{eq:pL0-6}
    b_{\sigma} = b_{\tau} \hspace{15mm} {\rm where\ } \left. \sigma
    \right|_F = \tau\ {\rm or\ } \bar \tau.
  \end{equation}

  Moreover, equation (\ref{eq:pL0-5}) implies that if $\tau \in \B{F}$
  and $\rho(\tau)=\r$, then 
  \begin{equation} \label{eq:pL0-6.1}
    \Phi(\hat \tau) = b_{\tau}^{-1}\, r.
  \end{equation}
  Summing over $\tau\in\G{F}$, and using the fact that $\rho$ is a
  bijection of $\B{F}$ and $\RsL$,
  \begin{equation} \label{eq:pL0-7}
    \sum_{\tau \in \G{F}} \Phi(\hat \tau) = \sum_{\tau \in \B{F}}
    \Phi(n_{\tau}\, \hat \tau) = \sum_{r \in \Rs} n_{\tau}\, b_{\tau}^{-1} r,
    \hspace{10mm}({\rm where\ } \rho(\tau)=\r).
  \end{equation}
  The left-hand side vanishes because $\sum_{\tau \in \G{F}} \hat
  \tau|_{\Vo{F}} = 0$. Therefore
  \begin{equation} \label{eq:pL0-8}
    \sum_{r \in \Rs} n_{\tau}\, b_{\tau}^{-1} r = 0, 
  \end{equation}
  while the unique linear dependence relation in equation
  (\ref{eq:omegar}) implies that there must be a constant $c$ such that
  $n_{\tau}^{-1} b_{\tau} \omega_r = c$ for all $r \in \Rs$. The
  constant $c$ is a positive integer, or one-half such, since $b_{\tau}$
  and $\omega_{r}$ are positive integers and $n_{\tau}=1$ or $2$. This
  implies part (1) of the Theorem.

  The equation that $\Phi$ must satisfy is, for all $\tau \in \B{F}$,
  \begin{equation} \label{eq:pL0-9}
    \Phi(\hat\tau) = \frac{1}{c n_{\tau}} \times \omega_r r \hspace{15mm}
        {\rm where\ } \r=\rho(\tau).
  \end{equation}
  Comparison with equation (\ref{eq:phi}) shows that $\Phi=c^{-1} \times
  \phi_{\rho}$, which is part (2) of the Theorem.

  \vspace{5mm}
  The claim that $\lax[2]=\lax[]+\lax[1]$ is an embedding is obvious.

\end{proof}

\begin{remark} \label{re:difficult}
{\rm Theorem \eqref{thm:lax} exploits the naturality of the
  constructions. In cases where the group $A$ is not of finite index
  in $\U_F$, one encounters the problem that there is no obvious Lax
  map. This is what makes question \ref{qu:4} difficult.

}
\end{remark}

\begin{remark} \label{re:wts}
{\rm Condition (1) implies that $b_{\sigma}$ depends only on
  $\left. \sigma \right|_F$. Condition (2) implies that $2c$ is
  divisible by all $\omega_r$, hence by their lcm,
  $\omega$. Therefore, there is a unique choice of $\Phi$ and integers
  $b_{\sigma}$ if one insists that $c$ be as small as possible and the
  $b_{\sigma}$ be even. In case $F$ is totally real, condition (1)
  implies that $c$ is divisible by
  $\omega=\lcm(\omega_r:r\in\Rs)$. When $c$ is chosen to be $2\omega$
  -- so that $b_\sigma=2 \omega/\omega_r$ is even --, condition (2)
  implies that $\Phi(\left.\hat{\sigma}\right|_F) = \frac{1}{2}w_r r$,
  where $w_r = \omega_r/\omega$, and $\rho(\sigma|_F)=\r$. This
  condition is stated in \cite[Lemma 7]{B7}, except for the factor of
  $\frac{1}{2}$ in $\Phi$.\footnote{The coefficients $b_\sigma$ in
    \cite{B7} are one-half those in the present paper.} This
  discrepancy is due to the choice of a slightly different Poisson
  structure in \cite[equation (9-10)]{B7}. With these choices, the
  Hamiltonian $\Htoda$ in equation (\ref{eq:HL0}) is equal to that in
  \cite[equation 15]{B7} when $E=F$ is totally real. In particular
  ({\it c.f.} equation \ref{eq:HL0}),
\begin{equation} \label{eq:Qm}
\Qm = \phi_{\rho}^* \cdot \kappa \cdot \phi_{\rho}.
\end{equation}

 In the event that $F$ has no real embeddings, this smallest choice is
    $c=\omega$ and $b_{\sigma} = 2 \omega/\omega_r$; in other events,
    the solution is somewhat more involved to state, as it depends on
    the bijection $\rho$ (see tables
    \ref{tab:ex3-1-lax-matrices-delta}--\ref{tab:ex3-3-lax-matrices-delta}
    in section \ref{sssec:ex3} for examples). It is possible to
    state }
\end{remark}

{
\def\r{{\sf r}}
\def\c{{\sf c}}
\begin{prop} \label{prop:quad}
Let $\r$ be the number of real embeddings of $F$, and $2\c$ the number
of non-real embeddings. If the Hamiltonian $\Htoda$ is fibre-wise
quadratic, then table \eqref{tab:eig} is true. In particular, if $\r >
4$ and $\c > 0$, then none of the Hamiltonians $\Htoda$ are fibre-wise
quadratic.
\end{prop}

\begin{proof}
From equation (\ref{eq:HL0}), it is clear that $\Htoda$ is fibre-wise
quadratic iff $b_\sigma=2$ for all $\sigma$. Since, for all roots
$r$, $\omega_r = n_\tau c / 2$ where $\tau = \left. \sigma \right|_F$
and $\rho(\tau)={\bf r},$ one has
\begin{align*}
\tau &\in \G{F}^c: & \tau &\in \G{F}^r: \\
\omega_r &= c & \omega_r &= c/2.
\end{align*}
This shows that each weight is $1$ or $2$. If $\c=0$, then $c=2$ and
$\omega_r=1$ for all roots. If $\r=0$, then $c=1$ and $\omega_r=1$ for
all roots. If $\r,\c>0$, then $c=2$ and $\RsL$ has $\r$ roots with
weight $1$ and $\c$ roots with weight $2$.  Inspection of the root
systems in figures (\ref{fig:rs1}--\ref{fig:rs23}) completes the
proof.
\end{proof}

}

\subsection{Quotients of the Lax Representations} \label{ssec:quot}

\begin{lemma} \label{lem:laxL0-1}
  There is a natural action of $\Delta = \U^+_F \star \O_F$---which factors
  through $\U_F^+$---on $\L^*_R$ such that the map defined in equation
  (\ref{eq:L0-1}) is $\Delta$-equivariant, hence induces a
  Poisson map
  $$
  \xymatrix{T^*\tilde \Sigma \ar[rr]^{\lax[]} \ar[d]^{\tilde\Pi} && \L^*_R \ar[d]\\
    T^* \Sigma \ar[rr]^{\lax[]} && \L^*_R/\U^+_F.  
  }
  $$
  The action of $\U^+_F$ on $\im \lax[] \subset \L_R^*$ is free and
  proper.
\end{lemma}
\begin{proof}
  Define the action of $g=(u,\alpha) \in \U^+_F \star \O_F$ on $\L^*_R$ by 
  \begin{equation} \label{eq:act}
    g\cdot\e_{\r} = \left\{ \begin{array}{lcl} 
      |\sigma(u)|^{-b_{\sigma}} \cdot \e_{\r} 
      & {\rm if\ } & \r \in \RsL, \rho(\sigma)=\r\\
      |\sigma(u)|^{b_{\sigma}} \cdot \e_{\r} 
      & {\rm if\ } & -\r \in \RsL, \rho(\sigma)=\r\\ 
      \e_r & & {\rm otherwise,}
    \end{array}
    \right.
  \end{equation}
  and $g|_{\L_0} = 1$. It is straightforward to see that $\lax[](g\cdot P)
  = g\cdot \lax[](P)$ for all $g$ and $P\in T^*\tilde\Sigma$. 

  Since the coefficients of $\e_{\r}$, $-\r\in\RsL$, do not vanish on $\im
  \lax[]$, one sees that the action of $\U^+_F$ is semi-conjugate to its
  action on $\Vo{F}$. Hence, it is free and proper.
\end{proof}

\begin{remark} \label{re:b}
  {\rm
    The preceding lemma implies that $\Htoda$ and
    all the ``spectral'' integrals of $\Htoda$ descend, but with some
    additional work. The alternative Lax matrix, equation (\ref{eq:L0-2}),
    gives us a simple proof of this fact.
  }
\end{remark}

\begin{lemma} \label{lem:laxL0-2}
  The map defined in equation (\ref{eq:L0-2}) is $\U_F^+ \star
  \O_F$-invariant, hence it induces a Poisson map
  $$
  \xymatrix{T^*\tilde \Sigma \ar[rr]^{\lax[]} \ar[d]^{\tilde\Pi} && \L^*_R\\
    T^* \Sigma \ar[urr]^{\lax[]}.  
  }
  $$ Consequently, if $h$ is a Casimir of $\L^*$, then $h\comp \lax[]$
  Poisson commutes with $\Htoda$.
\end{lemma}

\begin{proof}
  By equation~(\ref{eq:gamma}), one can write 
  \begin{equation} \label{eq:L0-gamma}
    \lax[](P) = \sum_{\sigma \in \G{E}} \e_{\r} \ \ + \ \ \Phi(\px) \ \ + \ \
    \frac{1}{2} \times \sum_{\sigma \in \G{E}} \gamma_{\sigma} \cdot
    \e_{-\r}.
  \end{equation}
  By Lemma~\ref{lem:gamma}, each function $\gamma_{\sigma}$ is
  $\U^+_F$-invariant, hence $\U_F^+ \star \O_F$-invariant.
\end{proof}

\subsection{Additional Integrals}
A consequence of Theorem~\ref{thm:lax} is that the function $P \mapsto
\py\ : \ T^*\hat \Sigma \to \V{E}^*$ is a first integral of any
function on $\L^*_R$ pulled-back to $T^*\hat \Sigma$ by the Lax matrix
$\lax[]$ (equation (\ref{eq:L0-1})). Unfortunately, this map is not
$\U^+_F$-invariant. However, one is able to construct a map, $\fint$,
from $\py$ which is $\U^+_F$-invariant. Naively, one might try to
define $\fint$ by means of equivariance. That is the task of this
section.

For each $\tau \in \G{F}$, let the subspace of $\V{E}$ spanned by $\left\{
\sigma\ :\ \sigma|_F = \tau \right\}$ be denoted by
$\V{\tau,E}$. One may define
\begin{equation} \label{eq:pytau}
  \py_{\tau} = \sum_{\sigma|_F = \tau} \py_{\sigma}\cdot \hat \sigma,
\end{equation}
for all $\tau \in \G{F}$. Since $\py_{\bar\sigma} = \bar\py_{\sigma}$
for all $\sigma$, it is clear that complex conjugation induces a real
linear isomorphism between $\V{\tau,E}^*$ and $\V{\bar
  \tau,E}^*$. This linear isomorphism maps $\py_{\tau} \to \py_{\bar
  \tau}$, which implies that as real vector spaces
\begin{equation} \label{eq:VE}
  \V{E}^* \cong \sum_{\tau\in\B{F}} \V{\tau,E}^*.
\end{equation}
In the sequel, this natural isomorphism is understood.

The group $\U^+_F$ acts on $\V{E}^*$ by $u\cdot \hat \sigma =
\sigma(u)^{-1}\cdot \hat \sigma$. Since $u \in F$, this action is (\cf
equation \ref{eq:up})
\begin{equation} \label{eq:UFact}
  u\cdot \py = \sum_{\tau\in\B{F}} \tau(u)^{-1} \cdot \py_{\tau}.
\end{equation}
\begin{lemma} \label{lem:UVact}
  Let
  \begin{equation} \label{eq:VE0}
    \V{E,0}^* = \left\{ \py \in \V{E}^*\ :\ \forall \tau\in\B{F},
    \py_{\tau}\neq 0   \right\}.
  \end{equation}
  The set $\V{E,0}^*$ is $\U^+_F$-invariant and $\V{E,0}^*/\U^+_F$ is a
  smooth manifold of dimension $\dim \V{E}^*$.
\end{lemma}

\begin{proof}
{\def\tnohat{{\mathrm t}}
  Inspection of equation (\ref{eq:UFact}) shows the invariance of
  $\V{E,0}^*$. To prove that the action of $\U^+_F$ is free and
  proper, define $\hat\q : \V{E,0}^* \to \V{o,F}/\logu{F}$ by
  \begin{align}
    \hat\q(\py) &= \sum_{\tau \in \G{F}} \ln |\py_{\tau}| \cdot \tau -
    \sum_{\tau \in \G{F}} \ln |\py_\tau| \cdot \tnohat & \bmod
    \logu{F}, \label{eq:q}\\ \intertext{where} \tnohat &=
    \frac{1}{|\G{F}|}\sum_{\tau \in \G{F}} \tau.\notag
  \end{align}
 From
  equation (\ref{eq:UFact}), one sees that $u^* |\py_\tau| =
  -\ln|\tau(u)| + |\py_\tau|$, so $\hat\q$ is $\U^+_F$-invariant,
  hence it defines a continuous map $\q : \V{E,0}^*/\U^+_F \to
  \V{o,F}/\logu{F}$. The action of $\U^+_F$ is therefore both free and
  proper, since $\q$ maps cosets onto cosets.
}
\end{proof}

Define a function $g_\tau : T^*\hat \Sigma \to \R$ by
\begin{equation} \label{eq:g}
  g_{\tau}(P) = |\py_\tau|^2 = \sum_{\left. \sigma \right|_F = \tau}
  |\py_{\sigma}|^2.
\end{equation}
These functions are first integrals of $\Htoda$ (see below), but they
are not $\U_F^+$-invariant. However, their product is invariant:
\begin{equation} \label{eq:caspb}
  \caspb = \prod_{\tau\in\G{F}} g_{\tau} = \prod_{\tau\in\G{F}}
  \sum_{\left. \sigma \right|_F = \tau} |\py_{\sigma}|^2.
\end{equation}
From $\caspb$ one obtains the important subspaces
\begin{equation} \label{eq:U}
  \UU = \left\{ P \in T^*\Sigma \ : \ \caspb(P) \neq 0  \right\},
  \hspace{10mm} \ZZ = \left\{ P \in T^*\Sigma \ : \ \caspb(P)=0  \right\}.
\end{equation}
It is clear from the definition of $\caspb$ that $\ZZ$ is the union
$\cup_{\tau\in\G{F}} \ZZ_\tau$, where $\ZZ_\tau = g_\tau^{-1}(0)$
(although $g_{\tau}$ is not $\U^+_F$-invariant, its zero set is). It
is also clear that $\UU$ is an open and dense analytic submanifold of
$T^*\Sigma$, $\ZZ=\Sigma \times \Vo{F}^* \times \V{E,0}^*$, and that
$\ZZ$ is an analytic sub-variety.

\begin{lemma} \label{lem:fint}
  Define the map $\fint : \UU \to \V{E,0}^*/\U^+_F$ by, for all $P \in \UU$,
  \begin{equation} \label{eq:fint}
    \fint(P) = \py \cdot \U^+_F
  \end{equation}
  where the action of $\U^+_F$ is given by equation (\ref{eq:UFact}). Then $\fint$ is an analytic submersion.
\end{lemma}
\begin{proof}
  This is clear.
\end{proof}

\setcounter{listctr}{1}
\begin{remark} \label{re:a}
  {\rm
    (\listenum) $\UU$ is a union of regular Liouville tori
    and singular tori (see below). The singular set $\ZZ$ has co-dimension
    equal to $[E:F]$. Therefore, when $F \subsetneq E$, the co-dimension is
    two or more. In this case, the set of regular Liouville tori is
    connected. (\listenum) The topological structure of $\V{E,0}^*/\U^+_F$ is
    interesting. The map $\q$ is a submersion whose typical fibre is
    diffeomorphic to the Cartesian product of the unit spheres in
    $\V{\tau,E}^*$, for $\tau\in\B{F}$, with the positive real
    numbers. The bundle is generally non-trivial, since the action of
    $\U^+_F$ twists the fibres. Indeed, one sees that the map $\q \times
    \caspb$ is a proper submersion with
    \begin{equation} \label{eq:veo}
      \xymatrix{
	\prod_{\tau\in\B{F}} S^{d_{\tau}-1} \ar@{^{(}->}[rr] &&
	\V{E,0}^*/\U^+_F \ar@{->>}[rr]^{\q \times \caspb} && \V{o,F}/\logu{F}
	\times \R^+
      }
    \end{equation} where we identify $\caspb$ with a function defined on $\V{E}^*$
    and $d_{\tau} = [E:F]$. This also exhibits $\V{E,0}^*/\U^+_F$ as a
    compact manifold times $\R^+$. The compact manifold is something
    like a torus bundle over a torus. In particular, the ends of
    $\V{E,0}^*/\U^+_F$ are quite uncomplicated. (\listenum) Let us relate the
    preceding discussion to that in the introduction, \cf diagram
    \eqref{eq:mmap} and figure \ref{fig:orbits}. Let $\sim$ be the
    equivalence relation on $\V{E}^*$ that is generated by defining
    $\py \sim 0$ if $\py_\tau = 0$ for some $\tau \in \B{F}$ and $\py
    \sim u\cdot \py$ for all $u \in \U^+_F$. The topological space
    $\V{E}^*/\sim$ is a quotient of $\V{E}^*/\U^+_F$ where one
    collapses the set $\{ \py\ :\ \prod_{\tau \in \B{F}} \py_\tau =
    0 \}$ to a point. We have the following
    commutative diagram:
    {
      %% a la p. of rotating environment examples
      \newsavebox{\inclboxa}
      \savebox{\inclboxa}{\parbox[b]{5mm}{{\tiny\text{incl.}\vspace{1mm}}}}
      \def\inclusiona{{\begin{rotate}{32}\usebox{\inclboxa}\end{rotate}}}
      \newsavebox{\inclboxb}
      \savebox{\inclboxb}{\parbox[b]{5mm}{{\tiny\text{incl.}\vspace{1mm}}}}
      \def\inclusionb{{\begin{rotate}{-32}\usebox{\inclboxb}\end{rotate}}}

      \begin{equation} \label{eq:mmap-2}
	\xymatrix@!C@!R{
	  & \hat{\UU} \ar[rr]^(.7){\py} \ar[ld]_{\inclusiona}\ar[dd]^(.7){/\U^+_F}|\hole && \V{E,0}^* \ar[dd]^(.7){/\U^+_F}
	  \ar[ld]^{\inclusiona}\\
	  T^*\hat{\Sigma} \ar[dd]_(.7){/\U^+_F} \ar[rr]^(.7){\hat{f}=\py} && \V{E}^*
	  \ar[dd]^(.7){/\U^+_F}\\
	  & \UU \ar[rr]^(.7){\fint}|(.5){\text{ \ \ }} \ar[dl]^{\inclusiona} && \V{E,0}^*/\U^+_F  \ar[rd]\ar[dr]^{\inclusionb}\ar[dl]^{\inclusiona}\\ 
	  T^* \Sigma  \ar[rr] \ar@/_8mm/[rrrr]^{f} && \V{E}^*/\U^+_F
	  \ar[rr]^{\text{collapse}} && \V{E}^*/\sim
	}
      \end{equation}
    }% % where $\hat{f}$ and $f$ are analogues of the same-named maps
    in \eqref{eq:mmap} and $\hat{f}=\py$ is the momentum-map of the
    torus $\V{E}/\Eta{E}$ acting on $T^*\hat{\Sigma}$. One can see
    that $\V{E,0}^*/\U^+_F$ is the complement of the coset of $0$ in
    $\V{E}^*/\sim$ and that the first-integral map $f$ is the natural
    extension of the map $\fint$ from $\UU$ to $T^*\Sigma$. From the
    diagram \eqref{eq:veo}, one can see that $\V{E}^*/\sim$ is
    homeomorphic to the cone on $\R^+\backslash\V{E,0}^*/\U^+_F$,
    where $\R^+$ acts by scalar multiplication. The diagram
    \eqref{eq:veo} also shows that when $F=E$, the fibres of $\q
    \times \caspb$ are disconnected, so that $\V{E}^*/\sim$ is a union
    of disjoint cones pinched at the cone point as in figure
    \eqref{fig:orbits}. When $F$ is a proper subfield of $E$, then the
    fibres of $\q \times \caspb$ are connected and $\V{E}^*/\sim$ is a
    cone on a connected space. (\listenum) There are globally-defined,
    $\U^+_F$-invariant functions on $\V{E}^*$. The most natural
    construction is a generalisation of the quadratic Casimir from the
    $3$-dimensional $Sol$ manifolds and the Casimir in the totally
    real case~\cite{BT:2000a,B7}. Each copy of $\V{\tau,E}$ may be
    naturally identified with $\V{E/F}$,%
    \footnote{Note that $\V{E/E} = \R$.}
     and similarly for the dual spaces. One can therefore define the
    map
    \begin{align} \label{al:casimir}
      \casimir(\py) &= \prod_{\tau \in \G{F}} \py_{\tau}, && \casimir
      : \V{E}^* \to S^d(\V{E/F})
    \end{align}
    where $d=[E:F]$ and $S^*(\V{E/F})$ is the vector space of polynomial
    functions on the vector space $\V{E/F}$. It is clear that $\casimir$ is
    $\U^+_F$ invariant. A simple computation shows that $\q \times
    \casimir$ is a submersion on $\V{E,0}^*/\U^+_F$. (\listenum) One might
    want to use the map
    $$
    \py \mapsto \sum_{\tau\in\B{F}} \frac{\py_{\tau}}{|\py_{\tau}|},
    \hspace{10mm} \V{E,0}^* \to  \prod_{\tau\in\B{F}} S^{d_{\tau}-1}
    \subset \sum_{\tau\in\B{F}} \V{\tau,E}^*
    $$ to ``split'' the fibre bundle in \eqref{eq:veo}. In general,
    however, the map induced by equivariance is not
    well-defined. Rather, to obtain a well-defined map by
    equivariance, the set $U_{\tau} = \left\{ \tau(u)/|\tau(u)|\ :\
    u\in\U^+_F \right\}$ needs to be finite for all $\tau \in \G{F}$;
    if one of these sets is not finite, then the co-domain of the
    induced map is not a manifold; if all the sets are finite, then
    the induced map's co-domain is a product of lens spaces so it does
    not split the fibre bundle, but it does split a suitable finite
    covering. Finiteness fails in many important cases: if $F$
    possesses a unit of infinite order on the unit circle, for
    example. (\listenum) The map $\fint$ induces a sub-algebra of
    $C^{\infty}(T^*\Sigma)$ by
    \begin{equation} \label{eq:R}
      \RR = \left\{ \fint^* h\ : \ h \in C^{\infty}(\V{E,0}^*/\U^+_F),
      \ h \text{ has compact support }\right\}.
    \end{equation}
    The sub-algebra $\RR$ is the substitute on $T^*\Sigma$ for the
    momentum map $P \mapsto \py$ on the level of algebras of
    functions.

}
\end{remark}

\section{Complete Integrability} \label{sec:ci}
Let $Z^{\infty}(\L^*)$ be the set of smooth Casimirs of $\L^*$ with
its standard Poisson bracket. This section proves that

\begin{thm} \label{thm:cas}
  Let $h \in Z^{\infty}(\L^*)$ be a Casimir and let $\lax[] : T^*\Sigma
  \to \L^*_R$ be the Lax matrix of equation (\ref{eq:L0-1})
  \begin{equation*} 
    \lax[](P) = \sum_{\sigma \in \G{E}} \e_{\r} \ \ + \ \ \Phi(\px) \ \ +
    \ \ \sum_{\sigma \in \G{E}} |\py_{\sigma}|^{b_{\sigma}}\cdot \exp(
    b_{\sigma} \cdot \left\langle \hat \sigma , \x \right\rangle )\cdot
    \e_{-\r},
  \end{equation*}
  where $\Phi : \V{o,F}^* \to \h^*$ satisfies the conclusions of Theorem
  (\ref{thm:lax}). Then, the following are true
  \begin{enumerate}
  \item $H := \lax[]^* h$ is a completely integrable
    Hamiltonian with smooth integrals;
  \item the algebras $\La:=\lax[]^*Z^{\infty}(\L^*)$ and $\RR$ form a dual pair;
  \item the singular set is an analytic variety.
  \end{enumerate}
\end{thm}

\begin{prooflist}
\item[(1-2)]\ Let $\hat\RR = \hat\Pi^* \RR$ be the pullback of $\RR$ to
  $T^*\hat \Sigma$. By the construction of $\RR$, $\hat\RR \subset
  \py^* C^{\infty}(\V{E}^*)$ and their functional dimension is equal
  on $\hat\UU$. 

  A Casimir $h$ of $\L^*$ is, {\it a fortiori}, invariant under the
  co-adjoint action of $\L_0$. Therefore, $\left. h
  \right|_{\L_{-1}^*+\L_0^*+\L_{+1}^*}$ must be functionally dependent
  on the co-adjoint invariants of $\L_0$, $\e_{\r}\cdot \e_{-\r}$, $\r
  \in \RsL$ and $x\in\L_0$. From the formula for $\lax[]$, the function
  $H=\lax[]^* h$ must therefore be a function of $\gamma_{\sigma} =
  |\py_{\sigma}|^{b_{\sigma}}\, \exp(b_{\sigma}\left\langle \hat
  \sigma,\x \right\rangle)$ and $\px$. These functions, and therefore
  $H$, are involution with $\hat\RR$.

  This proves that $\La$ and $\RR$ are commuting algebras of functions
  whose sum $\La+\RR$ is also abelian.

  Let $\reg \subset \e+\L_0^*+\L_{+1}^*$ be the set of regular points
  of the algebra $Z^{\infty}(\L^*)$ restricted to the subspace
  $\e+\L_0^*+\L_{+1}^*$, where $\e = \sum_{ \r \in \RsL }
  \e_{\r}$. This regular-point set is an open and dense real-analytic
  subset of $\e+\L_0^*+\L_{+1}^*$. Since $\left. \lax[] \right|_{\UU}$
  is an analytic submersion whose image is open in
  $\e+\L_0^*+\L_{+1}^*$, $\lax[]^{-1}( \reg )$ is an open and dense
  analytic subset of $\UU$.

  Therefore, for all $P \in \lax[]^{-1}(\reg)$,
  $$\dim \d \La_P = {\rm rank}\, \Rs = \dim \V{o,F}, \hspace{10mm} \dim \d
  \RR_P = \dim \V{E},$$
  while it is clear that
  $$\d \La_P \cap \d \RR_P = \{0\}.$$ Since $\dim \Sigma = \dim \V{o,F}
  + \dim \V{E}$, this proves (1-2).

\item[(3)]\ The singular set of $\RR + \La$ is the union of
  $\lax[]^{-1}(\reg^c)$ and $\ZZ = \caspb^{-1}(0)$. Both are
  real-analytic subsets of $T^*\Sigma$, hence their union is, too.
\end{prooflist}

\begin{proof}[Theorem \ref{thm:1}]
Let $A$ be maximal in the sense of definition \ref{de:max}. This
implies that $\Z^b$ is an irreducible $A$-module. Let $T \in
\GL{b;\C}$ be a matrix that conjugates $A$ to a subgroup of the set of
diagonal matrices in $\GL{b;\C}$ and let $\Gamma=T^{-1}AT$ and
$M=T^{-1}\Z^b$. Let $F$ be the extension field of $\Q$ that is
generated by the $(1,1)$-entries of $\gamma \in \Gamma$; since $A$ is
maximal, $F/\Q$ has degree $b$. The map $\delta$, defined for each
$\gamma \in \Gamma$ by,
\begin{align*}
\delta(\gamma) &:= \gamma_{11} & \delta : \Gamma \to \U_F
\end{align*}
is a group homomorphism. Indeed, the maps
$\delta_j(\gamma)=\gamma_{jj}$ are group homomorphisms into the group
of units of the $j$-th conjugate of $F$.

It is clear that the first column of the matrix $T$ can be supposed to
have entries in $\O_F$ and the $j$-th column of $T$ can be supposed to
be the $j$-th conjugate of the first column. It is claimed that $\det
T = q \cdot d$ where $q$ is a non-zero integer $q$ and $d$ is the
different of $F$. By definition, $d = \det U$ where the entries of the
first column of $U$ form a $\Z$-basis of $\O_F$ and the remaining
columns are the conjugates of the first column. Let $v$ be the first
column of $T$. If the entries of $v$ do not rationally span $F$, then
there is a non-zero $t \in \Z^b$ such that $\left\langle t , v
\right\rangle =0$. One can take the conjugates of this linear equation
and conclude that $t$ is orthogonal to each column of $T$ and
therefore $t=0$. Absurd. One concludes that the entries of $v$
generate a finite index subgroup of $\O_F$. The index of this subgroup
is $\det T/d$. This proves the claim. Therefore, for all $m \in M$,
the $j$-th entry of $qd \times m$ lies in the $j$-th conjugate of
$\O_F$. Define the map $\delta$ for each $m \in M$ by
\begin{align*}
\delta(m) &:= qd\cdot m_1& \delta : M \to \O_F,
\end{align*}
where $m_1$ is the first entry of $m$. It is clear that $\delta$ is a
morphism of modules that faithfully intertwines the representation of
$\Gamma$ on $M$ with that of $\delta(\Gamma)$ on $\delta(M)$; or,
$\delta$ extends to a group embedding $\xyinto[]{M \star \Gamma}{\O_F
\star \U_F}$ whence there is a group embedding $\xyinto[]{ \Z^b \star
A}{\O_F \star \U_F}$.

Because $\Z^b$ is an irreducible $A$-module, the degree of $F/\Q$ is
$b$ so $\Z^b$ is embedded as a finite index subgroup of $\O_F$. Since
$A$ is maximal, $A$ is embedded as a torsion-free, finite-index
subgroup of $\U_F$. Since $\delta(\Gamma)$ is torsion-free, there is a
choice of $\U^+_F$ such that $\delta(\Gamma) \subset
\U^+_F$. Therefore, one has obtained an embedding $\xyinto[]{ \Z^b
\star A}{\O_F \star \U^+_F}$ which is of finite index. This proves
that $\Sigma_A$ is a finite covering of the manifold $\Sigma$
constructed in lemma \eqref{lem:sigma} with $E=F$.

The proof of the theorem follows now by virtue of theorem
\ref{thm:cas} and the fact that the covering map $T^* \Sigma_A \to T^*
\Sigma$ is a local symplectomorphism.
\end{proof}

\subsection{Examples} \label{ssec:examples}
Let us illustrate the results of this section with two examples. 

\subsubsection{A non-normal cubic extension} \label{sssec:ex1}
To illustrate the construction behind Theorem~\ref{thm:1} take the
case where $A \lhd \GL{3;\Z}$ is the group generated by
\begin{align} \label{al:ex1-gen}
{\mathcal A}_1 =
\begin{bmatrix}
0 & 1 & 0\\
3 & 0 & 1\\
1 & 2 & 0
\end{bmatrix}, &&
{\mathcal A}_2 =
\begin{bmatrix}
-2 & 1  & 0\\
3  & -2 & 1\\
1  & 2  & -2
\end{bmatrix}.
\end{align}
$A$ is conjugate by a $T \in \SL{3;\R{}}$ to the group
$\Gamma$ generated by
\begin{align} \label{al:ex1-gen2}
{\mathcal B}_1 =
\begin{bmatrix}
\alpha_1 & 0        & 0\\
0       & \alpha_2 & 0\\
0       &  0       & \alpha_3
\end{bmatrix}, && 
{\mathcal B}_2 =
\begin{bmatrix}
\alpha_4 & 0        & 0\\
0       & \alpha_5 & 0\\
0       &  0       & \alpha_6
\end{bmatrix},
\end{align}
 where $\alpha_j$ for $j=1,2,3$ are the roots of the cubic
$f(x)=x^3-5x-1$ and $\alpha_j=\alpha_{j-3}-2$ for $j=4,5,6$. For
definiteness, one can take $T$ to be the matrix
\begin{equation} \label{eq:ex1-T}
T = 
\begin{bmatrix}
4 &  3 \alpha_3 + \alpha_2 &   3 \alpha_3 + \alpha_2\\
6 &  5 \alpha_2 + \alpha_1 &   5 \alpha_2^2 + \alpha_1^2\\
1 &     \alpha_3 &          \alpha_3^2
\end{bmatrix},
\end{equation}
whence $\det T=\sqrt{473}$, which is the different of $f$ and the
number field $F=\Q[\alpha_1]$. Let $M= T^{-1}(\Z^{3})$ and $\Delta = M
\star\Gamma$ so that $T^*\Sigma_A = T^*( \Delta \backslash \R^{3}
\times \R^{2})$. 

To define the Lax matrix in \eqref{eq:L0-2}, it is convenient to embed
$A$ by
{
\def\logA{{\begin{bmatrix}
\log |\alpha_{3i+1}| &&\\
& \log |\alpha_{3i+2}|&\\
&&\log |\alpha_{3i+3}|
\end{bmatrix}}}
\[
\xymatrix{
A \ar[rr]^{\log \circ \Ad{T^{-1}}} && \h, & {\mathcal A}_{i+1}^2
\ar@{{|}->}[r] & \logA
}
\]
}%
for $i=0,1$, where $\h \cong \R^2$ is the Cartan subalgebra of
$\SL{3;\R}$ consisting of trace zero diagonal $3\times 3$
matrices. This embeds $A$ as a lattice in $\h$. One can define the
coordinates for $P=(\py,\y,\px,\x) \in T^* \ucSigma=T^*\R^3 \times
T^*\h$ and thereby obtain the Lax matrix
\begin{align} \label{al:ex1-lax}
\lax[](P) = 
\begin{bmatrix}
0 & 0 & \lambda^{-1}\\
1 & 0 & 0\\
0 & 1 & 0
\end{bmatrix}
+
\begin{bmatrix}
\px_1 &    0 & 0\\
0    & \px_2 & 0\\
0    &    0 & \px_3
\end{bmatrix}
+
\frac{1}{2} \times
\begin{bmatrix}
0    & \delta_1 & 0\\
0    & 0        & \delta_2\\
\lambda\delta_3 & 0 & 0
\end{bmatrix}
\end{align}
where $\delta_i=|\py_i|^2\, \exp(2\x_i-2\x_{i+1})$ and $\sum \px_i=\sum
\x_i=0$. One obtains the two Poisson-commuting functions
\begin{align} \label{al:ex1-int}
\Htoda &= \frac{1}{2} \times \trace{\lax[]^2} = \frac{1}{2} \times
\left( \px_1^2 + \px_2^2 + \px_3^2 \right) +
\frac{1}{2} \times ( \delta_1 + \delta_2 + \delta_3 )\\
\mathbf{F} &= \frac{1}{3} \times \trace{\lax[]^3} \equiv \frac{1}{3} \times \left( \px_1^3 + \px_2^3 + \px_3^3
\right) - \frac{1}{2} \times \left( \delta_1 \px_3 + \delta_2 \px_1 +
\delta_3 \px_2  \right) %+ \frac{1}{8} \lambda + \lambda^{-1}
\end{align}
that are in involution with $\py$. 

One may permute the indices $i$; it is clear that a cyclic permutation yields
the same $\Htoda$ and it is not difficult to see that transpositions yield
equivalent hamiltonians (remark \ref{re:cubic}).

\subsubsection{A non-normal cubic extension and $\Z^6$} \label{sssec:ex2}
To illustrate the construction behind Theorem~\ref{thm:1} take the
case where $A \lhd \GL{6;\Z}$ is the group generated by
\begin{align} \label{al:ex2-gen}
{\mathcal A}_1 =
\begin{bmatrix}
0 & 2 & -4 & 0 & 1 & -2\\
0 & 0 & 0 & -2 & 2 & 1\\
-1 & 0 & 0 & -1 & 0 & 1\\
0 & 0 & -1 & 0 & 0 & -2\\
0 & 1 & -1 & 0 & 0 & 1\\
0 & 0 & 0 & -1 & 1 & 0
\end{bmatrix}, &&
{\mathcal A}_2 =
\begin{bmatrix}
-2 & 2 & -4 & 0 & 1 & -2\\
0 & -2 & 0 & -2 & 2 & 1\\
-1 & 0 & -2 & -1 & 0 & 1\\
0 & 0 & -1 & -2 & 0 & -2\\
0 & 1 & -1 & 0 & -2 & 1\\
0 & 0 & 0 & -1 & 1 & -2
\end{bmatrix}.
\end{align}
$A$ is conjugate by a $T \in \SL{6;\R{}}$ to the group
$\Gamma$ generated by
\begin{align} \label{al:ex2-gen2}
{\mathcal B}_1 =
\begin{bmatrix}
\alpha_1I_2 & 0        & 0\\
0       & \alpha_2I_2 & 0\\
0       &  0       & \alpha_3I_2
\end{bmatrix}, && 
{\mathcal B}_2 =
\begin{bmatrix}
\alpha_4I_2 & 0        & 0\\
0       & \alpha_5I_2 & 0\\
0       &  0       & \alpha_6I_2
\end{bmatrix},
\end{align}
where $I_2$ is the $2\times 2$ identity matrix and $\alpha_j$ for
$j=1,2,3$ are the roots of the cubic $f(x)=x^3-5x-1$ and
$\alpha_j=\alpha_{j-3}-2$ for $j=4,5,6$. One notes that the matrix
${\mathcal A}_1$ is the matrix of the root $\alpha_1$ acting on the
integers of $\O_E$, where $E=\Q[\alpha_1,\sqrt{473}]$ is the normal
closure of the field $F$ of the previous example. There is not a
simple expression for such a matrix $T$, because unlike the previous
example ${\mathcal A}_1$ is not conjugate over $\Z$ to its companion
matrix. In all events, let $M= T^{-1}(\Z^{6})$ and $\Delta = M
\star\Gamma$ so that $T^*\Sigma_A = T^*( \Delta \backslash \R^{6}
\times \R^{2})$.

To define the Lax matrix in \eqref{eq:L0-2}, it is convenient to embed
$A$ by
{
\def\logA{{\begin{bmatrix}
\log |\alpha_{3i+1}| I_2 &&\\
& \log |\alpha_{3i+2}| I_2&\\
&&\log |\alpha_{3i+3}| I_2
\end{bmatrix}}}
\[
\xymatrix{
A \ar[rr]^{\log \circ \Ad{T^{-1}}} && \h, & {\mathcal A}_{i+1}
\ar@{{|}->}[r] & \logA
}
\]
}% for $i=0,1$, where $\h \cong \R^5$ is the Cartan subalgebra of $\SL{6;\R}$
consisting of trace zero diagonal $6\times 6$ matrices. This embeds $A$ as a
lattice in the subspace $\h_2 \subset \h$ consisting of matrices which are of
the form $B \otimes I_2$ for a $3 \times 3$ diagonal, trace zero matrix
$B$. One can define the coordinates for $P=(\py,\y,\px,\x) \in T^*
\ucSigma=T^*\R^6 \times T^*\h_2$ and thereby obtain the Lax matrix
\begin{align} \label{al:ex2-lax}
\lax[](P) = 
\begin{bmatrix}
0 & 0 & \lambda^{-1} I_2\\
I_2 & 0 & 0\\
0 & I_2 & 0
\end{bmatrix}
+
\begin{bmatrix}
\px_1 I_2 &    0 & 0\\
0    & \px_2 I_2 & 0\\
0    &    0 & \px_3 I_2
\end{bmatrix}
+
\frac{1}{2} \times
\begin{bmatrix}
0    & \delta_1 I_2 & 0\\
0    & 0        & \delta_2 I_2\\
\lambda\delta_3 I_2 & 0 & 0
\end{bmatrix}
\end{align}
where $\delta_i=|\py_i|^2\, \exp(2\x_i-2\x_{i+1})$ and $\sum \px_i=\sum
\x_i=0$. One obtains the two Poisson-commuting functions
\begin{align} \label{al:ex2-int}
\Htoda &= \frac{1}{4} \times \trace{\lax[]^2} = \frac{1}{2} \times
\left( \px_1^2 + \px_2^2 + \px_3^2 \right) +
\frac{1}{2} \times ( \delta_1 + \delta_2 + \delta_3 )\\
\mathbf{F} &= \frac{1}{6} \times \trace{\lax[]^3} \equiv \frac{1}{3} \times \left( \px_1^3 + \px_2^3 + \px_3^3
\right) - \frac{1}{2} \times \left( \delta_1 \px_3 + \delta_2 \px_1 +
\delta_3 \px_2  \right) %+ \frac{1}{8} \lambda + \lambda^{-1}
\end{align}
that are in involution with $\py$ ($\equiv$ indicates equality modulo
functions of $\py$). 

\subsubsection{A non-normal quartic extension} \label{sssec:ex3}
{
To illustrate the construction behind Theorem~\ref{thm:1} take the
case where $A \lhd \GL{4;\Z}$ is the group generated by
\begin{align} \label{al:ex3-1-gen}
{\mathcal A}_1 =
\begin{bmatrix}
1 & -1 & -1 & 1\\
1 & 0 & 0 & 1\\
0 & -1 & 0 & 0\\
0 & 1 & 0 & 1\\
\end{bmatrix}, &&
{\mathcal A}_2 =
\begin{bmatrix}
0 & 1 & 1 & 1\\
0 & 1 & 0 & 1\\
0 & 0 & 0 & -1\\
1 & 0 & -1 & 1\\
\end{bmatrix}.
\end{align}
$A$ is conjugate by a $T \in \SL{4;\R{}}$ to the group
$\Gamma$ generated by
\begin{align} \label{al:ex3-1-gen2}
{\mathcal B}_1 = \diag{\alpha_1,\ldots,\alpha_4},
&& 
{\mathcal B}_2 =  \diag{\alpha_5,\ldots,\alpha_8},
\end{align}
{%
\def\nume{{\left(1+s\sqrt{2} + t\sqrt{(1+s\sqrt{2})^2-4}\right)}}%
where $\alpha_j$ for $j=1,\ldots,4$ are the roots of the palindromic
quartic $f(x)=x^4-2x^3+x^2-2x-1$ and
$\alpha_j=\alpha_{j-4}^3-\alpha_{j-4}^2-1$ for $j=5,\ldots,8$. The
roots $\alpha_j$ equal $\ds{\frac{1}{2} \times \nume}$ where $s,t \in
\{\pm 1\}$, $j=1,\ldots,4$. This gives two real reciprocal roots that
are approximately $1.883$ and $0.531$ and two conjugate complex roots
on the unit circle that are approximately $0.207 \pm 0.978
\sqrt{-1}$. Since $f$ is $\Q$-irreducible, the complex roots are not
roots of unity, which also implies that the largest positive root is a
Salem number. One notes that the matrix ${\mathcal A}_1$ is the matrix
of the root $\alpha_1$ acting on the integers of $\O_F$, where
$F=\Q[\alpha_1]$.

As with example \ref{sssec:ex1}, one can compute a straightforward
representation of $T$
\begin{equation} \label{eq:ex3-T}
\begin{bmatrix}
-1 & \alpha_4-\alpha_3+\alpha_2-2\,\alpha_1 & \alpha_4^2-\alpha_3^2+\alpha_2^2-2\,\alpha_1^2 & \alpha_4^3-\alpha_3^3+\alpha_2^3-2\,\alpha_1^3\\
 0 & \alpha_4-\alpha_3+\alpha_2-\alpha_1  & \alpha_4^2-\alpha_3^2+\alpha_2^2-\alpha_1^2 & \alpha_4^3-\alpha_3^3+\alpha_2^3-\alpha_1^3 \\
 1 & -\alpha_4+\alpha_3+\alpha_1 & -\alpha_4^2+\alpha_3^2+\alpha_1^2 & -\alpha_4^3+\alpha_3^3+\alpha_1^3\\
 2 & \alpha_3+\alpha_1 & \alpha_3^2+\alpha_1^2 & \alpha_3^3+\alpha_1^3\\
\end{bmatrix}
\end{equation}
and one can verify that $\det T=-8\sqrt{-7}$, which is the different
of $F$. In all events, let $M= T^{-1}(\Z^{4})$ and $\Delta = M
\star\Gamma$ so that $T^*\Sigma_A = T^*( \Delta \backslash \R^{4}
\times \R^{2})$. 
}

To define a Lax matrix as in \eqref{eq:L0-2}, it is convenient to
embed $A$ into the Cartan subalgebra $\h \cong \R^2$ of the real
symplectic group of $4 \times 4$ matrices
\[
\h = \left\{ 
\diag{a,b,-a,-b} :\ a,b\in \R
\right\}
\]
by the embedding
{%
\def\logA{{\diag{\log |\alpha_{4i+1}|,\log |\alpha_{4i+2}|,\log |\alpha_{4i+4}|,\log |\alpha_{4i+3}|}}}
\[
\xymatrix{
A \ar[rr]^{\log \circ \Ad{T^{-1}}} && \h, & {\mathcal A}_{i+1}
\ar@{{|}->}[r] & \logA
}
\]
}%%%%%%%%%%%%%%%%%%%%%%%%%%%%%%%%%%%%%%%%%%%%%%%%%%%%%%%%%%%%%%%%%%%%%%%%%%%%%%%%%%%%%%%%%%%%%%%%%%%%%%%%%%%%%%%%%%%%%%%%%%
for $i=0,1$. To make this an embedding, one must stipulate that the roots
$\alpha_j$ and $\alpha_{5-j}$ must be reciprocals for $j=1,2$; it is also
supposed that $\alpha_1$ (resp. $\alpha_2$) has positive imaginary part
(resp. is the largest real root of $f$). This embeds $A$ as a lattice in
$\h$. One can define the coordinates for $P=(\py,\y,\px,\x) \in T^*
\ucSigma=T^*\R^4 \times T^*\h$ and thereby obtain the Lax matrix $\lax[](P)$
\begin{equation} \label{al:ex3-1-lax}
\begin{bmatrix}
0 & 1 & 0 & 0\\
0 & 0 & 0 & 1\\
\lambda^{-1} & 0 & 0 & 0\\
0 & 0 & -1 & 0
\end{bmatrix}
+
\overbrace{
\begin{bmatrix}
a_1\px_1 & 0 & 0 & 0\\
0    & a_2\px_2 & 0 & 0\\
0    &    0 & -a_1\px_1 & 0\\
0    &    0 & 0 & -a_2\px_2
\end{bmatrix}
}^{\Phi(\px)}
+
\frac{1}{2} \times
\begin{bmatrix}
0 & 0 & \lambda \delta_3 & 0\\
\delta_1 & 0 & 0 & 0\\
0 & 0 & 0 & -\delta_1\\
0 & \delta_2 & 0 & 0
\end{bmatrix}
\end{equation}
where $\delta_i,a_i$ are determined in Table \ref{tab:ex3-1-lax-matrices-delta}. One obtains the two Poisson-commuting functions
\begin{align}
\Htoda &= \frac{1}{4} \times \trace{\lax[]^2} = \frac{1}{2} \times
\left( a_1^2\px_1^2 + a_2^2\px_2^2 \right) +
\frac{1}{2} \times ( \delta_1 + \frac{1}{2} \delta_2 + \frac{1}{2} \delta_3 )  \label{al:ex3-1-int-h}\\
\mathbf{F} &= \det \lax[] \equiv
\frac{\delta_{2} \delta_{3}}{4}  + \frac{\delta_{1}^2}{4} - \delta_{1} a_1a_2\px_{1} \px_{2} + \frac{a_2^2\px_{2}^2 \delta_{3}}{2} + \frac{a_1^2\px_{1}^2 \delta_{2}}{2} + a_1^2a_2^2\px_{1}^2 \px_{2}^2 \label{al:ex3-1-int-f}
%%
%%\mathbf{F} &= \frac{1}{2} \times \trace{\lax[]^4} \equiv 
%%\frac{\delta^2_1}{2} + \frac{\delta^2_2}{4} + \frac{\delta^2_3}{4} + \delta_1 \delta_2 + \delta_1 \delta_3 + X_{0}^2 \left(\delta_3 + 2 \delta_1\right) + X_{1}^2 \left(\delta_2 + 2 \delta_1\right) + 2 X_{0} \delta_1 X_{1} + X_0^4 + X_1^4%%-\frac{\delta^2_1 \delta_2 \delta_3 \lambda}{8}-\frac{2}{\lambda} + 
\end{align}
that are in involution with $\py$ ($\equiv$ indicates equality modulo
functions of $\py$).  }

To explain the following choices for the functions $\delta_i$, one defines the
embeddings of the number field $F$ by $\tau_i(\alpha_1) = \alpha_i$, so that
$\B{F}=\{\tau_1,\tau_2,\tau_3\}$ and $\tau_4=\bar{\tau}_1$. A bijection $\rho
: \B{F} \to \RsL$ is identified as a permutation $s$ of $\{1,2,3\}$ under the
convention that $\rho(\tau_i)=\r_{s(j)}$. Only three choices are listed since
the remaining three are obtained by permuting $\py_2$ and $\py_3$ in the
formulae below (these unlisted choices are also conjugate to the listed
choices, since this permutation induces an analytic symplectomorphism of
$T^*\Sigma$).
{
\def\a{{\phantom{1}}}
\begin{center}
\begin{table}[htb]
\newcolumntype{R}{>{$}r<{$}}
\newcolumntype{C}{>{$}c<{$}}
\setlength{\extrarowheight}{2pt}
\begin{tabular}{|*{3}{R|}*{1}{C|}*{3}{R|}}
\hline
c & \rho & b_\tau & a_1,a_2 & \delta_1 & \delta_2 & \delta_3\\\hline\hline
2 & (1)    & 2,2,2 & 1,1                     & 2|\py_1|^2e^{2\x_1-2\x_2} &  |\py_2|^2e^{4\x_2\a} &  |\py_3|^2e^{-4\x_1\a} \\
4 & (2\,1) & 8,2,4 & \frac{1}{2},\frac{1}{4} &  |\py_2|^2e^{4\x_1-8\x_2} & 2|\py_1|^8e^{16\x_2}  &  |\py_3|^4e^{-8\x_1\a} \\
4 & (3\,1) & 8,4,2 & \frac{1}{4},\frac{1}{2} &  |\py_3|^2e^{8\x_1-4\x_2} &  |\py_2|^4e^{8\x_2\a} & 2|\py_1|^8e^{-16\x_1}\\\hline
\end{tabular}
\caption{Choices for the Lax matrix $\lax[]$; $\y_i$ ($\py_i$) is a coordinate on the $\alpha_i$-eigenspace with
  $\y_1=\bar{\y}_4$ ($\py_1=\bar{\py}_4$). See Theorem \ref{thm:lax}. } \label{tab:ex3-1-lax-matrices-delta}
\end{table}
\end{center}
}%%
With these choices of $\delta_i$, \eqref{al:ex3-1-lax} gives a Lax representation
of the hamiltonian vector field of $\Htoda$ \eqref{al:ex3-1-int-h} with the integral
$\mathbf{F}$ \eqref{al:ex3-1-int-f}. Although fibrewise convex for all choices,
the hamiltonian $\Htoda$ is only fibre-wise quadratic for the first choice.

\subsubsection*{Additional Lax Representations} 
One can define additional Lax representations with the aid of the remaining
rank $2$ affine Kac-Moody algebras.

\subsubsection*{$A^{(1)}_2$} 
Embed $A$ into the Cartan subalgebra $\h \cong \R^2$ of $\SL{3;\R}$ via
{%
\def\logA{{\diag{2\log |\alpha_{4i+1}|,\log |\alpha_{4i+2}|,\log |\alpha_{4i+3}|}}}
\[
\xymatrix{
A \ar[rr]^{\log \circ \Ad{T^{-1}}} && \h, & {\mathcal A}_{i+1}
\ar@{{|}->}[r] & \logA
}
\]
}%%%%%%%%%%%%%%%%%%%%%%%%%%%%%%%%%%%%%%%%%%%%%%%%%%%%%%%%%%%%%%%%%%%%%%%%%%%%%%%%%%%%%%%%%%%%%%%%%%%%%%%%%%%%%%%%%%%%%%%%%%
where the roots $\alpha_j$ are labelled as above. This embeds $A$ as a lattice
in $\h$. One can define the coordinates for $P=(\py,\y,\px,\x) \in T^*
\ucSigma=T^*\R^4 \times T^*\h$ and thereby obtain the Lax matrix
\begin{align} \label{al:ex3-2-lax}
\lax[](P) = 
\begin{bmatrix}
0 & 0 & \lambda^{-1}\\
1 & 0 & 0\\
0 & 1 & 0
\end{bmatrix}
+
\begin{bmatrix}
a_1\px_1 &    0 & 0\\
0    & a_2\px_2 & 0\\
0    &    0 & a_3\px_3
\end{bmatrix}
+
\frac{1}{2} \times
\begin{bmatrix}
0    & \delta_1 & 0\\
0    & 0        & \delta_2\\
\lambda\delta_3 & 0 & 0
\end{bmatrix}
\end{align}
where $\sum a_i\px_i=\sum a_i^{-1}\x_i=0$ and $\delta_i$ is defined
below. One obtains the two Poisson-commuting functions $\Htoda =
\frac{1}{2} \times \trace{\lax[]^2}$ and $\mathbf{F} = \det \lax[]$ where
\begin{align}
\Htoda &= a_{1}^2 \px_{1}^2 + a_{1} \px_{1} a_{2} \px_{2} + a_{2}^2 \px_{2}^2 + \frac{1}{2} \times \left(\delta_1+\delta_2+\delta_3 \right) \label{al:ex3-2-int-h}\\
\mathbf{F} &\equiv  - a_{1} \px_{1} a_{2}^2 \px_{2}^2 - a_{1}^2 \px_{1}^2 a_{2} \px_{2} + \frac{1}{2}\, a_{1} \px_{1}  \left(\delta_{1}
- \delta_{2}\right) + \frac{1}{2}\, a_{2} \px_{2} \left(\delta_{1} - \delta_{3}\right),\label{al:ex3-2-int-f}
\end{align}
where the functions $\delta_i$ are determined in table
\ref{tab:ex3-2-lax-matrices-delta}, following the conventions in table \ref{tab:ex3-1-lax-matrices-delta}.
\begin{center}
\begin{table}[htb]
\newcolumntype{R}{>{$}r<{$}}
\newcolumntype{C}{>{$}c<{$}}
\setlength{\extrarowheight}{2pt}
\begin{tabular}{|*{3}{R|}*{1}{C|}*{3}{R|}}
\hline
c & \rho & b_\tau & a_1,a_2 & \delta_1 & \delta_2 & \delta_3\\\hline\hline

2 & (1)    & 2,2,4 & \frac{1}{2},1 & 2|\py_{1}|^2e^{2\x_{2}-4\x_{1}} &  |\py_{2}|^2e^{-4\x_{2}-4\x_{1}} &  |\py_{3}|^4e^{2\x_{2}+8\x_{1}}\\
2 & (1\,2) & 2,4,2 & 1,\frac{1}{2} &  |\py_{2}|^4e^{4\x_{2}-2\x_{1}} & 2|\py_{1}|^2e^{-8\x_{2}-2\x_{1}} &  |\py_{3}|^2e^{4\x_{2}+4\x_{1}}\\
2 & (1\,3) & 4,2,2 & 1,1           &  |\py_{3}|^2e^{2\x_{2}-2\x_{1}} &  |\py_{2}|^2e^{-4\x_{2}-2\x_{1}} & 2|\py_{1}|^4e^{2\x_{2}+4\x_{1}}\\\hline
\end{tabular}
\caption{Choices for the Lax matrix $\lax[]$; $\y_i$ ($\py_i$) is a coordinate on the $\alpha_i$-eigenspace with
  $\y_1=\bar{\y}_4$ ($\py_1=\bar{\py}_4$). See Theorem \ref{thm:lax}. } \label{tab:ex3-2-lax-matrices-delta}
\end{table}
\end{center}

\subsubsection*{$G_2^{(1)}$}
One proceeds as above and obtains the hamiltonian 
\begin{equation} \label{eq:ex3-3-int-h}
\Htoda =\frac{1}{24}\times\left(a_{1}^2\px_{1}^2+3a_{1}\px_{1}a_{2}\px_{2}+3a_{2}^2\px_{2}^2\right)+16\times(3\delta_{1}+\delta_{2}+\delta_{3})
\end{equation}
where $\delta_i$ is defined by
\begin{center}
\begin{table}[htb]
{
\def\a{\phantom{1}}
\def\b{\phantom{112\x_{1}}}
\newcolumntype{R}{>{$}r<{$}}
\newcolumntype{C}{>{$}c<{$}}
\setlength{\extrarowheight}{2pt}
\begin{tabular}{|*{3}{R|}*{1}{C|}*{3}{R|}}
\hline
c & \rho & b_\tau & a_1,a_2 & \delta_1 & \delta_2 & \delta_3\\\hline\hline
12 & (1)       & 8,6,12 & \frac{1}{4},\frac{1}{3} & 2|\py_{1}|^8e^{16\x_{1}-6\x_{2}\a} &  |\py_{2}|^{6\a}e^{12\x_{2}-24\x_{1}}  &  |\py_{3}|^{12} e^{-6\x_{2}\a} \\
12 & (3\,2)    & 8,12,6 & \frac{1}{4},\frac{1}{3} & 2|\py_{1}|^8e^{16\x_{1}-12\x_{2}}  &  |\py_{3}|^{6\a}e^{24\x_{2}-24\x_{1}}  &  |\py_{2}|^{12} e^{-12\x_{2}} \\
24 & (3\,2\,1) & 24,4,6 & \frac{1}{3},\frac{1}{12}&  |\py_{3}|^6e^{48\x_{1}-4\x_{2}\a} & 2|\py_{1}|^{24} e^{8\x_{2}-72\x_{1}\a} &  |\py_{2}|^{4\a}e^{-4\x_{2}\a} \\
12 & (2\,1)    & 6,2,6  & 1,\frac{1}{3}           &  |\py_{2}|^2e^{12\x_{1}-2\x_{2}\a} & 2|\py_{1}|^{6\a}e^{4\x_{2}-18\x_{1}\a} &  |\py_{3}|^{6\a}e^{-2\x_{2}\a} \\
12 & (2\,3\,1) & 6,6,2  & \frac{1}{3},1           &  |\py_{2}|^6e^{12\x_{1}-6\x_{2}\a} &  |\py_{3}|^{2\a}e^{12\x_{2}-18\x_{1}}  & 2|\py_{1}|^{6\a}e^{-6\x_{2}\a} \\
24 & (3\,1)    & 24,6,4 & \frac{1}{2},\frac{1}{3} &  |\py_{3}|^4e^{48\x_{1}-6\x_{2}\a} &  |\py_{2}|^{6\a}e^{12\x_{2}-72\x_{1}}  & 2|\py_{1}|^{24} e^{-6\x_{2}\a} \\
\hline
\end{tabular}
}
\caption{Choices for the Lax matrix $\lax[]$; $\y_i$ ($\py_i$) is a
  coordinate on the $\alpha_i$-eigenspace with $\y_1=\bar{\y}_4$
  ($\py_1=\bar{\py}_4$) and $\x_i$ is the coordinate on $\h$ induced
  by the simple
  coroots~\cite[p. 346]{FH}.} \label{tab:ex3-3-lax-matrices-delta}
\end{table}
\end{center}

\ifthenelse{\shortfile=3}{\end{document}}{}

\section{The singular set and gradient flows} \label{sec:gradientflow} 
Two prefatory comments: first, the fibre bundle structure
\[
\xyexact[\V{E}/\O_E]{\Sigma}{\Vo{F}/\logu{F}}{\p}{}
\]
induces the sub-bundle $\Vert = \ker \d\p \subset T \Sigma$ and its
annihilator $\Hstar \subset T^* \Sigma$. The sub-bundle $\Hstar$ is
naturally isomorphic to $\Sigma \times \Vo{F}^*$. Second, recall that
the {\em stable manifold} of a point $p$ is the set of points whose
orbits converge to that of $p$'s as time goes to $\infty$; the {\em
unstable manifold} is defined symmetrically as time goes to $-\infty$;
the stable and unstable manifolds of a set are the union of the stable
and unstable manifolds of each point in the set. In this section it is
shown that

\begin{thm} \label{thm:nh}
  $\Hstar$ is an invariant set for the Hamiltonian flow of $\Htoda$
  (equation \ref{eq:HL0}). The stable and unstable manifolds of
  $\Hstar$, $\W{}^{\pm}(\Hstar)$, coincide and
  \begin{equation} \label{eq:wpm}
    \W{}^{\pm}(\Hstar) = \ZZ \hspace{10mm}(=\caspb^{-1}(0)).
  \end{equation}
\end{thm}

Before proceeding with the proof, let us explain why theorem
\ref{thm:nh} is natural from the perspective of \BT\  lattices. It is a
well-known result that the open \BT\  lattices undergo scattering:
the particles interact over some time interval and then separate and
proceed off to infinity. The net result of the interaction is that the
momenta of the particles may be permuted from $t=-\infty$ to
$t=\infty$; in terms of the Lax matrix, $\lax[](-\infty)$ and
$\lax[](\infty)$ are diagonal matrices which differ by the action of
some element in the Weyl group. Since the open \BT\  lattices are
obtained from the periodic \BT\  lattices by turning off the potential
term associated to the root $\bminrt$, it is plausible that when other
potential terms are turned off, the system should still exhibit such
scattering behaviour. To confirm this, one must develop the
double-bracket or gradient representation of these systems.

\subsection{Double-bracket and gradient representations} \label{ssec:db} 
Let us recall the constructions of \cite{BG}, where it is demonstrated
that the open \BT\ lattices may be viewed as gradient flows. Let $\g$
be a semi-simple Lie algebra with Cartan-Killing form $\kappa =
\left\langle\langle , \right\rangle\rangle$. For $x \in \g^*$ let
$\orb{x}$ denote the co-adjoint orbit of $x$, let $\g_x$ be the
stabiliser algebra of $x$ and let $\gp_x$ be the $\kappa$-orthogonal
complement of $\g_x$. The map $v \mapsto \ad{v}x$ is a linear
isomorphism of $\gp_x$ with $T_x\orb{x}$.

\begin{defn} \label{de:nm}
  The {\em normal metric}, $\nm$, on $\orb{x}$ is defined at
  $T_x\orb{x}$ by
  \begin{equation} \label{eq:nm}
    \forall u,v\in\gp_x: \hspace{10mm}\nm(\ad{u}x,\ad{v}x) =
    \left\langle\langle u, v \right\rangle\rangle
  \end{equation}
\end{defn}

\begin{lemma} \label{lem:nm}
  If $H \in C^{\infty}(\g^*)$, then the gradient vector field of
  $H|\orb{x}$ at $x$ is
  \begin{equation} \label{eq:gnm}
    \gradnm H(x) = - [x,[x,y]] \in T_x \orb{x}
  \end{equation}
  where $y = \nabla H(x)$ is the $\kappa$-gradient of $H$.
\end{lemma}

For a proof, see \cite{BG}.

\subsection{\BT\  lattices and double brackets} \label{ssec:db2} 
Let us specialise the construction of the previous section. The
semi-simple Lie algebra is the loop algebra $\L$ or its twisted
counterpart of section \ref{ssec:split}. Let $\RsL_0 \subsetneq \RsL$
be a proper subset obtained by removing a single root from $\RsL$. Let
\begin{xalignat}{3} \label{eq:x}
  x &= h + \sum_{\r \in \RsL_0} x_{\r} (\e_{\r} + \e_{-\r}) \in
  \L^*, &
  m &= \sum_{\r \in \RsL_0} x_{\r} (\e_{\r} - \e_{-\r}), & 
  X(x) &= [x,m],
\end{xalignat}
where $h\in\h$. The vector field $X$ is a \BT-like vector field
associated to the splitting of $\Lnot \subset \L$ as in section
\ref{ssec:car}.

\begin{lemma} \label{lem:gtoda}
  $X$ is a gradient vector field relative to the normal metric, hence
  $X$ is tangent to $\orb{x}$.
\end{lemma}

\begin{proof}
  It suffices to determine a $y \in \h$ such that $X = \gradnm H$ where
  $H(x) = \left\langle\langle x , y \right\rangle\rangle$. To do so, it
  suffices to determine $y$ such that $m = -[x,y]$. This reduces to the
  solubility of the equations 
  \begin{equation} \label{eq:xr}
    \forall \r \in \RsL: \qquad x_{\r} = x_{\r} \left\langle r , y \right\rangle.
  \end{equation}
  Since at least one of the $x_{\r}$ vanishes, and any subset of $\RsL$
  of cardinality $\#\RsL - 1$ restricts to a basis of $\h^*$, there is
  always a solution to (\ref{eq:xr}). 
\end{proof}

The vector field $X$ is equivalent to the differential equations
\begin{equation} \label{eq:X}
  -\dot{h} = \sum_{\r\in\RsL_0} 2 x_{\r}^2\, h_{\r}, \qquad{\rm and}\qquad \forall
  \r\in\RsL: \quad \dot{x}_{\r} = x_{\r} \left\langle r , h \right\rangle,
\end{equation}
where $h_{\r} = [\e_{\r},\e_{-\r}]$. In particular, $X$ is tangent to
$x_{\r}=0$ for any $\r$. It is also clear that $X$ vanishes at $x$ iff 
\begin{equation} \label{eq:van}
  \forall \r[s] \in\RsL: \qquad x_{\r[s]}\left\langle s , h \right\rangle=0
  \quad {\rm and} \quad \sum_{\r\in\RsL_0} 2 x_{\r}^2\,
  \left\langle\langle s , r \right\rangle\rangle = 0,
\end{equation}
where the identity $\left\langle s, h_{\r} \right\rangle =
\left\langle\langle s , r \right\rangle\rangle$ has been used. Since
the matrix $\left[ \left\langle\langle s ,r \right\rangle\rangle
  \right]_{r,s\in\RsL_0}$ has full rank, the second part of
(\ref{eq:van}) implies that $x_{\r}=0$ for all $\r\in\RsL_0$ and
therefore for all $\r\in\RsL$. This proves that

\begin{lemma} \label{lem:Xzero}
  $X$ vanishes at $x$ iff $x\in\h$.
\end{lemma}

It remains to prove that all orbits of $X$ limit onto $\h$. Since
$\dot{H} = \left\langle\langle y , -[x,[x,y]] \right\rangle\rangle =
\left\langle\langle \ad{y}x,\ad{y}x \right\rangle\rangle$, and
$\ad{y}x = -\sum_{\r\in\RsL} x_{\r}\,\left\langle r , y
\right\rangle\,(\e_{\r}-\e_{-\r})$ one concludes from (\ref{eq:xr})that
\begin{equation} \label{eq:hdot}
  \dot{H} = -2 \sum_{\r\in\RsL}x_{\r}^2\, \leq 0
\end{equation}
with equality iff $X=0$. Thus, the $\omega$-limit set of every point
$x$ lies in $\h$, hence $\orb{x}\cap \h$. The latter is a finite set
and since $X$ is a gradient vector field on $\orb{x}$, the
$\omega$-limit set is a single point. Let $h_0 \in \orb{x} \cap \h$ be
this point and let $\RsL_1 = \{\r\ :\ x_{\r} = 0\}$. Let us linearise
$X$ about $h_0$ subject to the condition that $x_{\r}=0$ for all
$\r\in\RsL_1$:
\begin{equation} \label{eq:DX}
  -\delta\dot{ h} = 0, \qquad{\rm and}\qquad \forall\r\not\in\RsL_1: \quad \delta\dot{ x}_{\r} = \delta x_{\r} \left\langle r , h_0 \right\rangle,
\end{equation}
where $\delta x, \delta h$ denote variations. It is clear that a
necessary condition for stability of $h_0$ is that $\left\langle r ,
h_0 \right\rangle \leq 0$ for all $\r\not\in\RsL_1$. A simple argument
involving the transitivity of the action of the Weyl group on the Weyl
chambers, shows that such an $h_0$ must exist. This proves
\begin{lemma} \label{lem:X}
  For each $x$ of the form in equation (\ref{eq:x}), the $\omega$-limit
  set of $x$ under the gradient flow of $X = \gradnm H$ is a point $h_0
  \in \orb{x}\cap \h$ that satisfies $\left\langle r ,h_0 \right\rangle
  \leq 0$ for all $\r\in\RsL_1$.
\end{lemma}

A similar statement is true for the $\alpha$-limit set, too. It should
be observed that while $\h$ contains the $\omega$-limit set of every
point $x$, $\h$ is not a normally hyperbolic manifold. One can see this
from (\ref{eq:DX}): when $\left\langle h_0 ,r \right\rangle = 0$, one
loses hyperbolicity.

\begin{proof}[Theorem \ref{thm:nh}]
  For each $\tau\in\G{F}$ the Hamiltonian vector field of $\Htoda$ in
  (\ref{eq:HL0}), when restricted to the invariant set
  $g_{\tau}^{-1}(0)$ (equation \ref{eq:g}), is semi-conjugate to a
  vector field of the form of $X$ in (\ref{eq:x}). The semi-conjugacy is
  provided by the Lax representation in equation (\ref{eq:L0-3}). Lemma
  \ref{lem:X} implies that the $\omega$-limit set of a point $P\in
  g_{\tau}^{-1}(0)$ lies in $\Hstar$. Similarly for the $\alpha$-limit
  set of $P$. Since $\caspb^{-1}(0) =
  \cup_{\tau\in\G{F}}g_{\tau}^{-1}(0)$, this proves the theorem.
\end{proof}

\ifthenelse{\shortfile=4}{\end{document}}{}

\section{Uniqueness up to Energy-Preserving Topological Conjugacy} \label{sec:ex}

\subsection{Marked Homology Spectrum of a Flow}
Two flows $\phi : M\times\R \to M$ and $\varphi : N\times\R \to N$ are
topologically conjugate if there is a homeomorphism $h : M \to N$ such
that $h \phi_t = \varphi_t h$ for all $t\in\R$. Let $\PerO_{\phi}$ be
the set of periodic points of the flow $\phi$. For each periodic orbit
$\gamma$ of $\phi$, let the homology class of $\gamma$ be denoted by
$\bar{\gamma}$ and its period by $\Period(\gamma)$. Let
$\PerO_{\phi,\bar \gamma, T}$ denote the union of periodic orbits of
$\phi$ whose homology class is $\bar \gamma$ and period is $T$. The
number of connected components of $\PerO_{\phi,\bar \gamma, T}$ is
denoted by $\beta_{\phi,\bar \gamma, T}$. The following two
definitions originate in Schwartzman's work~\cite{Schwartzman}.

\begin{definition} \label{def:mhs}
  Let $\HSpec_{\phi} = \{ (\bar{\gamma},\Period(\gamma),\beta_{\phi,\bar
    \gamma,\Period(\gamma)})\ :\ \gamma \in \PerO_{\phi} \}$. We call
  $\HSpec_{\phi}$ the {\em marked homology spectrum} of $\phi$.
\end{definition}

\noindent
The marked homology spectrum is a subset of $H_1(M;\Z{}) \times \R
\times \N$ that is an invariant of topological conjugacy in the
following sense: if $\phi$ and $\varphi$ are topologically conjugate
then $$(h_* \times id_{\R} \times id_{\N})(\HSpec_{\phi}) =
\HSpec_{\varphi},$$ where $h_* : H_1(M;\Z{}) \to H_1(N;\Z{})$ is the
obvious isomorphism.

\medskip
\bigskip
\begin{example} \label{ex:1} {\rm 
    Let $v \in \Vo{F}$ and define the flow $\phi^v : \Sigma \times \R \to
    \Sigma$ by
    \begin{equation} \label{eq:ex1}
      \phi_t^v(\y,\x) = (\y,\x + t v) \bmod \Delta.
    \end{equation}
    A point $(\y,\x)\in \Sigma$ is periodic of period $T$ for $\phi^v$
    iff $Tv = \ell(u)$ for some $u \in \U^+_F$ and $u\cdot \y = \y
    \bmod \Eta{E}$.

    The map $u : \V{E}/\Eta{E} \to \V{E}/\Eta{E}$ is a toral automorphism. The
    number of fixed points of $u$ is, up to sign, the degree of the map
    $u-1$. The latter is $\det(u-1) = \prod_{\sigma\in\G{E}} \sigma(u-1)$,
    which is also the norm of $u-1\in E$. But since $u-1\in F$, this norm
    equals $N_F(u-1)^{[E:F]}$.

    Thus,
    \begin{equation} \label{eq:phiv}
      \HSpec_{\phi^v} = \left\{ (\ell(u),T, |N_F(u-1)|^{[E:F]})\ :\ \forall
      u\in \U^+_F\ \&\ T\in\R^+\ {\rm s.t.}\ Tv = \ell(u)  \right\}
    \end{equation}

  }
\end{example}

\begin{example} \label{ex:2} {\rm
    Let $\Qm : \V{o,F}^* \to \V{o,F}$ be a linear isomorphism and $M =
    T^*(\V{o,F}/\logu{F}) = \V{o,F}^* \times \V{o,F}/\logu{F}$. Let
    $\phi_t(\px,\x) = ( \px, \x + t \Qm \cdot \px\ \bmod
    \logu{F})$. Clearly, $\eta^{\pm}(\px,\x) = \{ \pm \Qm \cdot \px
    \}$ for all $(\px,\x) \in M$.

    Let ${\mathcal V}_1 = \{ (\px,\x) \in M\ :\ \langle \Qm \cdot \px,
    \px \rangle = 1 \}$ be the unit-sphere bundle, $\phi^1 = \phi |
    {\mathcal V}_1$ and $|m|_{\Qm} = \sqrt{ |\langle \Qm^{-1}m,m
    \rangle| }$ for all $m \in \V{o,F}$. The marked homology spectrum
    of $\phi^1$ is easily seen to equal
            \begin{equation} \label{eq:ms}
              \HSpec_{\phi^1} = \{ (\ell(u),|\ell(u)|_\Qm, 1 )\ :\ u \in
              \U^+_F \}.
            \end{equation}
  }
\end{example}

\begin{example} \label{ex:3} {\rm
    The fibre-bundle structure
    $\xyexact[\V{E}/\Eta{E}]{\Sigma}{\Vo{F}/\logu{F}}{\p}{}$ allows one to
    pullback the unit-sphere bundle ${\mathcal V}_1$ and the flow $\phi^1$
    of the previous example. Let $\varphi^1$ be the pulled-back flow on
    $\p^* {\mathcal V}_1$. The previous two examples show that the marked
    homology spectrum of $\varphi^1$ is
    \begin{equation} \label{eq:ms2}
      \HSpec_{\phi^1} = \{ (\ell(u),|\ell(u)|_\Qm, |N_F(u-1)|^{[E:F]} )\ :\ u \in
      \U^+_F \}.
    \end{equation}
    The marked homology spectrum is especially interesting because it
    contains information about both the quadratic form restricted to the
    Dirichlet lattice, and it contains information about the periodic
    points of the toral automorphisms $u : \V{E}/\Eta{E} \to \V{E}/\Eta{E}$ for
    $u\in\U^+_F$. In \cite{B7}, this extra information about the
    fixed points of the toral automorphisms was not noticed. It turns out
    that this information is extremely important.
  }
\end{example}

\medskip

\subsection{Asymptotic Homology of a Flow}
Let $\pi : \hat{M} \to M$ be the universal abelian covering of
$M$. The flow $\phi$ is covered by a flow $\hat{\phi} :
\hat{M}\times\R \to \hat{M}$. Let $F \subset \hat{M}$ be a fundamental
domain for the group of deck transformations ${\rm Deck}(\pi)$. For
each $p \in M$, choose $\hat{p} \in F \cap \pi^{-1}(p)$. For each $t$
there is a $g \in {\rm Deck}(\pi)$ such that $\hat{\phi}_t(p) \in
g.F$; let $g_t(p)$ be one such element and let $\frac{1}{t} g_t(p) \in
{\rm Deck}(\pi) \otimes_{\Z{}} \R{}$. Recall that ${\rm Deck}(\pi)
\otimes_{\Z{}} \R{} \simeq H_1(M;\R{})$.

\begin{definition} \label{def:ah}
  Let
  $$\eta_{\phi}(p) := \bigcap_{T \geq 0}\ \overline{ \left\{
    \frac{1}{t} g_t(p)\ :\ t \geq T\right\} }$$ be the {\em asymptotic
    homology} of $p \in M$. Let $\eta^{\pm}_{\phi} = \eta_{\phi^{\pm}}$
  where $\phi^{\pm}_t = \phi_{\pm t}$.
\end{definition}

One can show that $\eta_{\phi}(p)$ is independent of the choice of
representatives and if $M$ is compact then $\eta_{\phi}(p)$ is
non-empty for all $p$. It is also clear that if there is a
semi-conjugacy $h$ with $h\comp \phi = \varphi \comp h$, then $h_*
\eta^{\pm}_{\phi}(p) = \eta^{\pm}_{\varphi}(h(p))$.

\begin{lemma} \label{lem:asymphomology1}
  Let $\Htoda$ be a Hamiltonian defined by equation~\ref{eq:HL0}, and
  let $\varphi : T^* \Sigma \times \R \to T^* \Sigma$ be its Hamiltonian
  flow. Let $\UU_{\tau} = \{ g_{\tau} \neq 0 \}$ for each $\tau \in
  \G{F}$. If $P \in \UU_{\tau}$, then
  $$\langle \eta^{\pm}_{\varphi}(P), \hat{\tau} \rangle \leq 0.$$
\end{lemma}

\noindent
{\em Remark.} This lemma is very close in spririt to lemma \ref{lem:X}.

\begin{proof} 
  Let $\hat{P} = (\py,\y+\Eta{E},\px,\x) \in \hat{\UU}_{\sigma}$ and
  let $P = \Pi(\hat{P})$, \cf \eqref{eq:cot_sigma}. Since
  $g_{\tau}(\hat P) \neq 0$, $\py_{\tau} \neq 0$. If $v \in
  \eta^{\pm}_{\varphi}(P)$, then there is a sequence $T_k \to \pm
  \infty$ such that
  $$v = \lim_{k \to \infty}\ \frac{1}{|T_k|}( \x( T_k ) - \x(0) ),$$
  where $\hat{\varphi}_t(\y+\Eta{E},\py,\px,\x) =
  (\py(t),\y(t)+\Eta{E},\px(t),\x(t))$ and $\hat{\varphi}_t$ is the
  lift of $\varphi_t$ to $T^*\hat \Sigma$. Thus:
  $$\langle v, \hat{\tau} \rangle = \lim_{k \to \infty}\ \frac{1}{|T_k|}
  \langle \x( T_k ), \hat{\tau} \rangle.$$ On the other hand
  $\hat{\Htoda}$ and $g_{\tau}$ are first integrals of
  $\hat{\varphi}_t$. Inspection of equation~\ref{eq:HL0} shows that
  $\hat{\Htoda}(\hat{P}) \geq g_{\tau}^{b_{\tau}/2} \exp(
  b_{\sigma} \langle \x(T), \hat{\tau} \rangle)$ for all $T$. Since
  $b_\sigma, b_{\tau} > 0$ and $g_{\tau} \neq 0$, this inequality implies
  that
  $$\frac{1}{|T_k|}\langle \x( T_k ), \hat{\tau} \rangle
  \leq\frac{1}{|T_k| b_\sigma } \left( \ln \hat{\Htoda} - \frac{b_{\tau}}{2} \, \ln
  g_{\tau} \right) \stackrel{k \to \infty}{ \longrightarrow } 0.$$
  Since $v \in \eta^{\pm}_{\varphi}(P)$ was arbitrary, this proves the
  lemma.
\end{proof}

As noted above, the fibre bundle structure
$\xyexact[\V{E}/\Eta{E}]{\Sigma}{\Vo{F}/\logu{F}}{\p}{}$ of $\Sigma$
induces the sub-bundle $\Vert = \ker \d\p \subset T \Sigma$ and its
annihilator $\Hstar \subset T^* \Sigma$. The sub-bundle $\Hstar$ is
the intersection of $\ZZ_{\tau} = g_{\tau}^{-1}(0)$ over all
$\tau\in\B{F}$; it is also isomorphic to $\Sigma \times \Vo{F}^*$.

\begin{lemma} \label{lem:asymphomology}
  Let $\Htoda_1, \Htoda_2$ be defined by equation~\ref{eq:HL0} with root
  bases $\Psi_1, \Psi_2$ . If $h : T^* \Sigma \to T^* \Sigma$ conjugates
  their Hamiltonian flows, then
  $$h(\Hstar) = \Hstar.$$
\end{lemma}

\begin{proof}
  Let $U$ be the set of points in $\Hstar$ that are mapped out of
  $\Hstar$ under $h$. Since $P \not\in \Hstar$ iff $\exists \tau \in
  \B{F}$ such that $g_{\tau}(P) \neq 0$, one sees that 
  $$U = h^{-1}(\cup_{\tau \in \B{F}}\ \UU_\tau) \cap \Hstar.$$ It
  suffices to prove that $U$ is empty, since a symmetric argument
  applies to $h^{-1}$. Therefore, it suffices to prove that $U_{\tau}
  = h^{-1}( \UU_{\tau} ) \cap \Hstar$ is empty for all $\tau$. Since
  $\UU_{\tau}$ is open, $U_{\tau}$ is an open subset of $\Hstar$, so
  to prove that it is empty, it suffices to show that $U_{\tau}$ is
  nowhere dense. As noted above, $\Hstar$ is naturally isomorphic to
  $\Sigma \times V^*_o$. Let $\pi_o : \Hstar \to V^*_o$ denote the
  projection onto the second factor. Clearly, $\pi_o$ is an open map
  and $\pi_o(P) = \px$ where $P=\Pi(0,\y,\px,\x) \in \Hstar$. It
  suffices to show that $\pi_o(U_{\tau})$ lies in a hyper-plane to
  prove the lemma.

  Let $\varphi^i$ be the Hamiltonian flow of $\Htoda_i$, and $\Qm_i$ the
  quadratic form used to define $\Htoda_i$ (Equation~\ref{eq:HL0}). If $P \in
  U_{\tau}$, then $P \in \Hstar$ so $$\eta^{\pm}_{\varphi^1}(P) = \{
  \pm \Qm_1 \cdot \px \},$$ while $h(P) \in \UU_{\tau}$, so from the previous
  lemma $$\langle \eta^{\pm}_{\phi^2}(h(P)), \hat{\tau} \rangle \leq
  0.$$ Since $\varphi^2_t h = h \varphi^1_t$,
  $$\eta^{\pm}_{\varphi^2}(h(P)) = h_* \eta^{\pm}_{\varphi^1}(P)$$
  which implies that $$\pm \langle h_* \Qm_1 \px, \hat{\tau} \rangle
  \leq 0.$$ Therefore, $\langle h_* \Qm_1 \px, \hat{\tau} \rangle$
  vanishes. Since $h_* \Qm_1$ is non-degenerate, $\px = \pi_o(P)$ lies
  in a fixed hyper-plane. Thus, $\pi_o(U_{\tau})$ lies in a
  hyper-plane. Since $\pi_o$ is an open map, $U_{\tau}$ is empty.
\end{proof}

\begin{remark} \label{re:convex}
{\rm

Lemmas \ref{lem:asymphomology1} and \ref{lem:asymphomology} can be
reformulated and shown to hold in much greater generality. Let
$\Sigma_A$ be defined as in \ref{ssec:cf} and let $H : T^*\Sigma_A \to
\R$ be a smooth, fibre-wise convex hamiltonian that is
left-invariant. Left-invariance implies that $H$ enjoys the integral
$f$ \eqref{eq:mmap}. In particular, if one defines the function
$\gamma_i(P) = |p_{y_i} \exp(\left\langle \ell_i , x \right\rangle)|$,
then the properness of $H$ implies that there is a function $c=c(H)$
such that $0 \leq \gamma_i(P) \leq c(H(P))$ for all $P \in
T^*\Sigma_A$. The proof of lemma \ref{lem:asymphomology} applies to
show that if $\gamma_i(P) \neq 0$, then $\left\langle \ell_i , v
\right\rangle \leq 0$ for all $v \in \eta^{\pm}(P)$. This implies that
the asymptotic homology of a point $P$ with $\prod_i \gamma_i(P) \neq
0$ is trivial and that a topological conjugacy of two such hamiltonian
flows must map $\Hstar$ to itself.

}
\end{remark}

\begin{definition} \label{de:ep}
A homeomorphism $h : T^* \Sigma \to T^* \Sigma$ is {\em energy-preserving}
if $h(\{\Htoda_1 = \frac{1}{2} \}) = \{ \Htoda_2 = \frac{1}{2} \}$. 
\end{definition}

We use the notation of Lemma~\ref{lem:asymphomology} and its proof:

\begin{thm} \label{thm:conj}
  Let $\Htoda_1,\Htoda_2$ be defined by Equation~\ref{eq:HL0} corresponding to root
  bases $\Psi_1, \Psi_2$. If $h \in {\rm Homeo}(T^* \Sigma)$ is an
  energy-preserving conjugacy of $\varphi^1$ with $\varphi^2$, then 
  \begin{enumerate}

  \item $h_* : H_1(T^*\Sigma) \to H_1(T^*\Sigma)$ induces automorphisms
    of $\logu{F}$ and $\U^+_F$ such that the following commutes
    $$
    \xymatrix{
      \U^+_F \ar[r]^{\alpha} \ar[d]^{\ell} & \U^+_F \ar[d]^{\ell}\\
      \logu{F} \ar[r]^{f} & \logu{F};
    } \eqno(*)
    $$
  \item $f$ is an isometry of $(\logu{F},\Qm_2)$ with $(\logu{F},\Qm_1)$;
  \item $\alpha$ preserves the number of fixed points of $u \in \U^+_F$
    acting on $\V{E}/\Eta{E}$:
    $$
    |N_F(\alpha(u)-1)| = |N_F(u-1)| \hspace{15mm} \forall u \in \U^+_F.
    $$
  \end{enumerate}
\end{thm}

\begin{prooflist}
\item The map $h_*$ on $H_1$ induces an automorphism $f$ of $\logu{F}$. The
  isomorphism $\ell$ allows the definition of $\alpha$ as an
  automorphism of $\U^+_F$ and shows that (*) commutes.

\item Let ${\mathcal V}_i = \Hstar \cap
  \Htoda_i^{-1}(\frac{1}{2})$. Since $h$ is energy preserving,
  Lemma~\ref{lem:asymphomology1} implies that $h({\mathcal V}_1) =
  {\mathcal V}_2$. Let $\varphi^i | {\mathcal V}_i$ be denoted by
  $\Phi^i$ and let $h | {\mathcal V}_1$ continue to be denoted by
  $h$. Examples~\ref{ex:2} and~\ref{ex:3} show that
  \begin{align*}
    \HSpec_{\Phi^i} &= \{ (\ell(u),|\ell(u)|_{\Qm_i},
    |N_F(u-1)|^{[E:F]})\ :\ u \in \U^+_F, u\neq \pm1 \}
  \end{align*} 
  for $i=1,2$. 

\item Finally, by hypothesis $h \Phi^1$ equals $\Phi^2 h$, so
  from the identity $\HSpec_{\Phi^2} = (h_* \times id_{\R} \times
  id_{\N})\HSpec_{\Phi^1}$ one sees that
  \begin{align}
    |\ell(u)|_{\Qm_1} &= |f \comp \ell(u)|_{\Qm_2} =
    |\ell(\alpha(u))|_{\Qm_2}, \label{eq:isom} \\ 
    |N_F(u-1)| &= |N_F(\alpha(u)-1)| \label{eq:norm}
  \end{align}
  for all $u \in \U^+_F$. Equation (\ref{eq:isom}) shows that $f$ is an
  isometry, while equation (\ref{eq:norm}) shows that $\alpha$ preserves
  the number of fixed points.
\end{prooflist}

\medskip

Let us dualise Theorem~\ref{thm:conj}. Let $\phi_i$ be a linear
isomorphism $\Vo{F}^* \to \h^*_i$ induced by a bijection $\rho_i :
\B{F} \to \Rs$ (see Definition~\ref{def:bi}). The norms $| \cdot
|_{\Qm_i}$ on $\logu{F}$ are equivalent modulo $\Aut{\logu{F}}$ iff the
dual norms $| \cdot |^*_{\Qm_i}$ on $\logu[*]{F}$ are equivalent modulo
$\Aut{\logu[*]{F}}$. Since, by Theorem~\ref{thm:lax}, there is a $c_i
\in \N$ such that $| \px |^*_{\Qm_i} = c_i^{-1} \sqrt{ \llangle
  \phi_i( \px ),\phi_i( \px ) \rrangle_i }$, Theorem~\ref{thm:conj}
implies

\begin{corollary} \label{thm:conjd}
  If $\varphi^1$ and $\varphi^2$ are topologically conjugate by an
  energy-preserving homeomorphism, then there exists $\mu \in {\rm
    Isom}(\h^*_2;\h^*_1)$ and $g=f^* \in \Aut{\logu[*]{F}}$ such that
  \begin{equation} \label{eq:mug}
    \mu = \frac{c_2}{c_1} \times \phi_1 g \phi_2^{-1}.
  \end{equation}
\end{corollary}

\begin{remark} \label{re:c}
  {\rm
    One might attempt to use Corollary~\ref{thm:conjd} to try to
    determine the topological conjugacy classes of Hamiltonian
    flows. This is the approach taken in~\cite{B7}. However, this
    approach leads to some very delicate and long-outstanding issues in
    transcendence and algebraic-independence theory. This paper skirts
    those difficulties by employing all the information in the marked
    homology spectrum.
  }
\end{remark}

\subsection{Periodic points of toral automorphisms}
Part (3) of Theorem \ref{thm:conj} has a useful corollary: the number
of period-$k$ periodic points of the automorphisms $u$ and $\alpha(u)$ of
the torus $\V{E}/\Eta{E}$ are equal for all $k$. Therefore, their
asymptotic rates of growth are equal. Define the function $\hr : \Vo{F}
\to \R$ by
\begin{equation} \label{eq:h}
  \hr(v) = \sum_{\tau\in\B{F}} n_{\tau}\, \left\langle \hat \tau , v
  \right\rangle^+
\end{equation}
for all $v \in \Vo{F}$, where $\bullet^+ = \max(\bullet,0)$. Since the
growth rate of the number of period-$k$ periodic points of $u \in
\U^+_F$ is $[E:F] \times \hr(\ell(u))$, this proves
\begin{lemma} \label{lem:htop}
  Under the hypotheses of Theorem \ref{thm:conj}, the automorphism $f :
  \logu{F} \to \logu{F}$ satisfies
  $$\hr = \hr \comp f.$$
\end{lemma}

The function $\hr$ is piecewise linear. One can characterise the sets
on which $\hr$ is linear as follows. For $J \subset \B{F}$, let
\begin{equation} \label{eq:VoJ}
  \Vo{F}^J := \left\{ v \in \Vo{F}\ :\ \forall \tau\in J,\ \left\langle
  \hat \tau , v \right\rangle > 0\ \&\ \forall \tau\not\in J,\
  \left\langle \hat \tau , v \right\rangle < 0\ \right\}.
\end{equation}
Note that if $J=\emptyset$ or $J=\B{F}$, then $\Vo{F}^J$ is empty;
otherwise $\Vo{F}^J$ is an open set that is closed under addition and
multiplication by positive scalars. Since $\Vo{F}^J$ is open and it is
closed under positive dilations, it contains balls of arbitrarily
large diameter and hence it contains points in $\logu{F}$. Therefore, 
\begin{equation} \label{eq:LJ}
  \logu[J]{F} := \logu{F} \cap \Vo{F}^J
\end{equation}
is a non-empty subset of $\logu{F}$, for all $J \subset \B{F}$ with $J\neq
\emptyset$ and $\B{F}$.

To return to $\hr$: for all $J \subset \B{F}$, define 
\begin{equation} \label{eq:rJ}
  r_J := \sum_{\tau\in J} n_{\tau}\, \hat \tau.
\end{equation}

\begin{lemma} \label{lem:rJ}
  The following is true:
  \begin{enumerate}
  \item if $v \in \Vo{F}^J$, then $\hr(v) = \left\langle r_J , v \right\rangle$;
  \item if $v \in \logu[J]{F}$ and $f(v) \in \logu[I]{F}$, then $r_J = f^* r_I$;
  \item for each $J \subset \B{F}$ with $J\neq
    \emptyset$ and $\B{F}$, there is a unique $I \subset \B{F}$ such that
    $f(\logu[J]{F}) \subset \logu[I]{F}$;
  \item $f$ induces a permutation $\perm$ of the power set $\pwrset{\B{F}}$ that
    satisfies 
    \begin{enumerate}
    \item $\perm(\emptyset)=\emptyset$ and $\perm(\B{F})=\B{F}$;
    \item $\perm(J) = I$ iff $f(\logu[J]{F}) \subset \logu[I]{F}$.
    \end{enumerate}
  \end{enumerate}
\end{lemma}
\begin{prooflist}
\item $\hr$ may be characterised as: $\hr(v) = \max_{I \subset \B{F}}
  \left\langle r_I , v \right\rangle$. On the set $\Vo{F}^J$, this
  maximum is achieved uniquely at $I=J$. This proves that $\hr = r_J$
  on $\Vo{F}^J$.

\item Let $v \in \logu[J]{F}$ and $f(v) \in \logu[I]{F}$. Lemma
  (\ref{lem:htop}) implies that 
  $$\left\langle r_J , v \right\rangle = \hr(v) = \hr(f(v)) =
  \left\langle f^* r_I , v \right\rangle.$$ It is clear that the set
  $\logu[J]{F} \cap f^{-1}(\logu[I]{F})$ is an intersection of Zariski dense
  subsets of $\Vo{F}$, hence is Zariski dense since it is
  non-empty. Therefore $r_J$ must equal $f^* r_I$ on $\Vo{F}$.

\item Let $v_i \in \logu[J]{F}$ and assume that $f(v_i) \in
  \logu[{I_i}]{F}$. Therefore, from the previous step $f^*r_{I_1} = r_J =
  f^*r_{I_2}$. Since $f$ is an automorphism $r_{I_1} = r_{I_2}$. Since
  the map $I \mapsto r_I\ :\ \pwrset{\B{F}} \to \Vo{F}^*$ is injective
  except at $\emptyset$ and $\B{F}$ (both are sent to $0$), one
  concludes that $I_1=I_2$.

\item From step (3), the properties (a-b) uniquely define a map $\perm
  : \pwrset{\B{F}} \to \pwrset{\B{F}}$ because $\Vo{F}^J \neq \emptyset$---hence
  $\logu[J]{F} \neq \emptyset$ --- for all $J \subset \B{F}$, $J \neq
  \emptyset, \B{F}$. This map $\perm$ is invertible because $f$ is
  induced by the homeomorphism $h$: one can equally start with
  $h^{-1}$, get $f^{-1}$ and define $\perm'$ thusly. Step (3) shows
  that $\perm' = \perm^{-1}$.

\end{prooflist}

Let us be more precise about the nature of $f$. Lemma \ref{lem:fnat}
should be compared with \cite[Theorem 7]{B7}, where the Gel'fond
conjecture \cite{Gelfond} is invoked to obtain the weaker conclusion
that $f^* \in \Aut{\logu[*]{F}} \cap \Aut{\Vo{F,\Q}^*}.$

\begin{lemma} \label{lem:fnat}
  Let $\Vo{F,\Z}^*$ be the $\Z$-module spanned by $\left\{ n_\tau
  \hat\tau|_{\Vo{F}} : \tau \in \B{F} \right\}$ and $\logu[*]{F} = {\rm
    Hom}(\logu{F},\Z)$.  Then
  \begin{equation} \label{eq:fnat}
    f^* \in \Aut{\logu[*]{F}} \cap \Aut{\Vo{F,\Z}^*}
  \end{equation}
\end{lemma}

\begin{proof}
  Note that $\perm$ is defined by $r_J = f^* r_{\perm(J)}$ for all
  $J$. If $J=\perm^{-1}\left\{ \tau \right\}$, then
  \begin{equation} \label{eq:frJ}
    f^*(n_\tau  \hat \tau) = r_J \in \Vo{F,\Z}^*,
  \end{equation} 
  since $r_{\left\{ \tau \right\}} = n_\tau \hat \tau$. On the other
  hand, if $J = \left\{ \tau \right\}$, then $$(f^*)^{-1}(n_\tau \hat
  \tau) = r_{\perm(J)} \in \Vo{F,\Z}^*.$$ This proves that $f^* \in {\rm
    Aut}(\Vo{F,\Z}^*)$, and since $f \in \Aut{\logu{F}}$, the lemma is
  proven.
\end{proof}

\begin{lemma} \label{lem:disj}
  Let $\perm : \pwrset{\B{F}} \to \pwrset{\B{F}}$ be the permutation defined in
  Lemma \ref{lem:rJ}. If $I,J \subset \B{F}$ are disjoint sets, then
  $$\perm( I \sqcup J ) = \perm(I) \sqcup \perm(J),$$
  $\sqcup = $ disjoint union. Consequently, $\perm$ is induced by a
  permutation of $\B{F}$.
\end{lemma}
\begin{proof}
  Since $I \cap J = \emptyset$, $r_{I \sqcup J} = r_I + r_J$. Therefore
  \begin{equation} \label{eq:r}
    r_{\perm(I \sqcup J)} = (f^*)^{-1} r_{I \sqcup J} = (f^*)^{-1}( r_I
    + r_J) = r_{\perm(I)} + r_{\perm(J)}.
  \end{equation}
  Assume that $\perm(I)$ and $\perm(J)$ are not disjoint. Then, there is
  a $\tau \in \perm(I) \cap \perm(J)$. The coefficient on $\hat \tau$ in
  the right-hand side of (\ref{eq:r}) is therefore $2 n_\tau$. The
  coefficient on $\hat \tau$ in the left-hand side of (\ref{eq:r}) is at most
  $n_\tau$, however. Absurd. Therefore $\perm(I) \cap \perm(J)$ must be
  empty.

  Consider the $\#\B{F}+1$ subsets of $\B{F}$ that contain at most $1$
  element. This is the largest family of pairwise disjoint subsets of
  $\B{F}$. Therefore, $\perm$ must be map this family to itself. Since
  $\perm(\emptyset) = \emptyset$, $\perm$ maps the singleton sets to
  singletons. 
\end{proof}

Let the permutation of $\B{F}$ induced by $\perm$ be denoted by
$\perm$, too. Equation (\ref{eq:frJ}) is thereby simplified to
\begin{align}
  \forall \tau\in\B{F}: &  \hspace{5mm}  n_\sigma\, f^*\hat\sigma =
  n_\tau\, \hat\tau \qquad
  \iff \qquad \perm(\tau) = \sigma. \label{eq:frJ1}
\end{align}
Intuitively, one wants to say that $\perm$ should not mix up the real
and non-real embeddings, so the coefficients on both sides of
(\ref{eq:frJ1}) ought to be equal. To prove this, observe that
(\ref{eq:frJ1}) implies that
\begin{align} \label{al:fu}
  \forall u\in\U^+_F: & \hspace{5mm} f \comp \ell(u) =
  \sum_{\tau\in\G{F}}  \frac{n_\tau }{n_{\perm(\tau)}}\, \ln|\tau(u)|
  \cdot \tau. 
\end{align}
Since $f \in \Aut{\logu{F}}$, the right-hand side lies in $\logu{F} \subset
\Vo{F}$ for all $u$. Let $\xi = \sum_{\tau\in\G{F}} \frac{n_\tau
}{n_{\perm(\tau)}}\, \hat \tau \in \V{F}^*$; one sees that
$\left\langle \xi , \ell(u)\right\rangle = 0$ since $f\comp \ell(u)
\in \Vo{F}$. Since $\logu{F}$ spans $\Vo{F}$, this shows that $\xi \in
\Vo{F}^{\perp}$. Since
\begin{equation} \label{eq:vofp}
  \Vo{F}^{\perp} = \Span \left\{ \tau-\hat \tau, \epsilon\ :\ \tau
  \in \G{F}^c, \epsilon = \sum_{\tau\in\G{F}}\hat \tau \right\},
\end{equation}
and the coefficients $n_\tau / n_{\perm(\tau)}$ are constant under
the involution $\tau \mapsto \bar \tau$, one sees that $\xi$ must
be a multiple of $\epsilon$. Therefore, $n_\tau / n_{\perm(\tau)}$
must be independent of $\tau$. Since $\perm$ is a permutation, this
forces $n_\tau / n_{\perm(\tau)}$ to be identically equal to
unity. This proves

\begin{lemma} \label{lem:f}
  The permutation $\perm$ of $\G{F}$ preserves the type of each
  embedding. In particular,
  \begin{align}
    \forall \tau\in\B{F}: &  \hspace{5mm}  f^*\hat\sigma = \hat\tau \qquad
    \iff \qquad \perm(\tau) = \sigma. \label{eq:frJ2}
  \end{align}
\end{lemma}

\begin{lemma} \label{lem:alpha}
  For each $\tau\in\B{F}$, there exists a homomorphism $\zeta_{\tau} :
  \U^+_F \to S^1$ such that
  \begin{enumerate}

  \item for all $u\in\U^+_F$, $\tau(\alpha(u)) = \zeta_{\tau}(u) \cdot
    \sigma(u)$ where $\perm(\sigma)=\tau$;
  \item $\zeta_{\tau}$ maps $\U^+_F$ into $S^1 \cap \U_K$ where $K$ is
    the normal closure of $F$.

  \end{enumerate}
\end{lemma}
\begin{proof}
  The equation $f(\ell(u))=\ell(\alpha(u))$ implies, via equation
  (\ref{eq:frJ2}), that $|\tau(\alpha(u))| = |\sigma(u)|$ when
  $\sigma=\perm^{-1}(\tau)$. Therefore, there is a unit modulus number
  $\zeta = \zeta_{\tau}(u)$ such that $\tau(\alpha(u)) = \zeta \cdot
  \sigma(u)$. The number $\zeta$ is a ratio of numbers in conjugates of
  $F$, hence it lies in the smallest field containing all conjugates of
  $F$, $K$. Moreover, one sees that $\zeta_{\tau}$ is a ratio of two
  homomorphisms, hence it is a homomorphism. Finally, since $\zeta$ is a
  ratio of units of $K$, it is a unit of $K$.
\end{proof}

\subsection{Strictly Hyperbolic Number Fields} Lemma \ref{lem:alpha}
shows that, if one can force $\zeta_{\tau}$ to be trivial, then
$\alpha$ is an automorphism of $F$ and $\perm$ is induced by right
composition by $\alpha^{-1}$. One expects that this is always the
case: the symmetries of the number field $F/\Q$ ought to appear as
symmetries (=topological conjugacies) of the Hamiltonian system, and
vice versa. However, when $K$ contains infinite order elements in
$S^1$, it is difficult to say anything meaningful about
$\zeta_{\tau}$. This is quite likely related to the fact that if $u
\in \U_K \cap S^1$ has infinite order, then the induced automorphism
of the torus $\V{K}/\O_K$ is {\em partially hyperbolic}.

\begin{definition} \label{def:hyperb}
  A unit $u \in \U_F$ is {\em hyperbolic} if none of its conjugates
  have unit modulus. $F$ is {\em hyperbolic} if its only
  non-hyperbolic units are roots of unity. $F$ is {\em strictly
  hyperbolic} if its normal closure, $K$, is hyperbolic.
\end{definition}

In other words, $F$ is hyperbolic iff
\begin{equation} \label{eq:sh}
  \# \U_F \cap S^1 < \infty.
\end{equation}
If $F$ is hyperbolic, then $\U^+_F$ acts on the torus $\V{F}/\Eta{F}$
  as a group of Anosov automorphisms; if $F$ is {\em strictly
  hyperbolic}, then the `closure' of $\U^+_F$, $\U^+_K$, acts on the
  torus $\V{K}/\O_K$ as a group of Anosov automorphisms.

Strict hyperbolicity is a property of the normal closure $K$: $K$
itself is strictly hyperbolic and so, therefore, are all its
subfields. Examples of strictly hyperbolic number fields are legion;
there also appear to be many hyperbolic but not strictly hyperbolic
number fields.

\begin{examples}
\item $F$ is {\em totally real} if all its conjugates are real. In
  this case, its normal closure is also totally real and so $\U_K \cap
  S^1=\left\{ \pm 1 \right\}$. Thus, all totally real number fields
  are strictly hyperbolic.

\item Let $\zeta$ be a $p$-th root of unity for some odd prime
  $p$. The field $K=\Q(\zeta)$ has the totally real subfield
  $F=\Q(\zeta+\zeta^{-1})$ of index $2$. The Dirichlet theorem on the
  group of units implies that $\U_F$ is of finite index in
  $\U_K$. Since $F$ is totally real, $K$ is strictly hyperbolic.

\item More generally, let $K/\Q$ be a non-real, normal extension of
  $\Q$. If $K$ has a totally real subfield $F$ of index $2$, then, as
  above, $\U_F$ is a finite-index subgroup of $\U_K$, hence $K$ is
  strictly hyperbolic.

\item A penultimate, concrete example: let $F=\Q(a)$ where $a$ is the
  unique real root of $p(x) = x^3 + 3x-1$. The discriminant of $p$ is
  $d = -27 \times 5$, so $\sqrt{d}\not\in\Q$, which implies that $F$ is
  not a normal extension of $\Q$ ($p$'s roots are approximately
  $0.3222, -0.1611 \pm 1.7544 \sqrt{-1}$, which also implies $F$
  cannot be normal). Therefore, the normal closure of $F$ is a degree
  $6$ extension $K$. The group $\U^+_K$ has rank $2$ since $K$ has no
  real embeddings, while $a$ and one of its conjugates are
  multiplicatively independent units in $\U^+_K$, neither of which
  lies on $S^1$. This means that $\U_K \cap S^1$ must be finite, so
  $K$ and $F$ are strictly hyperbolic.

\item Let us end with an example of a hyperbolic number field that is
  not strictly hyperbolic. Let $a,b,c$ be the roots of $p(x)=x^3+3x-1$
  where $a$ is the real root as in the previous example. It is clear
  that $|b|=1/\sqrt{a}$. Let $E=\Q(\sqrt{a})$, which is a real,
  degree $6$ extension of $\Q$ and let $E'=\Q(\sqrt{b})$ and
  $E''=\Q(\sqrt{c})$ be the conjugates of $E$. It is claimed that $E$
  is hyperbolic, that is, if $u \in \U_E$ has a conjugate $v$ of unit
  modulus, then $u = \pm 1$. To verify this claim, let $u \in E$ have
  a conjugate of unit modulus. Without loss of generality, this
  conjugate can be assumed to be some $v \in E'$. Since $\bar{b}=c$,
  one sees that $\bar{v} \in E''$ and that $v\bar{v}=1$ implies that
  $v,\bar{v} \in E' \cap E''$. The field $E' \cap E''$ is of degree
  $1,2,3$ or $6$. It cannot be $6$, since $E' \neq E''$, so its degree
  is $1,2$ or $3$. The degree of $E' \cap E''$ cannot be $3$ so it
  must be $1$ or $2$. %
%
% If the degree were three, since it is closed under complex conjugation, the field must be a normal extension of $\Q$, whence $u,v,\bar{v}$ all lie the field. Therefore, $u \in E\cap E' \cap E'' =\Q$. This proves $u=\pm 1$.
%
If the degree is $1$, then the claim is proved; if the degree is $2$,
  then $v$ is a unit in a complex quadratic number field, hence $v$ is
  a root of unity. This implies that $u$ is a root of unity in the
  real field $E$, hence $u=\pm 1$ as claimed.

On the other hand, the normal closure $L$ of $E$ contains $\sqrt{a}$
  and $b$ and therefore the unit modulus number $\eta=b\sqrt{a}$. If
  $\eta$ were an $n$-th root of unity, then $1 = \eta^{4n} =
  b^{4n}a^{2n}$; but $a$ and $b$ are multiplicatively independent in
  $K=\Q(a,b,c)$, so $n=0$. This shows that $\eta \in \U_L \cap S^1$
  has infinite order and completes the proof that $E$ is hyperbolic
  but not strictly hyperbolic.
\end{examples}

%\noindent
Let us turn to a theorem which demonstrates the importance of strictly
hyperbolic number fields. The choice of the set $\B{F}$ involves an
arbitrariness which it has been possible to avoid up to this point. To
work around this arbitrariness, let the map $\perm$ be extended to a
map of $\G{F}$ by
\begin{equation} \label{eq:perm}
  \perm(\tau) = \overline{\perm(\bar \tau)} \hspace{10mm} \forall
  \tau\not\in\B{F}.
\end{equation}

\begin{thm} \label{thm:sh}
  If $F$ is strictly hyperbolic, then there is a $\beta \in \Aut{F/\Q}$
  such that
  \begin{enumerate}
  \item the induced maps $\xymatrix{ \U_F/\kerl_F
    \ar[r]^{\alpha}_{\beta} & \U_F/\kerl_F }$ coincide;\\
  \item $\perm(\tau) = \tau \comp \beta^{-1} \hspace{15mm} \forall \tau\in\G{F};$\\
  \item $f = \left. R_{\beta^{-1}} \right|_{\Vo{F}}$ where $R_{\beta} :
    \V{F} \to \V{F}$ is the linear transformation induced by
    precomposition with $\beta\in\Aut{F/\Q}$.
  \end{enumerate}
\end{thm}

Recall that $\kerl_F$ is the set of units in $\U_F$ all of whose
conjugate lie on $S^1$. If $F$ is strictly hyperbolic, then $\kerl_F =
\U_F \cap S^1$.

\begin{proof}
  For the purposes of this proof, it is convenient to extend
  $\alpha\in\Aut{\U^+_F}$ to an automorphism of $\U_F = \U^+_F \oplus
  \kerl_F$ by extending $\alpha$ as the identity on $\kerl_F$. The
  choice of extension of $\alpha$ is immaterial. The extension of
  $\alpha$ permits the extension of the homomorphism $\zeta_{\tau}$
  (Lemma \ref{lem:alpha}), too. Since $F$ is strictly hyperbolic, all
  conjugates of $\U_F \cap S^1$ lie in $S^1$. Since $\alpha$ maps $\U_F
  \cap S^1$ to itself, this implies that the extended homomorphism
  $\zeta_{\tau}$ maps $\U_F$ into $S^1$.

  Let $\U^1_F = \cap_{{\tau\in\B{F}}} \ker \zeta_{\tau}$. Since $\U_K
  \cap S^1$ is finite, $\ker \zeta_{\tau}$ is a finite-index subgroup of
  $\U^+_F$ for all $\tau$; thus $\U^1_F$ is a finite-index
  subgroup. Lemma \ref{lem:alpha}.1 implies that
  \begin{equation} \label{eq:u1}
    \forall u\in\U^1_F,\ \tau\in\B{F}:\ \hspace{10mm} \sigma(\alpha(u)) =
    \tau(u) \hspace{10mm} {\rm where\ } \perm(\tau)=\sigma.
  \end{equation}
  This implies that $\sigma(\U^1_F) \subset \tau(\U_F)$; and since
  $\sigma,\tau$ are injective, the group $\sigma(\U^1_F)$ is a
  finite-index subgroup of $\tau(\U_F)$. Therefore, $\tau(F) \cap
  \sigma(F)$ contains elements that are of degree $\deg F$. Thus, the two
  fields coincide:
  \begin{equation} \label{eq:coincide}
    \forall \tau\in\B{F}:\ \hspace{10mm} \tau(F)=\sigma(F)
    \hspace{10mm} {\rm where\ } \perm(\tau)=\sigma.
  \end{equation}
  Fix $\sigma,\tau \in \G{F}$ with $\perm(\tau)=\sigma$ and
  define
  \begin{equation} \label{eq:b}
    \beta_{\sigma} := \sigma^{-1} \comp \tau.
  \end{equation}
  Then, $\left. \beta_{\sigma} \right|_{\U^1_F} = \left. \alpha
  \right|_{\U^1_F}$ and $\beta_{\sigma} \in \Aut{F/\Q}$. Because $\U^1_F$
  is a finite-index subgroup of $\U_F$ it contains elements of degree
  $\deg F$. It is clear that two automorphisms of $F/\Q$ which coincide
  on an element of degree $\deg F$, coincide on $F$. Therefore, there is
  a single $\beta \in \Aut{F/\Q}$ such that $\beta_{\sigma} = \beta$ for
  all $\sigma$.

  Moreover, from (\ref{eq:coincide}) and the remarks in the first
  paragraph, one knows that $\zeta_{\sigma}$ maps $\U_F$ into $\U_F \cap
  S^1$. Consequently, $\sigma^{-1} \comp \zeta_{\sigma}$ maps $\U_F$
  into $\U_F \cap S^1$. Since
  \begin{equation} \label{eq:b0}
    \alpha(u) = \sigma^{-1}\left( \zeta_{\sigma}(u)  \right) \cdot
    \beta(u) \hspace{10mm} \forall u \in \U_F,
  \end{equation}
  one sees that the invariance of $S^1$ under embeddings of $F$ implies
  that the induced maps
  $
  \xymatrix{
    \U_F/\U_F \cap S^1 \ar[r]^{\alpha}_{\beta} & \U_F/\U_F \cap S^1
  }
  $ are equal. Since $\U^+_F$ is a non-canonical lifting of $\U_F/\U_F
  \cap S^1$ to $\U^+_F$, one may declare that $\alpha =
  \beta|_{\U^+_F}$. 

  Finally, equation (\ref{eq:b}) implies that 
  \begin{equation} \label{eq:bb}
    \forall \tau\in\B{F}: \qquad \perm(\tau) = \sigma \quad \iff \quad
    \sigma = \tau \comp \beta^{-1}.
  \end{equation}
  Therefore, the way in which $\perm$ is extended to $\G{F}$ shows that
  $\perm(\tau) = \tau \comp \beta^{-1}$ for all $\tau\in\G{F}$. This
  proves the theorem.
\end{proof}

\subsection{Topological conjugacy classes}
The results of the previous section afford the opportunity to classify
the Hamiltonian flows of the \BT-type Hamiltonians (equation
\ref{eq:HL0}) up to topological conjugacy --- at least in some
situations.

\medskip
\noindent{\bf Standing Hypothesis}: For the remainder of section
\ref{sec:ex}, unless explicitly stated otherwise, it is assumed that
$F$ is a strictly hyperbolic number field.
\medskip

\subsubsection{Root bases and Dynkin Diagrams} \label{sssec:rb}
Recall that for each root basis $\Rs$ there is a labelled graph
$\Dynkin{\Rs}$, called the Dynkin diagram, whose vertices are the
points of $\Rs$. A pair of distinct vertices $r,s$ have $4 \llangle
r,s \rrangle^2 / |r|^2 |s|^2$ edges connecting them, and if $|r| >
|s|$ then there is an arrow pointing from $r$ to $s$. The vertex $r$
has the label $\omega_r$. The Coxeter diagram is obtained from the
Dynkin diagram by erasing the labels and arrows. If $\Rs$ is a root
system other than $\Atwors{2n}$, then one says that a permutation
$\rho \in S(\Rs)$ is an {\em automorphism} of the Dynkin diagram
$\Dynkin{\Rs}$ iff the permutation leaves the Dynkin diagram unchanged
with the exception of the numbering of the roots.
$\Aut{\Dynkin{\Rs}}$ is the automorphism group of $\Dynkin{\Rs}$. Note
that $\rho \in \Aut{\Dynkin{\Rs}}$ iff $\omega_r = \omega_{\rho(r)}$
and $\llangle r,s \rrangle = \llangle \rho(r),\rho(s) \rrangle$ for
all $r,s \in \Rs$. For the root system $\Atwors{2n}$, one defines the
automorphism group, $\Aut{\Dynkin{\Atwors{2n}}}$, to be the group
generated by the permutation that maps $r_j \to r_{n+2-j}$ for all $j$
(see figures~\ref{fig:rs1}--\ref{fig:rs23}).

In the above discussion, one sees that the Cartan-Killing form must be
normalised. We adopt the following normalisation: the shortest roots
of $\Dtwors{n+1}$ and $\Atwors{2n}$ have length $1/\sqrt{2}$; all
other root systems' shortest roots have unit length. This
normalisation implies that the longest root(s) of $\Grs$ and
$\Dthreers$ have length $\sqrt{3}$, while all other root systems'
longest roots have length $\sqrt{2}$.

\begin{prop} \label{prop:isom}
  Assume that $F/\Q$ is strictly hyperbolic and $\#\B{F} > 2$. Let
  $\rho_i \in \Bi_i=\Bi(\B{F},\RsL_i)$ be bijections and let
  $\Htoda_i$ be defined by Equation \ref{eq:HL0} with Hamiltonian flow
  $\varphi^i$.  If there is an energy-preserving conjugacy of
  $\varphi^1$ with $\varphi^2$, then $\mu$ (equation \ref{eq:mug})
  induces $\nu:\RsL_2 \to \RsL_1$ which is an isomorphism of Coxeter
  diagrams. Thus,
  \begin{enumerate}
  \item[Case A: ] if $\RsL_1\not\in \left\{ \Crs{n},
    \Atwors{2n},\Dtwors{n+1}  \right\}$, then 
    \begin{enumerate}
    \item $\RsL_1 = \RsL_2$;
    \item the constants $c_1=c_2$ in the definition of $\Phi_i$ (theorem
      \ref{thm:lax});
    \item $\nu \in \Aut{\Dynkin{\Rs}}$.
    \end{enumerate}
  \item[Case B: ] if $\RsL_1 \in \left\{ \Crs{n},
    \Atwors{2n},\Dtwors{n+1}  \right\}$, then 
    \begin{enumerate}
    \item $\RsL_2 \in \left\{ \Crs{n},
      \Atwors{2n},\Dtwors{n+1}  \right\}$;
    \item the constants $c_1$ and $c_2$ in the definition of $\Phi_i$ (theorem
      \ref{thm:lax}) are related by the following diagram:
      \begin{center}
        \begin{tabular}{lp{3cm}}
          $
          \xymatrix{
            & \Crs{n} \ar[dl]^{\times 1} \ar@(ur,ul)[]_{\times 1} & \\
            \ar@(ul,dl)[]_{1 \times} \Atwors{2n} \ar[rr]^{\times  \frac{1}{2}} & & \Dtwors{n+1} \ar[ul]^{\times 2}
            \ar@(ur,dr)[]^{\times 1}
          }
          $
          &
          where the factor $\star$ yields $c_2 = c_1 \times \star$.
        \end{tabular}
      \end{center}
    \item if $\RsL_1=\RsL_2$, then $\nu \in \Aut{\Dynkin{\Rs}}$.
    \end{enumerate}
  \end{enumerate}
\end{prop}

\begin{proof}
  From equation (\ref{eq:frJ1}) and corollary \ref{thm:conjd}, one
  knows that $\mu$ in equation (\ref{eq:mug}) maps a root in $\h^*_2$ to
  a non-zero multiple of a root in $\h^*_1$. Let $\nu$ denote the
  induced bijection $\RsL_2 \to \RsL_1$. Since $\mu$ maps $r \in \Rs_2$
  to a scalar multiple of the root $\nu(r) \in \Rs_1$, one can write
  $\mu(r) = a_{r} \nu(r)$ for some coefficients $a_{r}$.

  To determine the coefficients $a_r$, note that $\phi_i(\tau) =
  n_{\tau}^{-1} \omega_{r_i} r_i$ where $\rho_i(\tau) = r_i$. One
  computes that
  \begin{xalignat*}{2}
    \mu(r) &= \frac{c_2}{c_1} \times \phi_1 \comp f^* \comp
    \phi^{-1}_2(r) && \hspace{10mm} \textrm{by definition of }\mu,\\
    &= \frac{c_2}{c_1} \times \frac{n_\tau}{\omega_r} \times
    \phi_1 \comp f^*(\hat\tau)
    && \hspace{10mm} {\rm where\ } \rho_2(\tau) = r,\\
    &= \frac{c_2}{c_1} \times  \frac{n_\sigma}{\omega_r} \times \phi_1(\hat\sigma)
    && \hspace{10mm} {\rm where\ } n_\sigma\,\hat\sigma = n_\tau\, f^* \hat\tau,\\ 
    &= \frac{c_2}{c_1} \times \frac{\omega_{\nu(r)}}{\omega_r} \times
    \nu(r)
    && \hspace{10mm} {\rm where\ } \rho_1(\sigma)=s, \phi_1(\hat
    \sigma)=\frac{\omega_s}{n_\sigma}\,  s\ \text{and}\
    \nu(r)=s.
  \end{xalignat*}
  Therefore, since $\mu$ is an isometry
  \begin{equation} \label{eq:mu1}
    \left\langle\langle r , s \right\rangle\rangle_2 = \left(
    \frac{c_2}{c_1} \right)^2  \times \frac{\omega_{\nu(r)}\,
      \omega_{\nu(s)}}{\omega_r\, \omega_s} \times
    \left\langle\langle \nu(r),\nu(s) \right\rangle\rangle_1
    \hspace{10mm} \forall r,s \in \Rs_2 .
  \end{equation}
  This implies that the Coxeter diagrams of $\RsL_1$ and $\RsL_2$ are
  isomorphic. Inspection of figures \ref{fig:rs1}--\ref{fig:rs23}
  shows that $\{\RsL_1,\RsL_2\}$ is contained in one of the following
  sets:
  \begin{align} \label{al:sets}
    &\left\{ \Ars{n}  \right\} &
    & \left\{ \Brs{n}, \Atwors{2n-1} \right\}  &  
    \left\{ \Crs{n}, \Atwors{2n},\Dtwors{n+1}  \right\}\notag\\
    &\left\{ \Drs{n}  \right\} &
    & \left\{ \Grs, \Dthreers  \right\}  & 
    \notag\\
    &\left\{ \Ers{n}  \right\}_{n=6,7,8} &
    & \left\{ \Frs, \Ers{6}  \right\}  & 
  \end{align}
  \begin{enumerate}
  \item[Case A. ] Suppose that we are in one of the cases covered by the
    first two columns of (\ref{al:sets}). Note that since $c_i \in \N$,
    one obtains that
    \begin{equation} \label{eq:ratio}
      \frac{|r|_2}{|\nu(r)|_1} = \frac{c_2}{c_1} \times
      \frac{\omega_{\nu(r)}}{\omega_r} \in \Q \hspace{10mm} \forall r\in\Rs_2.
    \end{equation}
    The possible ratios of root lengths is $1,\sqrt{2}$ and $\sqrt{3}$
    or ratios of the these three numbers. Therefore, the ratios are
    always $1$. This proves that $\nu$ is itself an
    isometry. Therefore $\RsL_1=\RsL_2$.

    To prove that $c_1=c_2$, note that since $\nu$ is a permutation of
    $\RsL$, it has finite order. If $r \in \Rs$ is a fixed point of $\nu^k$
    for some $k\geq 1$, then equation \ref{eq:ratio} implies that
    %  \begin{equation} \label{eq:ratio2}
    %    \frac{|r|}{|\nu(r)|} \times \frac{|\nu(r)|}{|\nu^2( r)|}
    %    \times \cdots \times \frac{|\nu^{k-1}(r)|}{|\nu^k(r)|} = \left( 
    %    \frac{c_2}{c_1}\right)^k \times \frac{\omega_{\nu(r)}}{\omega_r}
    %    \times \frac{\omega_{\nu^2(r)}}{\omega_{\nu(r)}} \times \cdots
    %    \times \frac{\omega_{\nu^k(r)}}{\omega_{\nu^{k-1}(r)}},
    %  \end{equation}
    \begin{equation} \label{eq:ratio2}
      1 = \left( 
      \frac{c_2}{c_1}\right)^k \times \frac{\omega_{\nu(r)}}{\omega_r}
      \times \frac{\omega_{\nu^2(r)}}{\omega_{\nu(r)}} \times \cdots
      \times \frac{\omega_{\nu^k(r)}}{\omega_{\nu^{k-1}(r)}},
    \end{equation}
    so $1 = \left( \frac{c_2}{c_1}\right)^k$.

  \item[Case B.]  In this case, following equation (\ref{eq:ratio}),
  the rational ratios that are possible are $1,2$ or $1/2$. A simple
  check shows that the natural Coxeter isomorphisms $\Crs{n} \to
  \Atwors{2n}$, $\Atwors{2n} \to \Dtwors{n+1}$ and $\Dtwors{n+1}
  \to \Crs{n}$ satisfy this constraint with $c_2/c_1$ equal to $1$,
  $1/2$ and $2$ respectively (see figure \ref{fig:obv}). These Coxeter
  isomorphisms are unique up to the action of the automorphism
  groups. If $\RsL_1=\RsL_2$, then these considerations imply that
  $\nu$ is a Dynkin diagram automorphism and $c_2=c_1$.
\begin{center}
\begin{figure}[!ht]
\input{obvious} 
\caption{The natural Coxeter isomorphisms $\nu : \RsL_2 \to
\RsL_1$.}\label{fig:obv}
\end{figure}
\end{center}
  \end{enumerate}
\end{proof}

\begin{remark}
{\rm
In \cite[Lemma 24]{B7} there is a simpler version of theorem
\ref{prop:isom}. It is assumed there that $\RsL_i \neq \Atwors{2n}$
for both $i$ and $c_2=c_1$. In this case, $\nu$ must be an
automorphism of the Dynkin diagram.
}
\end{remark}

\ifthenelse{\shortfile=5}{\end{document}}{}

\section{Topological Entropy} \label{sec:an} 
In the proof of the complete integrability, Theorem \ref{thm:cas}, one
sees that the singular set of the algebra $\La+\RR$ is the union of
$\lax[]^{-1}(\reg^c)$ and $\ZZ=\caspb^{-1}(0)$. Theorem \ref{thm:nh}
shows that the non-wandering set of the Hamiltonian flow $\varphi$ of
$\Htoda$, restricted to the invariant set $\ZZ=\W{}^{\pm}(\Hstar)$, is
$\Hstar$. What happens on the other part of the singular set,
$\lax[]^{-1}(\reg^c)$?

\subsection{The $\Ars{n}$ lattice} \label{ssec:an} 
Thanks to the work of Foxman and Robbins \cite{FM1,FM2}, this question
is answerable for the $\Ars{n}$ lattice.
\begin{thm} \label{thm:an}
  Let $\RsL = \Ars{n}$ and $\Htoda$ be a \BT-like
  Hamiltonian defined in equation \ref{eq:HL0}. Then
  $\lax[]^{-1}(\reg^c)$ is stratified by symplectic manifolds that are
  invariant under the Hamiltonian flow of $\Htoda$. Moreover, $\Htoda$
  restricted to each stratum is completely integrable.
\end{thm}

\begin{proof}
  Let $h=2 \kappa \in C^{\infty}(\e + \L_0^* + \L_{+1}^*)$ be the
  Cartan-Killing form, so that $\Htoda = \lax[]^*h$ (equation
  \ref{eq:HL0}). Foxman and Robbins \cite{FM1,FM2} proved that $h$
  admits action-angle variables with singularities, which means that
  for each point $p \in \e + \L_0^* + \L_{+1}^*$, there are
  coordinates $(x,y,u,v)$ on a neighbourhood of $p$ in $\orb{p}^R$
  such that the canonical symplectic form on $\orb{p}$ and $h$ take
  the form
  \begin{xalignat*}{2}
    & \sum_{i=1}^k \d x_i \wedge \d y_i + \sum_{i=k+1}^n \d u_i \wedge \d
    v_i, && h=h(x,\rho)\\
    & &&  \rho_i=u_i^2 + v_i^2, \ i=k+1,\ldots n, 
  \end{xalignat*}
  where $k=0,\ldots,n$ is the co-rank of the singularity (and if $k=n$, then
  there is no singularity). The set $\rho=0$ is an invariant symplectic
  submanifold and $h$ is completely integrable on this submanifold.

  Let $X_k \subset \e + \L_0^* +
  \L_{+1}^*$ be one of these symplectic sub-manifolds of dimension $2k$
  and co-dimension $2l$ where $n=k+l$. Let $Y_k = \lax[]^{-1}(X_k)$.

  Because $\lax[] | \caspb^{-1}(\R-0)$ is a submersion onto $\e +
  \L_0^* + \L_{+1}^*$, $Y_k$ is a submanifold of $T^*\Sigma$ of
  co-dimension $2l$. Moreover, since $\lax[]$ is a Poisson submersion,
  $Y_k$ is also a symplectic submanifold. Since $X_k$ is invariant
  under the hamiltonian flow of $h$, $Y_k$ is similarly invariant.

  From the above description of the singular action-angle variables, the
  algebra $\La | Y_k$ is equal to $\lax[]^*Z^{\infty}(\L^*)|X_k$, which
  contains $k$ functionally independent elements. On the other hand, the
  algebra $\RR | Y_k$ contains $\dim \V{E}$ functionally independent
  elements. Therefore, in total, there are $k+\dim \V{E}$ functionally
  independent integrals of $\Htoda$ at each point of $Y_k$. Since $\dim
  Y_k = 2(k+\dim \V{E})$, this proves the complete integrability of
  $\Htoda | Y_k$, which proves the theorem.
\end{proof}

\begin{remark}
  {\rm
    It is clear from the proof that $\Htoda | \UU$ is completely
    integrable with singular action-angle variables. This is the mildest
    kind of singularity that a completely integrable may have. It is a
    stark contrast with the sort of singularity that develops along
    $\ZZ=\caspb^{-1}(0)$.
    
    It is natural to conjecture that the Foxman-Robbins theorem is
    true for all \BT\  lattices. 
  }
\end{remark}

\begin{cor} \label{co:an}
  Let $\RsL = A_n^{(1)}$ and $\Htoda$ be a Toda-like hamiltonian defined
  in equation \ref{eq:HL0}. The topological entropy of $\varphi_1 |
  \Htoda^{-1}(\frac{1}{2})$, the time-$1$ map of the hamiltonian flow of
  $\Htoda$, equals
\begin{equation} \label{eq:ent}
h_{top} = \frac{[E:F]}{c} \times \sqrt{\floor{\frac{n+1}{2}}}
\end{equation}
where $n=\dim\Vo{F}$ and $c \in \frac{1}{2} \Z^+$ as in theorem
\ref{thm:lax}.
\end{cor}

\begin{proof}
Since $\varphi$ admits singular action-angle variables on
$\caspb^{-1}(\R-0)$, we see that the topological entropy of $\varphi$
is generated entirely in $\caspb^{-1}(0)=\W{}^\pm(\Hstar)$. The
non-wandering set of $\varphi|\W{}^\pm(\Hstar)$ is $\Hstar$ by theorem
\ref{thm:nh}. Thus
\begin{align} \label{al:entan}
h_{top}(\varphi|\Htoda^{-1}(\frac{1}{2})) = h_{top}(\varphi|\Hstaro) =
\frac{[E:F]}{c} \times \sqrt{\floor{\frac{n+1}{2}}} && \text{by table \ref{tab:entropy}.}
\end{align}

\end{proof}

\subsection{The remaining \BT\  lattices} \label{ssec:topent} 
As in \cite[Section 3]{B7}, the universal covering space
$\tilde{\Sigma}=\V{E} \times \Vo{F}$ admits the structure of a
solvable Lie group. The element $v \in \Vo{F}$ acts by right
translation by the one-parameter subgroup
\begin{equation} \label{eq:v}
\tilde{\phi}^v_t(\y,\x) = (\y+t\cdot v,\x).
\end{equation}
This flow descends to a flow $\phi^v$ on $\Sigma$. As in \cite[Lemma
  12]{B7}, 
\begin{equation} \label{eq:hv}
h_{top}(\phi^v) = \ds [E:F] \times \sum_{\tau \in \B{F}}
n_\tau \, \left\langle \hat{\tau} , v \right\rangle^+
\end{equation}
where $u^+=\max\{u,0\}$. 

Let $\Htoda$ be defined by equation (\ref{eq:HL0}) and let
\begin{equation} \label{eq:v1}
\Hstaro = \Hstar \cap \Htoda^{-1}(\frac{1}{2})
\end{equation}
where $\Hstar$ is defined in section \ref{sec:gradientflow}. If $v =
\Qm \cdot \px$ with $\px \in \Vo{F}^*$, and $\left\langle \Qm\cdot\px
, \px \right\rangle = 1$, then $\Delta \cdot (\y,\x,0,\px) \in
\Hstaro$. The topological entropy of the Hamiltonian flow $\varphi$ of
$\Htoda$ is therefore equal to
\begin{align} \label{al:htop}
&\frac{1}{[E:F]} \times h_{top}(\varphi|\Hstaro) \notag\\
&=\max_{\px:\left\langle \Qm\cdot\px ,\px  \right\rangle=1} \ 
\sum_{\tau\in\B{F}} n_\tau \,
\left\langle \hat{\tau},\Qm\cdot\px \right\rangle^+ \notag\\
&= \max_{\px:\left\langle \Qm\cdot\px ,\px  \right\rangle=1} \ 
\sum_{\tau\in\B{F}} n_\tau \, \left\langle\langle
\phi_\rho(\hat{\tau}),
\phi_\rho(\px)\right\rangle\rangle \notag \\
&= \max_{s\in\h: \left\langle\langle s,s\right\rangle\rangle=1} \
\sum_{\r\in\RsL} \frac{\omega_r}{c} \times \left\langle r,s\right\rangle^+ \hspace{4mm}\text{where
}\phi_{\rho}(\hat\tau)=\frac{\omega_r}{n_\tau c} r, s=\phi_{\rho}(\px)
\notag\\
&= c^{-1} \times \max_{I\subset \RsL} 
\left| \sum_{\r\in I} \omega_r\, r \right|
\end{align}
The right-hand side of \ref{al:htop} is computed in \cite[Lemma 13 and
  Theorem 3]{B7}. These results are summarised in table \ref{tab:entropy}.

\begin{table}[!ht]
{
\setlength{\extrarowheight}{4pt}
\newcolumntype{L}{>{$}l<{$}}
\newcolumntype{C}{>{$}c<{$}}
\begin{tabular}{|L|C||L|C||L|C|}
\multicolumn{6}{c}{\fbox{$h_{top}(\varphi|\Hstaro)=h \times \frac{[E:F]}{c}$}} \\ 
\multicolumn{6}{C}{ }\\ \hline
\RsL & h &
\RsL & h &
\RsL & h \\
\hline \hline
\Brs{n},\ n \geq 3    	   &  2\sqrt{ n-1 }  	&			   
\Atwors{2n-1},\ n \geq 3   &  \sqrt{ 2(n-1) } 	&	& \\ %\hline	   
\Grs,\ (n=2)    	   &  2\sqrt{ 3 }  	&			   
\Dthreers,\ (n=2)    	   &  2  	        &	&\\ %\hline			   
\Frs,\ (n=4)    	   &  2\sqrt{6}  	&				   
\Etwors{},\ (n=4)    	   &  2\sqrt{3}  	&	&\\ %\hline		   
\Crs{n},\ n \geq 2    	   &  \sqrt{ 2n }  	&			   
\Atwors{2n},\ n \geq 2     &  2\sqrt{ n }  	&			   
\Dtwors{n+1},\ n \geq 2    &  \sqrt{ n }  \\ \hline			   
\Ers{6},\ (n=6)    	   &  2\sqrt{3}  	&				   
\Ers{7},\ (n=7)    	   &  2\sqrt{6}  	&				   
\Ers{8},\ (n=8)    	   &  2\sqrt{15}  \\ %\hline			   
\Ars{n},\ n \geq 2    	   &  \sqrt{ \floor{ \frac{n+1}{2} } } 	&	   
\Drs{n},\ n \geq 4    	   &  \sqrt{ 2(n-2) }  	& %\hline		   
\Ars{2},\ (n=1)    	   &  \sqrt{2} \\ \hline                              
\end{tabular}
}
\caption{Entropies of the \BT-like systems.  The root
 systems in the first 4 rows have isomorphic Coxeter graphs; the root
 systems in the last 2 rows have unique Coxeter graphs. $n=\dim
 \Vo{F}$.} \label{tab:entropy}
\end{table}

Table \ref{tab:entropy} permits one to give lower bounds on the number
of \BT-like systems which are not energy-preserving topologically
conjugate. 

\begin{prop} \label{prop:topnc}
For each $n\geq 2$, table \ref{tab:topnc} displays \BT-like systems,
defined in (\ref{eq:HL0}), that are not topologically conjugate via an
energy-preserving conjugacy.
\begin{table}[!ht]
{
\setlength{\extrarowheight}{4pt}
\newcolumntype{L}{>{$}l<{$}}
\begin{tabular}{LLL}
n & {\rm Root Systems}      & {\rm Total}\\\hline
2 & \Ars{2}, \Crs{2}, \Grs, \Atwors{2\cdot 2} & 4\\
3 & \Ars{3}, \Crs{3}, \Atwors{2\cdot 3}, \Atwors{2\cdot 3 - 1} & 4\\
4 & \Ars{4}, \Brs{4}, \Atwors{2\cdot 4}, \Atwors{2\cdot 4-1} & 4\\
5 & \Ars{5}, \Brs{5}, \Crs{5}, \Drs{5}, \Atwors{2\cdot 5}, \Atwors{2\cdot 5-1} & 6\\
6 & \Ars{6}, \Brs{6}, \Drs{6}, \Atwors{2\cdot 6}, \Atwors{2\cdot 6-1} & 5\\
7 & \Ars{7}, \Brs{7}, \Crs{7}, \Drs{7}, \Atwors{2\cdot 7}, \Atwors{2\cdot 7-1} & 6\\
8 & \Ars{8}, \Brs{8}, \Crs{8}, \Drs{8}, \Ers{8}, \Atwors{2\cdot 8-1} & 6\\
\geq9 {\rm\ even} & \Ars{n}, \Brs{n}, \Drs{n}, \Atwors{2\cdot n}, \Atwors{2\cdot n-1} & 5\\
\geq9 {\rm\ odd} & \Ars{n}, \Brs{n}, \Crs{n}, \Drs{n}, \Atwors{2\cdot n}, \Atwors{2\cdot n-1} & 6\\
\end{tabular}
}
\caption{Minimal number of \BT-like systems that are not
  iso-energetically topologically conjugate.} \label{tab:topnc}
\end{table}
\end{prop}

\begin{proof}
Use table \ref{tab:entropy} to determine a list of root systems the
ratio of whose entropies do not lie in $\frac{1}{2} \Z$. Note that
this list is not unique.
\end{proof}

%X
\subsection{Summary} \label{ssec:sum}
If the results from table \ref{tab:entropy} are combined with
proposition \ref{prop:isom}, one obtains the much stronger result:
\begin{thm} \label{thm:summary}
  Let $F/\Q$ be a strictly hyperbolic number field with $n+1=\# \B{F} >
  2$. The number of iso-energetic topological conjugacy classes of
  Hamiltonian flows constructed from Equation (\ref{eq:HL0}) is {\em at
    least}
  \begin{equation} \label{eq:topclass}
    \sum_{\rank \RsL=n} \# \left(\Aut{\Dynkin{\RsL}}\backslash \Bi(\RsL)/\Aut{F/\Q}\right).
  \end{equation}
  where we sum over all rank $n$ root systems {\em except}
  $\Dtwors{n+1}$.
\end{thm}

\begin{proof}
  By Proposition \ref{prop:isom}, we know that if the \BT-like
  Hamiltonian flows $\varphi^i$ are conjugate by an energy-preserving
  conjugacy, then there are two possibilities
  \begin{enumerate}
  \item[Case A.] The root systems coincide, $c_1=c_2$ and the map
    $\nu=\mu$ is an automorphism of $\Dynkin{\RsL}$. The definition of
    $\mu$ (equation \ref{eq:mug} and \supra \ref{eq:mu1}) implies
    that the maps $\phi_1,\phi_2$ are related by
    \begin{align} \label{al:aut}
      \phi_1 = \mu \cdot \phi_2 \cdot R_{\beta}^* && \beta\in\Aut{F/\Q}
    \end{align}
    where theorem \ref{thm:sh} is used. Conversely, given any
    $\phi_2$, a $\phi_1$ defined as in equation (\ref{al:aut}) is
    induced by a bijection $\rho_1 \in \Bi$.

  \item[Case B.] The two root systems differ, as in Case B of
    Proposition \ref{prop:isom}. The topological entropy of $\varphi^i
    | \Hstaro \cap \Htoda_i(\frac{1}{2})$ is an invariant of
    energy-preserving conjugacy by Lemma
    \ref{lem:asymphomology}. Table \ref{tab:entropy} implies that the
    root systems must therefore be $\left\{ \RsL_1,\RsL_2 \right\} =
    \left\{ \Atwors{2n},\Dtwors{n+1} \right\}$ or $\left\{
    \Crs{n},\Dtwors{n+1} \right\}$. Since the sum (\ref{eq:topclass})
    counts the conjugacy classes from only one of these two root
    systems, there is no double counting. This proves the theorem.
  \end{enumerate}
\end{proof}

\begin{remark} \label{re:cubic}
{\rm

In \cite[Example 3, p. 541]{B7}, the case where $F=E=\Q(\alpha)$, with
$\alpha$ a root of the cubic $x^3-4x+2$, was considered (\cf example
\ref{sssec:ex1} \supra). $F$ is a cubic, totally-real, non-normal extension of
$\Q$. Thus, $\Aut{F/\Q}$ is trivial and $F$ is strictly hyperbolic. If one
sums over the rank $2$ root systems and divides out by the order of their
automorphism groups, then Theorem \ref{thm:summary} implies that there are at
least
\begin{align} \label{al:ex3}
1+3+6+3+6=19 &&\text{(summing over $\Ars{2}, \Crs{2}, \Grs,
  \Atwors{2\cdot 2}, \Dthreers$)}
\end{align}
iso-energetic topological conjugacy classes. In \cite[theorem 8]{B7},
the lower bound of $10$ was conjectured.\footnote{Inexplicably, only
  the first three root systems are included in that sum, so the
  conjectural lower bound ought to be $19$.} This lower bound depended
on Gel'fond's conjecture concerning the algebraic independence of
rationally-independent sets of logarithms of algebraic numbers. The
results of the present paper, using dynamical systems theory, has
proven this lower bound.

In a similar vein, if $F=E$ is a totally real quartic field with
$\Aut{F/\Q}=1$, then one has at least
\begin{align} \label{al:ex4}
3+4 \times 12=51 &&\text{(summing over $\Ars{3}, \Brs{3}, \Crs{3},
  \Atwors{2\cdot 3}, \Atwors{2\cdot 3-1}$)}
\end{align}
iso-energetic topological conjugacy classes.
}
\end{remark}

\begin{remark} \label{re:fine}
{\rm Theorem \ref{thm:summary} provides a means to compute a lower
bound on the number of iso-energetic topological conjugacy classes
when $\Aut{F/\Q}$ is non-trivial, too. Both $\RsL$ and $\B{F}$ are
unnaturally isomorphic to the set $\left\{ 1,\ldots,n+1
\right\}$. Theorem \ref{thm:sh}, part 2, shows that the representation
of $\Aut{F/\Q}$ in the group of permutations of $\B{F}$,
$\symmgp{\B{F}}$, is the natural right regular representation (one
should view $\B{F} = \G{F}/( \cdot \sim \bar{\cdot} )$). By
definition, the automorphism group of the Dynkin diagram is a subgroup
of the group of permutations of the roots, $\symmgp{\RsL}$. Therefore,
the unnatural isomorphisms of $\RsL$ and $\B{F}$ with $\left\{
1,\ldots,n+1 \right\}$ identify the set of bijections $\Bi(\RsL)$ with
the symmetric group of $\left\{ 1,\ldots,n+1  \right\}$,
$S_{n+1}$, with the resulting equivariant diagram (where left/right
arrows denote the standard left (resp. right) actions)
\begin{align}
\xymatrix{
\Aut{\Dynkin{\RsL}} \ar@{^{(}->}[r] \ar@{_{(}->}[rd]  \ar[d]_{\cong} & \symmgp{\RsL}
\ar[r] \ar[d]_{\cong} & \Bi(\RsL) \ar[d]^{\cong} & \symmgp{\B{F}}
\ar[l]\ar[d]^{\cong} & \Aut{F/\Q}
\ar@{_{(}->}[l] \ar@{^{(}->}[ld] \ar[d]^{\cong}\\
G \ar@{^{(}->}[r] & \symmgp{} \ar[r]^{{\rm id.}}  & \symmgp{} & \symmgp{}
\ar[l]_{{\rm id.}} & H. \ar@{_{(}->}[l] 
}
\end{align}
This implies that $\#(G\backslash \symmgp{} / H)$ equals
$\#(\Aut{\Dynkin{\RsL}} \backslash \Bi(\RsL) / \Aut{F/\Q})$. Table
\ref{ta:nums} shows the cardinality of each of these sets for $n\leq
9$. The table is computed by a C++ program written by the author; the
computations were checked using the GAP software package
\cite{GAP}. The source code and instructions are freely available from
the author's
\href{http://www.maths.ed.ac.uk/~lbutler/toda-c.html}{web-page}.

\newcolumntype{C}{>{$}c<{$}}
\newcolumntype{L}{>{$}l<{$}}
\begin{longtable}[c]{|L|CCCCCL|}
\caption{The minimum number of iso-energetic topological conjugacy
  classes of Bogoyavlenskij-Toda-like systems. The {\em Total} column
  is based on Theorem \ref{thm:summary} and Tables \ref{fig:rs1}--\ref{fig:rs23} of
  Coxeter graph automorphism groups. \newline {\small $\Z_n=\Z/n\Z$, $D_n=$ the dihedral group
  of order $2n$, $Q=$ the quaternion group of order $8$.}}  \label{ta:nums}
\\ \hline
& \multicolumn{5}{c}{$ \#
  \left(\Aut{\Dynkin{\RsL}}\backslash \Bi(\RsL)/\Aut{F/\Q}\right).$} &
  \\
{\rm rank} &&&&&&\\
{\rm Galois\ grp} & \multicolumn{5}{c}{Root systems (grouped with
  isomorphic Coxeter diagrams)} & {\rm Total}\\ \hline
\endfirsthead
\hline
\multicolumn{7}{|c|}{\small\slshape Table \ref{ta:nums}, continued from previous page} \\\hline
\endhead
\hline\multicolumn{7}{|c|}{\small\slshape continued next page} \\\hline
\endfoot
\hline
%\multicolumn{7}{|c|}{\small $\Z_n=\Z/n\Z$, $D_n=$ the dihedral group
%  of order $2n$, $Q=$ the quaternion group of order $8$.}\\\hline
\endlastfoot
\hline
\input{bicoset.tex}%/home/lbutler/c/set/test.out}
%\input{/home/lbutler/c/set/bicoset.out}
\end{longtable}
}
\end{remark}

\section{Conclusion} \label{sec:conclusion} 

The current paper shows that there is a rich family of completely
integrable Hamiltonian systems to be found on the cotangent bundles of
compact 2-step $Sol$-manifolds. In addition to the questions in the
introduction, let us mention the following question which arises from
lemma \ref{lem:alpha} and theorem \ref{thm:sh}. 
\begin{question} \label{qu:6}
Let $F$ be a number field that is not strictly hyperbolic. Assume that
there is an automorphism $\alpha$ of $\U_F$ and a permutation $\perm$
of $\G{F}$ such that
\begin{align}
\forall u\in\U_F,\ \forall\tau\in\G{F}:&\qquad |\tau(\alpha(u))| = |\sigma(u)|
\qquad \text{where}\ \perm(\tau)=\sigma, \text{ and} \label{qu:cond}\\
\forall\tau\in\G{F}:&\qquad \perm(\bar\tau)=\overline{\perm(\tau)}\notag
\end{align}
Is it true that there is an automorphism $\beta$ of $F/\Q$ such that
$\alpha = \left. \beta \right|_{\U_F}$? In other words, is it true
that $\cap_{\tau\in\G{F}} \ker \zeta_\tau$ is always a finite-index
subgroup of $\U_F$?
\end{question}

It appears the likely that the answer is {\em yes}. To explain: If
$u_i$ is a basis of $\U^+_F$ and $\alpha \in \Aut{\U_F}$, then
$\alpha(u_i) = \epsilon_i \times \prod_j u_j^{a_{ji}}$ for some
integer matrix $A=[a_{ji}]$ that is invertible over the integers, and
some root of unity $\epsilon_i \in \U_F$. From the condition
\eqref{qu:cond}, one knows that the system of linear equations
\begin{align}
\sum_j a_{ji}\, \ln |\perm{\sigma}(u_i)| = \ln |\sigma(u_i)| \label{al:dirichlet}
\end{align}
is satisfied for all $j=1,\ldots,\#\B{F}-1$ and embeddings $\sigma
\in \B{F}$. For a fixed permutation $\perm{}$, one can treat
\eqref{al:dirichlet} as a linear system that determines $A$. If there
is an integer solution, then this determines an automorphism $\alpha$;
if not, then there is no such automorphism.

Salem number fields are good candidates to investigate
question~\ref{qu:6} because these number fields have many infinite
order units of modulus one. By means of \maxima~\cite{Maxima}, it has
been numerically verified that the answer to the refined question is
{\em yes} for the 13 lowest degree number fields generated by the
`small' Salem numbers listed by
\href{http://www.cecm.sfu.ca/~mjm/Lehmer/lists/SalemList.html}{Mossinghoff},
based on \cite[Table 1]{Boyd} and \cite[Table 1]{Mossinghoff}.

%% Root systems:
\include{rs}

\end{document}

%% file: definitions.tex
\def\Ars#1{\ensuremath{A^{(1)}_{#1}}}
\def\Brs#1{\ensuremath{B^{(1)}_{#1}}}
\def\Crs#1{\ensuremath{C^{(1)}_{#1}}}
\def\Drs#1{\ensuremath{D^{(1)}_{#1}}}
\def\Atwors#1{\ensuremath{A^{(2)}_{#1}}}
\def\Dtwors#1{\ensuremath{D^{(2)}_{#1}}}
\def\Dthreers{\ensuremath{D^{(3)}_{4}}}
\def\Ers#1{\ensuremath{E^{(1)}_{#1}}}
\def\Etwors{\ensuremath{E^{(2)}_{6}}}
\def\Grs{\ensuremath{G^{(1)}_2}}
\def\Frs{\ensuremath{F^{(1)}_4}}

\def\BT{Bogoyavlenskij-Toda}

\def \dim{{\mbox {dim}}\,}

\newcommand{\diag}[1]{\mathrm{diag}(#1)}
\newcommand{\trace}[1]{\mathrm{Tr}(#1)}
\newcommand{\comp}{{\ensuremath{\,\tiny{\circ}\,}}}
\renewcommand{\comp}{{\circ}}

\renewcommand{\Vert}{{\mathbf V}}
\def\Hstar{{{\mathbf V}^{\perp}}}
\def\Hstaro{{{\mathbf V}_1^{\perp}}}

\def\ppi1{{\Delta}}

\def\Dynkin#1{{\Gamma{\rm (}#1{\rm )}}}

\def\Z{{\mathbb Z}}
\def\N{{\mathbb N}}
\def\R{{\mathbb R}}
\def\T{{\mathbb T}}

\def \Q{{\mathbb Q}}
\def \C{{\mathbb C}}
\def \O{{\mathcal O}}
\def \U{{\mathcal U}}

\def\Smat{{\bf S}}

\def\b{{\mathfrak b}}

\def\F{ {\mathcal F} }

\def\a{ {\mathfrak a} }

\def\f{{\bf f}}

\newcommand{\p} [0] { {\bf p} }

\newcommand{\SL} [1] { {\mathrm{SL(}}#1 {\mathrm )} }
\newcommand{\GL} [1] { {\mathrm{GL(}}#1 {\mathrm )} }

\newcommand{\spl} [1] { {\it{sl}} (#1) }

\newcommand{\ad} [1] { \mathrm{ad}_{#1} }
\newcommand{\Ad} [1] { \mathrm{Ad}_{#1} }

\newcommand{\Lie} [1] { {\rm Lie(}#1{\rm )} }

\newcommand \im { {\rm im}\, }

\newcommand \rank { {\rm rank}\, }

\def\Bi{{\mathfrak B}}

\def\Aut#1{{\rm Aut(}#1{\rm )}}
\def\llangle{{\langle \langle}}
\def\rrangle{{\rangle \rangle}}
\newcommand{\logu} [2][] {{\pmb{\mathfrak L}}_{#2}^{#1}}

\def\D{{\rm D}}

\newcommand{\G} [1] {{\bf G}_{#1}}
\newcommand{\V} [1] {{\rm V}_{#1}}
\newcommand{\W} [1] {{\rm W}_{#1}}
\newcommand{\Vo} [1] {{\rm V}_{o,{#1}}}
\newcommand{\B} [1] { {\bf B}_{#1} }

\newcommand{\htt} [1] {{ {\rm ht}(#1) }}
\newcommand{\slr} [1] {{ {\rm sl}_{#1}\R }}
\renewcommand{\r}[1][r]{{ \bf{#1} }}  % r is the default argument
\newcommand{\xyonto}[3][ ]{{\xymatrix{ {#2} \ar@{->>}[r]^{#1} & {#3}}}}
\newcommand{\xyinto}[3][ ]{{\xymatrix{ {#2} \ar@{^{(}->}[r]^{#1} & {#3}}}}
\newcommand{\onto}[3][ ]{{#2 \stackrel{#1}{\twoheadrightarrow} #3}}
\newcommand{\into}[3][ ]{{#2 \stackrel{#1}{\hookrightarrow} #3}}
\newcommand{\bij}[3][ ]{{#2 \stackrel{#1}{\cong} #3}}
\newcommand{\lax}[1][ ]{{ {\bf L}_{#1} }}

\newcommand{\xyexact}[5][ ]{{\xymatrix{ {#1} \ar@{^{(}->}[r]^{#5} &
      {#2} \ar@{->>}[r]^{#4} & {#3} }}}
\newcommand{\symmgp}[1] {\ifthenelse{\equal{#1}{}}{S_{n+1}}{S(#1)}}

\def\Tf#1{{\bf T}_{#1}}
\def\px{{\mathsf X}}
\def\py{{\mathsf Y}}
\def\x{{\mathsf x}}
\def\y{{\mathsf y}}
\def\kerl{{\mathcal R}}
\def\t{\hat{\rm t}}

\def\p{{\rm p}}
\def\d{{\rm d}}
\def\g{ {\mathfrak g} }
\def\gp{ {\mathfrak g}^{\perp} }
\def\h{ {\mathfrak h} }
\def\Rs{{\Psi}}
\def\RsL{{\pmb{\Psi}}}
\def\L{{\mathcal L}}
\def\Lnot{{}_0{\mathcal L}}
\def\e{{\rm e}}
\def\lcm{{\rm lcm}}
\def\minrt{{\eta}}
\def\bminrt{{\pmb{\eta}}}
\def\Htoda{{\bf H}}
\def\Qm{{\mathcal Q}}

\def\caspb{{\mathrm k}}
\def\casimir{{\mathfrak k}}
\def\UU{{\mathfrak U}}
\def\ZZ{{\mathfrak Z}}

\def\fint{{\mathfrak f}}
\def\q{{\mathfrak q}}
\def\RR{{\mathsf R}}
\def\La{{\mathsf L}}
\def\reg{{\mathfrak R}}
\def\hr{{\mathrm h}}
\def\perm{{\pmb{\pi}}}

\def\PerO{{\rm P}}
\def\Period{{\rm Period}}
\def\HSpec{{\mathcal M}}

\def\pwrset#1{{\bf 2}^{#1}}
\def\Span{{\rm span\, }}
\def\Eta#1{{\rm N}_{#1}}

\def\cf{{\em c.f.\, }}
\def\orb#1{{\mathcal O}_{#1}}
\def\nm{{\bf n}}
\def\gradnm{{\stackrel{\nm}{\nabla}}}
\def\ucSigma{{\tilde{\Sigma}}}
\def\Hom#1{{\rm Hom}(#1)}
\def\metric{{\bf g}}
\def\ds#1{{\displaystyle #1}}
\def\f{{\textsl f}}
\def\jj{{\text{\j}}}
\def\floor#1{{\rm floor}\left(#1\right)}

\def\cf{\textit{c.f.}\,}
\def\supra{\textit{supra}\,}

\def\maxima{\href{http://maxima.sourceforge.net/}{{\tt Maxima}}}

\newtheorem{theorem}{Theorem}%[section]
\newtheorem{thm}{Theorem}[section]
\newtheorem{prop}{Proposition}[section]
\newtheorem{lemma}[thm]{Lemma}
\newtheorem{defn}{Definition}[section]
\newtheorem{definition}{Definition}%[section]
\newtheorem{corollary}[thm]{Corollary}
\newtheorem{cor}[thm]{Corollary}

\newcounter{remark}
\newtheorem{remark}[remark]{Remark}
\numberwithin{remark}{section}

\newcounter{question}
\newtheorem{question}[question]{Question}
\newcounter{listctr}
\renewcommand{\thelistctr}{\arabic{listctr}}
\newcommand{\listenum}[0]{{\thelistctr\addtocounter{listctr}{1}}}

\newcounter{example}
\newenvironment{example}{\stepcounter{example}\medskip\noindent{\bf Example
    \theexample}.\ }{\medskip}
\renewcommand{\theexample}{\arabic{section}.\arabic{example}.}
\numberwithin{example}{section}
\numberwithin{equation}{section}

\newcounter{notes}
\newenvironment{notes}%
{\begin{list}{(\arabic{notes})\ }%
{\usecounter{notes}%
\setlength{\labelsep}{0pt}%
\setlength{\leftmargin}{0pt}%
\setlength{\labelwidth}{0pt}%
\setlength{\parindent}{0pt}%
\setlength{\parskip}{0pt}
\setlength{\topsep}{0pt}%
\setlength{\partopsep}{0pt}%
}}%
{\end{list}}

\newenvironment{examples}
{\medskip\noindent{\bf Examples.}
\begin{notes}}
{\end{notes}\medskip}

\newenvironment{prooflist}%
{\begin{proof}
\begin{notes}
}%
{
\end{notes}
\end{proof}
}

%% file: obvious.tex
\def\disc{{\circle*{.5}}}
\setlength{\unitlength}{.4cm}
\begin{picture}(10,7)(-1,-.5)
%\allinethickness{.5mm}
%C
\put(0,0){\disc}
\put(2,0){\disc}
\put(6,0){\disc}
\put(8,0){\disc}
\multiput(2.25,0)(.4,0){9}{\line(1,0){.2}}
\put(6.25,-.1){\line(1,0){1.5}}
\put(6.25,.1){\line(1,0){1.5}}
\put(7.25,.5){\line(-1,-1){.5}}
\put(7.25,-.5){\line(-1,1){.5}}
\put(.25,-.1){\line(1,0){1.5}}
\put(.25,.1){\line(1,0){1.5}}
\put(.75,.5){\line(1,-1){.5}}
\put(.75,-.5){\line(1,1){.5}}

%A
\put(0,3){\disc}
\put(2,3){\disc}
\put(6,3){\disc}
\put(8,3){\disc}
\multiput(2.25,3)(.4,0){9}{\line(1,0){.2}}
\put(6.25,2.9){\line(1,0){1.5}}
\put(6.25,3.1){\line(1,0){1.5}}
\put(7.25,3.5){\line(-1,-1){.5}}
\put(7.25,2.5){\line(-1,1){.5}}
\put(.25,2.9){\line(1,0){1.5}}
\put(.25,3.1){\line(1,0){1.5}}
\put(1.25,3.5){\line(-1,-1){.5}}
\put(1.25,2.5){\line(-1,1){.5}}

%D
\put(0,6){\disc}
\put(2,6){\disc}
\put(6,6){\disc}
\put(8,6){\disc}
\multiput(2.25,6)(.4,0){9}{\line(1,0){.2}}
\put(6.25,5.9){\line(1,0){1.5}}
\put(6.25,6.1){\line(1,0){1.5}}
\put(6.75,6.5){\line(1,-1){.5}}
\put(6.75,5.5){\line(1,1){.5}}
\put(.25,5.9){\line(1,0){1.5}}
\put(.25,6.1){\line(1,0){1.5}}
\put(1.25,6.5){\line(-1,-1){.5}}
\put(1.25,5.5){\line(-1,1){.5}}

%labels
\put(-2.5,-.2){$C^{(1)}_n$}
\put(-2.5,2.8){$A^{(2)}_{2n}$}
\put(-2.5,5.8){$D^{(2)}_{n+1}$}

\thinlines
\put(0, .5){\vector(0,1){2}}
\put(0,3.5){\vector(0,1){2}}
\put(2, .5){\vector(0,1){2}}
\put(2,3.5){\vector(0,1){2}}
\put(6, .5){\vector(0,1){2}}
\put(6,3.5){\vector(0,1){2}}
\put(8, .5){\vector(0,1){2}}
\put(8,3.5){\vector(0,1){2}}
\spline(-.2,.7)(-.75,3)(-.2,5.5) \put(-.2,.7){\vector(1,-3){.1}}
\spline(8.2,.7)(8.75,3)(8.2,5.5) \put(8.2,.7){\vector(-1,-3){.1}}
\spline(5.8,.7)(5.25,3)(5.8,5.5) \put(5.8,.7){\vector(1,-3){.1}}
\spline(2.2,.7)(2.75,3)(2.2,5.5) \put(2.2,.7){\vector(-1,-3){.1}}
\end{picture}

%% file: bicoset.tex
{\rm rank}=2 &&&&&& \\\Aut{F/\Q} & A^{(1)}_2 & C^{(1)}_2/A^{(2)}_{2\cdot 2}/D^{(2)}_{2+1} & G^{(1)}_2/D^{(3)}_4 &&& {\rm Total} \\\hline
1			& 1 	& 3 \times 2	& 6 \times 2	&&& 19 \\
\Z_3			& 1 	& 1 \times 2	& 2 \times 2	&&& 7
\\\hline
{\rm rank}=3 &&&&&& \\\Aut{F/\Q} & A^{(1)}_3 & C^{(1)}_3/A^{(2)}_{2\cdot 3}/D^{(2)}_{3+1} & B^{(1)}_3/A^{(2)}_{2\cdot 3-1} &&& {\rm Total} \\\hline
1			& 3 	& 12 \times 2	& 12 \times 2	&&& 51 \\
\Z_2\oplus\Z_2		& 3 	& 6 \times 2	& 3 \times 2	&&& 21 \\
\Z_4			& 2 	& 4 \times 2	& 3 \times 2	&&& 16
\\\hline
{\rm rank}=4 &&&&&& \\\Aut{F/\Q} & A^{(1)}_4 & C^{(1)}_4/A^{(2)}_{2\cdot 4}/D^{(2)}_{4+1} & B^{(1)}_4/A^{(2)}_{2\cdot 4-1} & D^{(1)}_4 &  F^{(1)}_4/E^{(2)}_6 & {\rm Total} \\\hline
1			& 12 	& 60 \times 2	& 60 \times 2	& 5 	& 120 \times 2	& 497 \\
\Z_5			& 4 	& 12 \times 2	& 12 \times 2	& 1 	& 24 \times 2	& 101
\\\hline
{\rm rank}=5 &&&&&& \\\Aut{F/\Q} & A^{(1)}_6 & C^{(1)}_6/A^{(2)}_{2\cdot 6}/D^{(2)}_{6+1} & B^{(1)}_6/A^{(2)}_{2\cdot 6-1} & D^{(1)}_6 && {\rm Total} \\\hline
1			& 60 	& 360 \times 2	& 360 \times 2	& 90 	&& 1\,590 \\
\Z_6			& 14 	& 64 \times 2	& 60 \times 2	& 17 	&& 279 \\
S_3			& 19 	& 72 \times 2	& 60 \times 2	& 21 	&& 304
\\\hline
{\rm rank}=6 &&&&&& \\\Aut{F/\Q} & A^{(1)}_6 & C^{(1)}_6/A^{(2)}_{2\cdot 6}/D^{(2)}_{6+1} & B^{(1)}_6/A^{(2)}_{2\cdot 6-1} & D^{(1)}_6 & E^{(1)}_6& {\rm Total} \\\hline
1			& 360 	& 2\,520 \times 2	& 2\,520 \times 2	& 630 	& 840 	& 11\,910 \\
\Z_7			& 54 	& 360 \times 2	& 360 \times 2	& 90 	& 120 	& 1\,704
\\\hline
{\rm rank}=7 &&&&&& \\\Aut{F/\Q} & A^{(1)}_7 & C^{(1)}_7/A^{(2)}_{2\cdot 7}/D^{(2)}_{7+1} & B^{(1)}_7/A^{(2)}_{2\cdot 7-1} & D^{(1)}_7 & E^{(1)}_7& {\rm Total} \\\hline
1			& 2\,520 	& 20\,160 \times 2	& 20\,160 \times 2	& 5\,040 	& 20\,160 	& 108\,360 \\
\Z_8			& 332 	& 2\,544 \times 2	& 2\,520 \times 2	& 642 	& 2\,520 	& 13\,622 \\
\Z_2^3			& 420 	& 2\,688 \times 2	& 2\,520 \times 2	& 714 	& 2\,520 	& 14\,070 \\
\Z_2\oplus\Z_4		& 362 	& 2\,592 \times 2	& 2\,520 \times 2	& 666 	& 2\,520 	& 13\,772 \\
Q			& 333 	& 2\,544 \times 2	& 2\,520 \times 2	& 642 	& 2\,520 	& 13\,623 \\
D_4			& 391 	& 2\,640 \times 2	& 2\,520 \times 2	& 690 	& 2\,520 	& 13\,921
\\\hline
{\rm rank}=8 &&&&&& \\\Aut{F/\Q} & A^{(1)}_8 & C^{(1)}_8/A^{(2)}_{2\cdot 8}/D^{(2)}_{8+1} & B^{(1)}_8/A^{(2)}_{2\cdot 8-1} & D^{(1)}_8 & E^{(1)}_8& {\rm Total} \\\hline
1			& 20\,160 	& 181\,440 \times 2	& 181\,440 \times 2	& 45\,360 	& 362\,880 	& 1\,154\,160 \\
\Z_9			& 2\,246 	& 20\,160 \times 2	& 20\,160 \times 2	& 5\,040 	& 40\,320 	& 128\,246 \\
\Z_3^2			& 2\,256 	& 20\,160 \times 2	& 20\,160 \times 2	& 5\,040 	& 40\,320 	& 128\,256
\\\hline
{\rm rank}=9 &&&&&& \\\Aut{F/\Q} & A^{(1)}_9 & C^{(1)}_9/A^{(2)}_{2\cdot 9}/D^{(2)}_{9+1} & B^{(1)}_9/A^{(2)}_{2\cdot 9-1} & D^{(1)}_9 && {\rm Total} \\\hline
1			& 181\,440 	& 1\,814\,400 \times 2	& 1\,814\,400 \times 2	& 453\,600 	&& 7\,892\,640 \\
\Z_{10}			& 18\,264 	& 181\,632 \times 2	& 181\,440 \times 2	& 45\,456 	&& 789\,864 \\
D_5			& 18\,724 	& 182\,400 \times 2	& 181\,440 \times 2	& 45\,840 	&& 792\,244
\\\hline
%{\rm rank}=10 &&&&&& \\\Aut{F/\Q} & A^{(1)}_10 & C^{(1)}_10/A^{(2)}_{2\cdot 10}/D^{(2)}_{10+1} & B^{(1)}_10/A^{(2)}_{2\cdot 10-1} & D^{(1)}_10 && {\rm Total} \\\hline
%1			& 1\,814\,400 

%% file: rs.tex
{
\newcommand{\disc}[0]{{\circle*{.5}}}
\def\Z{{\mathbf Z}}

\def\Ars{
\begin{picture}(10,4)(-1,-.5)
%\thicklines
\allinethickness{.5mm}
\multiput(0,0)(2,0){2}{\disc}
\multiput(6,0)(2,0){2}{\disc}
\put(4,2.5){\disc}
\put(.25,0){\line(1,0){1.5}}
\put(6.25,0){\line(1,0){1.5}}
\put(.25,0){\line(1,0){1.5}}
\put(.25,0){\line(3,2){3.6}}
\put(7.75,0){\line(-3,2){3.6}}
\multiput(2.25,0)(.4,0){9}{\line(1,0){.2}}

\put(-.1,0.4){$1$} \put(-.1,-.8){$1$}
\put(1.9,0.4){$2$} \put(1.9,-.8){$1$}
\put(5.3,0.4){$n-1$} \put(5.9,-.8){$1$}
\put(7.9,0.4){$n$} \put(7.9,-.8){$1$}
\put(3.3,2.9){$n+1$} \put(3.9,1.7){$1$}
\end{picture}
}

\def\Brs{
\begin{picture}(10,4)(-1,-.5)
\allinethickness{.5mm}
\put(0,0){\disc}
\put(0,2){\disc}
\put(2,1){\disc}
\put(6,1){\disc}
\put(8,1){\disc}
\multiput(2.25,1)(.4,0){9}{\line(1,0){.2}}
\put(6.25,.9){\line(1,0){1.5}}
\put(6.25,1.1){\line(1,0){1.5}}
\put(6.75,1.5){\line(1,-1){.5}}
\put(6.75,.5){\line(1,1){.5}}
\put(0,0){\line(2,1){1.8}}
\put(0,2){\line(2,-1){1.8}}

\put(-.6,0.4){$n+1$} \put(-.1,-.8){$1$}
\put(-.1,2.4){$1$} \put(-.1,1.2){$1$}
\put(1.9,1.4){$2$} \put(1.9,.2){$2$}
\put(5.2,1.4){$n-1$} \put(5.9,.2){$2$}
\put(7.9,1.4){$n$} \put(7.9,.2){$2$}
\end{picture}
}

\def\Crs{
\begin{picture}(10,2)(-1,-.5)
\allinethickness{.5mm}
\put(0,0){\disc}
\put(2,0){\disc}
\put(6,0){\disc}
\put(8,0){\disc}
\multiput(2.25,0)(.4,0){9}{\line(1,0){.2}}
\put(6.25,-.1){\line(1,0){1.5}}
\put(6.25,.1){\line(1,0){1.5}}
\put(7.25,.5){\line(-1,-1){.5}}
\put(7.25,-.5){\line(-1,1){.5}}
\put(.25,-.1){\line(1,0){1.5}}
\put(.25,.1){\line(1,0){1.5}}
\put(.75,.5){\line(1,-1){.5}}
\put(.75,-.5){\line(1,1){.5}}

\put(-.6,0.4){$n+1$} \put(-.1,-.8){$1$}
\put(1.9,.4){$1$} \put(1.9,-.8){$2$}
\put(5.3,.4){$n-1$} \put(5.9,-.8){$2$}
\put(7.9,.4){$n$} \put(7.9,-.8){$1$}

\end{picture}
}

\def\Drs{
\begin{picture}(10,4)(-1,-.5)
\allinethickness{.5mm}
\put(0,0){\disc}
\put(0,2){\disc}
\put(2,1){\disc}
\put(6,1){\disc}
\put(8,0){\disc}
\put(8,2){\disc}
\multiput(2.25,1)(.4,0){9}{\line(1,0){.2}}
\put(6,1){\line(2,1){1.8}}
\put(6,1){\line(2,-1){1.8}}
\put(0,0){\line(2,1){1.8}}
\put(0,2){\line(2,-1){1.8}}

\put(-.6,0.4){$n+1$} \put(-.1,-.8){$1$}
\put(-.1,2.4){$1$} \put(-.1,1.2){$1$}
\put(1.9,1.4){$2$} \put(1.9,.2){$2$}
\put(5.3,1.4){$n-2$} \put(5.9,.2){$2$}
\put(7.3,2.4){$n-1$} \put(7.9,1.2){$1$}
\put(7.9,.4){$n$} \put(7.9,-.8){$1$}
\end{picture}
}

\def\Grs{
\begin{picture}(10,2)(-1,.5)
\allinethickness{.5mm}
\put(0,1){\disc}
\put(2,1){\disc}
\put(4,1){\disc}
\put(.7,1){\line(1,1){.5}}
\put(.7,1){\line(1,-1){.5}}

\put(0,1){\line(1,0){4}}
\put(0,1.2){\line(1,0){2}}
\put(0,.8){\line(1,0){2}}

\put(-.1,1.4){$2$} \put(-.1,.2){$3$}
\put(1.9,1.4){$1$} \put(1.9,.2){$2$}
\put(3.9,1.4){$3$} \put(3.9,.2){$1$}
\end{picture}
}

\def\Ee{
{\setlength{\unitlength}{.5cm}
\begin{picture}(10,6)(-1,-.5)
\allinethickness{.5mm}
\put(0,0){\disc}
\put(2,0){\disc}
\put(4,0){\disc}
\put(6,0){\disc}
\put(8,0){\disc}
\put(4,2){\disc}
\put(4,4){\disc}
\put(0,0){\line(1,0){8}}
\put(4,0){\line(0,1){4}}

\put(-.1,.4){$1$} \put(-.1,-1.0){$1$}
\put(1.9,.4){$3$} \put(1.9,-1.0){$2$}
\put(3.5,.4){$4$} \put(4.1,-1.0){$3$}
\put(5.9,0.4){$5$} \put(5.9,-1.0){$2$}
\put(7.9,0.4){$6$} \put(7.9,-1.0){$1$}
\put(3.5,2.4){$2$} \put(4.1,1.1){$2$}
\put(3.6,4.4){$7$} \put(4.1,3.1){$1$}
\end{picture}
}
}

\def\Eee{
{\setlength{\unitlength}{.5cm}
\begin{picture}(12,4)(-1,-.5)
\allinethickness{.5mm}
\put(0,0){\disc}
\put(2,0){\disc}
\put(4,0){\disc}
\put(6,0){\disc}
\put(8,0){\disc}
\put(10,0){\disc}
\put(12,0){\disc}
\put(6,2){\disc}
\put(0,0){\line(1,0){12}}
\put(6,0){\line(0,1){2}}

\put(-.1,.4){$8$} \put(-.1,-.9){$1$}
\put(1.9,.4){$1$} \put(1.9,-.9){$2$}
\put(3.9,.4){$3$} \put(3.9,-.9){$3$}
\put(5.6,0.4){$4$} \put(6.1,-.9){$4$}
\put(7.9,0.4){$5$} \put(7.9,-.9){$3$}
\put(9.9,0.4){$6$} \put(9.9,-.9){$2$}
\put(11.9,0.4){$7$} \put(11.9,-.9){$1$}
\put(5.6,2.4){$2$} \put(6.1,1.1){$2$}
\end{picture}
}
}

\def\Eeee{
{\setlength{\unitlength}{.5cm}
\begin{picture}(12,4)(-1,-.5)
\allinethickness{.5mm}
\put(0,0){\disc}
\put(2,0){\disc}
\put(4,0){\disc}
\put(6,0){\disc}
\put(8,0){\disc}
\put(10,0){\disc}
\put(12,0){\disc}
\put(14,0){\disc}
\put(4,2){\disc}
\put(0,0){\line(1,0){14}}
\put(4,0){\line(0,1){2}}

\put(-.1,.4){$1$} \put(-.1,-.9){$2$}
\put(1.9,.4){$3$} \put(1.9,-.9){$4$}
\put(3.6,.4){$4$} \put(4.1,-.9){$6$}
\put(5.9,0.4){$5$} \put(5.9,-.9){$5$}
\put(7.9,0.4){$6$} \put(7.9,-.9){$4$}
\put(9.9,0.4){$7$} \put(9.9,-.9){$3$}
\put(11.9,0.4){$8$} \put(11.9,-.9){$2$}
\put(13.9,0.4){$9$} \put(13.9,-.9){$1$}
\put(3.6,2.4){$2$} \put(4.1,1.1){$3$}
\end{picture}
}
}

\def\Ff{
{%\setlength{\unitlength}{.8cm}
\begin{picture}(10,2)(-1,0)
\allinethickness{.5mm}
\put(0,0){\disc}
\put(2,0){\disc}
\put(4,0){\disc}
\put(6,0){\disc}
\put(8,0){\disc}
\put(0,0){\line(1,0){4}}
\put(6,0){\line(1,0){2}}
\put(4,.1){\line(1,0){2}}
\put(4,-.1){\line(1,0){2}}

\put(4.75,.5){\line(1,-1){.5}}
\put(4.75,-.5){\line(1,1){.5}}

\put(-.1,0.4){$5$} \put(-.1,-.7){$1$}
\put(1.9,0.4){$1$} \put(1.9,-.7){$2$}
\put(3.9,0.4){$2$} \put(3.9,-.7){$3$}
\put(5.9,0.4){$3$} \put(5.9,-.7){$4$}
\put(7.9,0.4){$4$} \put(7.9,-.7){$2$}
\end{picture}
}
}

\def\Aa{
{%\setlength{\unitlength}{.8cm}
\begin{picture}(10,2)(-1,0)
\allinethickness{.5mm}
\put(0,0){\disc}
\put(2,0){\disc}
\put(0,-.25){\line(1,0){2}}
\put(0,-.1){\line(1,0){2}}
\put(0,.1){\line(1,0){2}}
\put(0,.25){\line(1,0){2}}
\put(1.25,.5){\line(-1,-1){.5}}
\put(1.25,-.5){\line(-1,1){.5}}

\put(-.1,0.4){$1$} \put(-.1,-.8){$1$}
\put(1.9,0.4){$2$} \put(1.9,-.8){$2$}
\end{picture}
}
}

\def\Aon{
{%\setlength{\unitlength}{.8cm}
\begin{picture}(10,1)(-1,0)
\allinethickness{.5mm}
\put(0,0){\disc}
\put(2,0){\disc}
\put(0,-.25){\line(1,0){2}}
\put(0,-.1){\line(1,0){2}}
\put(0,.1){\line(1,0){2}}
\put(0,.25){\line(1,0){2}}

\put(-.1,0.4){$1$} \put(-.1,-.8){$1$}
\put(1.9,0.4){$2$} \put(1.9,-.8){$1$}
\end{picture}
}
}

\def\Aaa{
\begin{picture}(10,3)(-1,-.5)
\allinethickness{.5mm}
\put(0,0){\disc}
\put(2,0){\disc}
\put(6,0){\disc}
\put(8,0){\disc}
\multiput(2.25,0)(.4,0){9}{\line(1,0){.2}}
\put(6.25,-.1){\line(1,0){1.5}}
\put(6.25,.1){\line(1,0){1.5}}
\put(7.25,.5){\line(-1,-1){.5}}
\put(7.25,-.5){\line(-1,1){.5}}
\put(.25,-.1){\line(1,0){1.5}}
\put(.25,.1){\line(1,0){1.5}}
\put(1.25,.5){\line(-1,-1){.5}}
\put(1.25,-.5){\line(-1,1){.5}}

\put(-.1,0.4){$1$} \put(-.1,-.8){$2$}
\put(1.9,.4){$2$} \put(1.9,-.8){$2$}
\put(5.9,.4){$n$} \put(5.9,-.8){$2$}
\put(7.4,.4){$n+1$} \put(7.9,-.8){$1$}

\end{picture}
}

\def\Aaaa{
\begin{picture}(10,4)(-1,-.5)
\allinethickness{.5mm}
\put(0,0){\disc}
\put(0,2){\disc}
\put(2,1){\disc}
\put(6,1){\disc}
\put(8,1){\disc}
\multiput(2.25,1)(.4,0){9}{\line(1,0){.2}}
\put(6.25,.9){\line(1,0){1.5}}
\put(6.25,1.1){\line(1,0){1.5}}
\put(7.25,1.5){\line(-1,-1){.5}}
\put(7.25,.5){\line(-1,1){.5}}
\put(0,0){\line(2,1){1.8}}
\put(0,2){\line(2,-1){1.8}}

\put(-.6,0.4){$n+1$} \put(-.1,-.8){$1$}
\put(-.1,2.4){$1$} \put(-.1,1.2){$1$}
\put(1.9,1.4){$2$} \put(1.9,.2){$2$}
\put(5.2,1.4){$n-1$} \put(5.9,.2){$2$}
\put(7.9,1.4){$n$} \put(7.9,.2){$1$}
\end{picture}
}

\def\Ddrs{
\begin{picture}(10,2.5)(-1,-.5)
\allinethickness{.5mm}
\put(0,0){\disc}
\put(2,0){\disc}
\put(6,0){\disc}
\put(8,0){\disc}
\multiput(2.25,0)(.4,0){9}{\line(1,0){.2}}
\put(6.25,-.1){\line(1,0){1.5}}
\put(6.25,.1){\line(1,0){1.5}}
\put(6.75,.5){\line(1,-1){.5}}
\put(6.75,-.5){\line(1,1){.5}}
\put(.25,-.1){\line(1,0){1.5}}
\put(.25,.1){\line(1,0){1.5}}
\put(1.25,.5){\line(-1,-1){.5}}
\put(1.25,-.5){\line(-1,1){.5}}

\put(-.1,0.4){$1$} \put(-.1,-.8){$1$}
\put(1.9,.4){$2$} \put(1.9,-.8){$1$}
\put(5.9,.4){$n$} \put(5.9,-.8){$1$}
\put(7.4,.4){$n+1$} \put(7.9,-.8){$1$}

\end{picture}
}

\def\Etwo{
{%\setlength{\unitlength}{.8cm}
\begin{picture}(10,2)(-1,0)
\allinethickness{.5mm}
\put(0,0){\disc}
\put(2,0){\disc}
\put(4,0){\disc}
\put(6,0){\disc}
\put(8,0){\disc}
\put(0,0){\line(1,0){4}}
\put(6,0){\line(1,0){2}}
\put(4,.1){\line(1,0){2}}
\put(4,-.1){\line(1,0){2}}

\put(5.25,.5){\line(-1,-1){.5}}
\put(5.25,-.5){\line(-1,1){.5}}

\put(-.1,0.4){$5$} \put(-.1,-.8){$1$}
\put(1.9,0.4){$1$} \put(1.9,-.8){$2$}
\put(3.9,0.4){$2$} \put(3.9,-.8){$3$}
\put(5.9,0.4){$3$} \put(5.9,-.8){$2$}
\put(7.9,0.4){$4$} \put(7.9,-.8){$1$}
\end{picture}
}
}

\def\Dfour{
\begin{picture}(10,3)(-1,.5)
\allinethickness{.5mm}
\put(0,1){\disc}
\put(2,1){\disc}
\put(4,1){\disc}
\put(1.25,1){\line(-1,1){.5}}
\put(1.25,1){\line(-1,-1){.5}}

\put(0,1){\line(1,0){4}}
\put(0,1.2){\line(1,0){2}}
\put(0,.8){\line(1,0){2}}

\put(-.1,1.4){$1$} \put(-.1,.2){$1$}
\put(1.9,1.4){$2$} \put(1.9,.2){$2$}
\put(3.9,1.4){$3$} \put(3.9,.2){$1$}
\end{picture}
}

\def\ExpPic{
\begin{picture}(2,2)(0,0)
\put(0,0){\disc}
\put(-1.5,.4){root number} \put(-.9,-.8){weight}
\end{picture}
}

\def\lhs{{Root \ System}}
\def\mhs{{Dynkin Diagram \hfill \ExpPic}}
\def\rhs{{Automorphism Group}}

\setlength{\unitlength}{.55cm}

\begin{table}[!ht]
\begin{tabular}{p{1.2cm}p{8cm}p{2cm}}
\lhs & \mhs & \rhs \\\hline
$A^{(1)}_1$ & \Aon & $\Z_2$\\
$A^{(1)}_n$ & \Ars & $D_{n+1}$\hfill $(n\geq 2)$\\
$B^{(1)}_n$ & \Brs & $\Z_2$\hfill $(n \geq 3)$\\
$C^{(1)}_n$ & \Crs & $\Z_2$\hfill $(n \geq 2)$\\
$D^{(1)}_n$ & \Drs & \vspace{-6mm}$S_4$\hfill$(n=4)$\newline$\Z_2^3$\hfill $(n > 4)$\\
$G^{(1)}_2$ & \Grs & $1$\\
$F^{(1)}_4$ & \Ff  & $1$\\
$E^{(1)}_6$ & \Ee  & $D_3$\\
$E^{(1)}_7$ & \Eee & $\Z_2$\\
$E^{(1)}_8$ & \Eeee & $1$\\
\end{tabular}
\caption{Root systems, their Dynkin diagrams and automorphism
  groups. Symmetries are indicated by arrows. $D_n$ is the
  symmetry group of a regular $n$-gon.}
\label{fig:rs1}
\end{table}

\begin{table}[htb]
\begin{tabular}{p{1.2cm}p{8cm}p{2cm}}
\lhs & \mhs & \rhs \\\hline &&\\
$A^{(2)}_2$      & \Aa & $1$\\
$A^{(2)}_{2n}$   & \Aaa & $\Z_2$ \hfill(see text, $n \geq 2$)\\
$A^{(2)}_{2n-1}$ & \Aaaa & $\Z_2$ \hfill ($n\geq 3$)\\
$D^{(2)}_{n+1}$  & \Ddrs & $\Z_2$  \hfill ($n\geq 2$)\\
$E^{(2)}_6$      & \Etwo & $1$\\
$D^{(3)}_4$      & \Dfour & $1$\\
\end{tabular}
\caption{Root systems, their Dynkin diagrams and automorphism
  groups. The shortest roots of $D^{(2)}_{n+1}$ and $A^{(2)}_{2n}$
  have length $1/\sqrt{2}$; all other root systems' shortest roots
  have unit length. The longest root(s) of $G^{(1)}_2$ and $D^{(3)}_4$
  have length $\sqrt{3}$; all other root systems' longest roots have
  length $\sqrt{2}$.}
\label{fig:rs23}
\end{table}
}

%% file: toda-c.bbl
\begin{thebibliography}{aa}

\bibitem{AvM1}
  M. Adler and P. van Moerbeke.
  \newblock Completely integrable systems, Euclidean Lie algebras, and curves. 
  \newblock {\em Adv. in Math.} 38(3):267--317 (1980).

\bibitem{AvM2}
  M. Adler and P. van Moerbeke.
  \newblock Linearization of Hamiltonian systems, Jacobi varieties and representation theory. 
  \newblock {\em Adv. in Math.} 38(3):318--379 (1980).

\bibitem{AvM3}
  M. Adler and P. van Moerbeke.
  \newblock   Kowalewski's asymptotic method, Kac-Moody Lie algebras and regularization.
  \newblock {\em Comm. Math. Phys.} 83(1):83--106 (1982).

\bibitem{Bogo}
  O.~I. Bogoyavlenskij, 
  \newblock On perturbations of the periodic Toda lattice.
  \newblock {\em Comm. Math. Phys.} 51(3):201--209 (1976).

\bibitem{BolsinovJovanovic:2001a}
  A.~V. Bolsinov and B. \u{I}ovanovich. 
  \newblock Integrable geodesic flows on homogeneous spaces. 
  \newblock {\em Mat. Sb.} 192(7):21--40 (2001); translation in Sb. Math. 192(7-8):951--968 (2001).

\bibitem{BT:2000a}
  A.~V. Bolsinov and I.~A. Taimanov.
  \newblock Integrable geodesic flows with positive topological entropy.
  \newblock {\em Invent. Math.}, 140(3):639--650 (2000).

\bibitem{BT:2000b}
  A.~V. Bolsinov and I.~A. Taimanov.
  \newblock Integrable geodesic flows on suspensions of automorphisms of
  tori. 
  \newblock (Russian) {\em Tr. Mat. Inst. Steklova} 231 (2000), Din. Sist., Avtom. i Beskon. Gruppy,
  46--63; 
  \newblock translation in {\em Proc. Steklov Inst. Math.}
  231(4):42--58 (2000).

\bibitem{Boyd}
  D.~W. Boyd.
  \newblock Reciprocal Polynomials having Small Measure.
  \newblock {\em Math. Comp.}, 53(187):355-357 (1989).

\bibitem{BG}
  A. Bloch and M. Gekhtman.
  \newblock   Hamiltonian and gradient structures in the Toda flows.
  \newblock   J. Geom. Phys. 27(3-4):230--248 (1998).

\bibitem{B5}
  L. Butler.
  \newblock Invariant fibrations of geodesic flows.
  \newblock {\em Topology}, 44(4):769--789 (2005).

\bibitem{B7}
  L. Butler.
  \newblock Positive-entropy integrable systems and the Toda lattice.
  \newblock {\em Inv. Math.}, 158(3):515--549 (2004).

\bibitem{B8}
  L. Butler.
  \newblock  The Maslov cocycle, smooth structures and real-analytic complete integrability.
  \newblock {\em Am. J. Math.} to appear. \href{http://arxiv.org/abs/0708.3157v2}{arxiv:0708.3157v2}

\bibitem{DD}
  P. Dazord and T. Delzant.
  \newblock Le probleme general des variables actions-angles.
  \newblock {\em J. Differential Geom.} 26(2):223--251 (1987).

\bibitem{Dinaburg}
  E.~I. Dinaburg.
  \newblock A connection between various entropy characterizations of
  dynamical systems.
  \newblock {\em Izv. Akad. Nauk SSSR Ser. Mat.}  35  1971 324--366.

\bibitem{FM1}
  J.~A. Foxman and J.~M. Robbins.
  \newblock Singularities, Lax degeneracies and Maslov indices of the
  periodic Toda chain.
  \newblock {\em Nonlinearity} 18(6):2795--2813 (2005).

\bibitem{FM2}
  J.~A. Foxman and J.~M. Robbins
  \newblock The Maslov index and nondegenerate singularities of
  integrable systems.
  \newblock Nonlinearity 18(6):2775--2794 (2005).

\bibitem{FH}
  W. Fulton and J. Harris
  \newblock Representation theory. A first course.
  \newblock Graduate Texts in Mathematics, 129. Springer-Verlag, New York, 1991. 

\bibitem{GAP}
  The GAP group.
  \newblock {\em GAP --- Groups, Algorithms and Programming},
  \newblock Version 4.4.6; (2005). \url{http://www.gap-system.org}

\bibitem{Gelfond}
  A.~O. Gelfond,
  \newblock Transcendental and algebraic numbers. 
  \newblock Dover, New York 1960. trans. Leo F. Boron. 

\bibitem{KKS}
  A. Katok, S. Katok and K. Schmidt.
  \newblock Rigidity of measurable structure for ${\mathbf Z}\sp d$-actions by automorphisms of a torus.
  \newblock {\em Comment. Math. Helv.} 77(4):718--745 (2002).

\bibitem{Kostant}
  B. Kostant.
  \newblock The solution to a generalized {T}oda lattice and representation theory,
  \newblock {\em Adv. in Math.} 34(3):195--338 (1979).

\bibitem{Kozbook} 
  V.~V. Kozlov,
  \newblock {\em Symmetries, topology and resonances in Hamiltonian
    mechanics.} 
  \newblock  Translated from the Russian manuscript by S. V. Bolotin,
  D. Treshchev and Yuri Fedorov.
  \newblock Ergebnisse der Mathematik und ihrer
  Grenzgebiete (3), 31. Springer-Verlag, Berlin, 1996.

\bibitem{KT}
  V.~V. Kozlov and D.~V. Treshchev.
  \newblock Polynomial integrals of Hamiltonian systems with exponential
  interaction.
  \newblock (Russian) {\em Izv. Akad. Nauk SSSR Ser. Mat.} 53(3):537--556 (1989); 
  \newblock translation in {\em Math. USSR-Izv.} 34(3):555--574 (1990).

\bibitem{Maxima}
  Maxima.sourceforge.net.
  \newblock {\em Maxima, A Computer Algebra System.} Version 5.18.1
  \newblock \url{http://maxima.sourceforge.net/}

\bibitem{Mossinghoff}
  M.~J. Mossinghoff.
  \newblock Polynomials with Small Mahler Measure Polynomials with Small Mahler Measure.
  \newblock {\em Math. Comp.}, 67(224):1697--1705 (1998).

\bibitem{Raghunathan}
  M.~S. Raghunathan.
  \newblock {\em Discrete subgroups of Lie groups.} 
  \newblock Ergebnisse der Mathematik und ihrer Grenzgebiete, Band
  68. Springer-Verlag, New York-Heidelberg, 1972.

\bibitem{RSTS}
  A.~G. Reyman and A. Semenov-Tian-Shansky, 
  \newblock Group-Theoretical Methods in the Theory of Finite-Dimensional
  Integrable Systems,
  \newblock in {\em Dynamical systems. VII. Integrable systems,
    nonholonomic dynamical systems}. Encyclopaedia of Mathematical
  Sciences, 16. Springer-Verlag, Berlin, 1994.

\bibitem{Schwartzman}
  S. Schwartzman.
  \newblock Asymptotic cycles. 
  \newblock {\em Ann. of Math.} 66(2):270--284 (1957). 

\bibitem{Wall}
  C.~T.~C. Wall,
  \newblock {\em Surgery on compact manifolds.} Second edition. 
  \newblock Edited and with a foreword by A. A. Ranicki. Mathematical Surveys     
  and Monographs, 69. American Mathematical Society, Providence, RI, 1999.

\end{thebibliography}
